\documentclass[]{ant_tech_report}

\usepackage{latexsym}
\usepackage[utf8]{inputenc}
\usepackage{inconsolata}
\ifdefined\pdfimageresolution
\fi
\usepackage{siunitx}
\usepackage{url}
\usepackage{multicol}
\usepackage{colortbl}
\usepackage{wrapfig}
\usepackage{makecell}
\usepackage{adjustbox}
\usepackage{fancyvrb}
\usepackage{fvextra}
\usepackage{soul}
\usepackage{float}
\usepackage{tikz}
\usepackage{fontawesome5}
\titleformat*{\paragraph}{\bfseries\itshape}

\usepackage[utf8]{inputenc}
\usepackage[T1]{fontenc}
\usepackage{booktabs}
\usepackage{longtable}
\usepackage{amsfonts}
\usepackage{nicefrac}
\usepackage{microtype}
\usepackage{xcolor}
\usepackage{graphicx}
\usepackage{tikz}
\usepackage{subcaption}
\usepackage{xspace}
\usepackage{enumitem}
\usepackage{amsmath}
\usepackage{amssymb}
\usepackage{pifont}
\usepackage[ruled,linesnumbered]{algorithm2e}
\usepackage{listings}
\usepackage{tcolorbox}
\tcbuselibrary{breakable,listings}
\usepackage{textcomp}
\usepackage{upquote}
\usepackage{multirow}
\usepackage{colortbl}
\usepackage{makecell}
\usetikzlibrary{arrows.meta,positioning,fit}

\usepackage{array}
\newcolumntype{L}[1]{>{\raggedright\arraybackslash}p{#1}}
\newcolumntype{C}[1]{>{\centering\arraybackslash}p{#1}}
\newcolumntype{R}[1]{>{\raggedleft\arraybackslash}p{#1}}

\definecolor{forgeblue}{HTML}{E3F2FD}
\definecolor{forgelight}{HTML}{F5F7FA}
\definecolor{forgepurple}{HTML}{7E57C2}
\definecolor{forgegreen}{HTML}{43A047}
\definecolor{forgered}{HTML}{FB8C00}
\definecolor{forgegray}{HTML}{E0E0E0}
\definecolor{forgeprimary}{HTML}{1565C0}
\definecolor{forgeprimarydark}{HTML}{0D47A1}
\definecolor{forgecyan}{HTML}{00ACC1}
\definecolor{oursrowcolor}{HTML}{E3F2FD}
\definecolor{baselinerowcolor}{HTML}{F5F7FA}

\newcommand{\ourmethod}{\textsc{MobileForge}\xspace}
\newcommand{\mobilegym}{\textsc{MobileGym}\xspace}
\newcommand{\curriculum}{\textsc{MobileGym-Curriculum}\xspace}
\newcommand{\critic}{\textsc{MobileGym-Critic}\xspace}
\newcommand{\hifpo}{HiFPO\xspace}
\DeclareRobustCommand{\llmname}[1]{{\fontfamily{pcr}\selectfont #1}}
\newcommand{\rise}[1]{\textcolor{forgegreen}{#1}}
\newcommand{\drop}[1]{\textcolor{forgered}{#1}}

\setlist[itemize]{leftmargin=*, topsep=2pt, itemsep=1pt}
\setlist[enumerate]{leftmargin=*, topsep=2pt, itemsep=1pt}

\newcommand{\paperTable}{%
  \small
  \setlength{\tabcolsep}{4.5pt}%
  \renewcommand{\arraystretch}{1.12}%
}
\newcommand{\paperCompactTable}{%
  \footnotesize
  \setlength{\tabcolsep}{3.8pt}%
  \renewcommand{\arraystretch}{1.10}%
}

\newenvironment{paperFit}[1][\linewidth]
  {\begin{adjustbox}{max width=#1}}
  {\end{adjustbox}}
\newcommand{\paperfigwide}{0.98\linewidth}
\newcommand{\paperfigmain}{0.94\linewidth}


\newcommand{\obsbox}[1]{%
  \begin{tcolorbox}[colframe=black!50, colback=forgeblue!45, boxrule=1pt, arc=2mm,
    top=4pt, bottom=4pt, left=6pt, right=6pt, boxsep=2pt, fontupper=\itshape]
    #1
  \end{tcolorbox}
}

\newcounter{insight}
\newcommand{\insight}[1]{%
  \par
  \refstepcounter{insight}%
  \addvspace{0.35em}%
  \noindent
  \begingroup
  \setlength{\fboxsep}{0pt}%
  \makebox[\linewidth][c]{%
    \colorbox{forgeblue}{%
      \parbox{0.97\linewidth}{%
        \vspace{0.35em}
        \hspace{0.65em}%
        \textcolor{forgeprimary}{\rule{1.4pt}{1.05em}}%
        \hspace{0.55em}%
        \parbox[t]{0.88\linewidth}{%
          \textbf{\textsc{Insight}~\theinsight.}~#1%
        }%
        \vspace{0.35em}
      }%
    }%
  }%
  \endgroup
  \par\addvspace{0.35em}%
}

\newtcolorbox[auto counter]{promptbox}[1][]{
    top=15pt,
    bottom=15pt,
    left=20pt,
    right=20pt,
    colback=gray!5,
    colframe=black,
    fonttitle=\bfseries,
    coltitle=white,
    title=Prompt~\thetcbcounter: #1,
    breakable,
    fontupper=\ttfamily\small,
}

\newtcblisting[auto counter]{promptlistingbox}[1][]{
    top=15pt,
    bottom=15pt,
    left=20pt,
    right=20pt,
    colback=gray!5,
    colframe=black,
    fonttitle=\bfseries,
    coltitle=white,
    title=Prompt~\thetcbcounter: #1,
    breakable,
    listing only,
    listing options={
        basicstyle=\ttfamily\footnotesize,
        breaklines=true,
        breakatwhitespace=false,
        columns=fullflexible,
        keepspaces=true,
        showstringspaces=false,
        upquote=true,
        literate=*{_}{\_}{1}
    }
}

\newcommand{\correspenvelope}{%
  \textnormal{\raisebox{0.02ex}{\scalebox{0.84}{\textcolor{black!70}{\faIcon[regular]{envelope}}}}}%
}

\renewcommand{\authorfont}{\fontsize{10.8}{11.8}\selectfont}

\newenvironment{paperresources}
  {\par\noindent\begin{minipage}{0.92\textwidth}
    \centering\footnotesize\sffamily\color{black!82}
    \setlength{\parindent}{0pt}\setlength{\parskip}{2pt}}
  {\end{minipage}\par\vspace{1.7mm}}
\newcommand{\resourceprojecticon}{\textcolor{antblue}{\faGlobe}}
\newcommand{\resourcegithubicon}{\textcolor{antblue}{\faGithub}}
\newcommand{\resourcehficon}{%
  \raisebox{0.3ex}{%
    \tikz[baseline=-0.5ex,scale=0.105]{%
      \fill[antblue] (0,0) circle (1);
      \fill[antblue] (-0.8,0.68) circle (0.34);
      \fill[antblue] (0.8,0.68) circle (0.34);
      \fill[white] (-0.34,0.22) circle (0.11);
      \fill[white] (0.34,0.22) circle (0.11);
      \draw[white,line width=0.12pt,line cap=round] (-0.36,-0.26) .. controls (0,-0.55) .. (0.36,-0.26);
    }%
  }%
}
\newcommand{\paperresourceicon}[4]{%
  #1\ \textbf{#2:}\ \href{#3}{{\ttfamily\fontsize{8}{9}\selectfont #4}}%
}

\renewcommand{\titlefont}{\fontsize{14.4}{17}\selectfont}
\title{MobileForge: Annotation-Free Adaptation for Mobile GUI Agents with Hierarchical Feedback-Guided Policy Optimization}

\author{%
\parbox{\textwidth}{\centering
Guangyi Liu$^{1,2,*}$,
Pengxiang Zhao$^{1,2,*}$,
Gao Wu$^{1,2,*}$,
Yiwen Yin$^{2,3,*}$,
Mading Li$^{2,\dagger}$
Liang Liu$^{1}$,\\[1mm]
Congxiao Liu$^{1,2}$,
Zhang Qi$^{2}$,
Mengyan Wang$^{2}$,
Liang Guo$^{2}$,
Jiangning Zhang$^{1,\correspenvelope}$, 
Yong Liu$^{1,\correspenvelope}$ \\
}}

\affiliation{%
\parbox{\textwidth}{\centering\small
$^1$Zhejiang University \quad $^2$Kuaishou Technology \quad $^3$Tsinghua University
}}

\renewcommand{\contributionlist}{\vspace{-2mm}}
\newcommand{\authorfootnotes}{%
{\scriptsize
\begin{tabular}{@{}l@{}}
\rule{47mm}{0.4pt}\\[-0.2ex]
$^*$Equal contribution.\quad $^\dagger$Project lead.\\[-0.2ex]
 \correspenvelope Corresponding author.
\end{tabular}}}
\fancypagestyle{firststyle}{
    \fancyhf{}
    \fancyhead[R]{
        \raisebox{-12.5mm}[0pt][0pt]{%
            \vbox{\vskip 3mm\hbox{%
                \includegraphics[height=12mm]{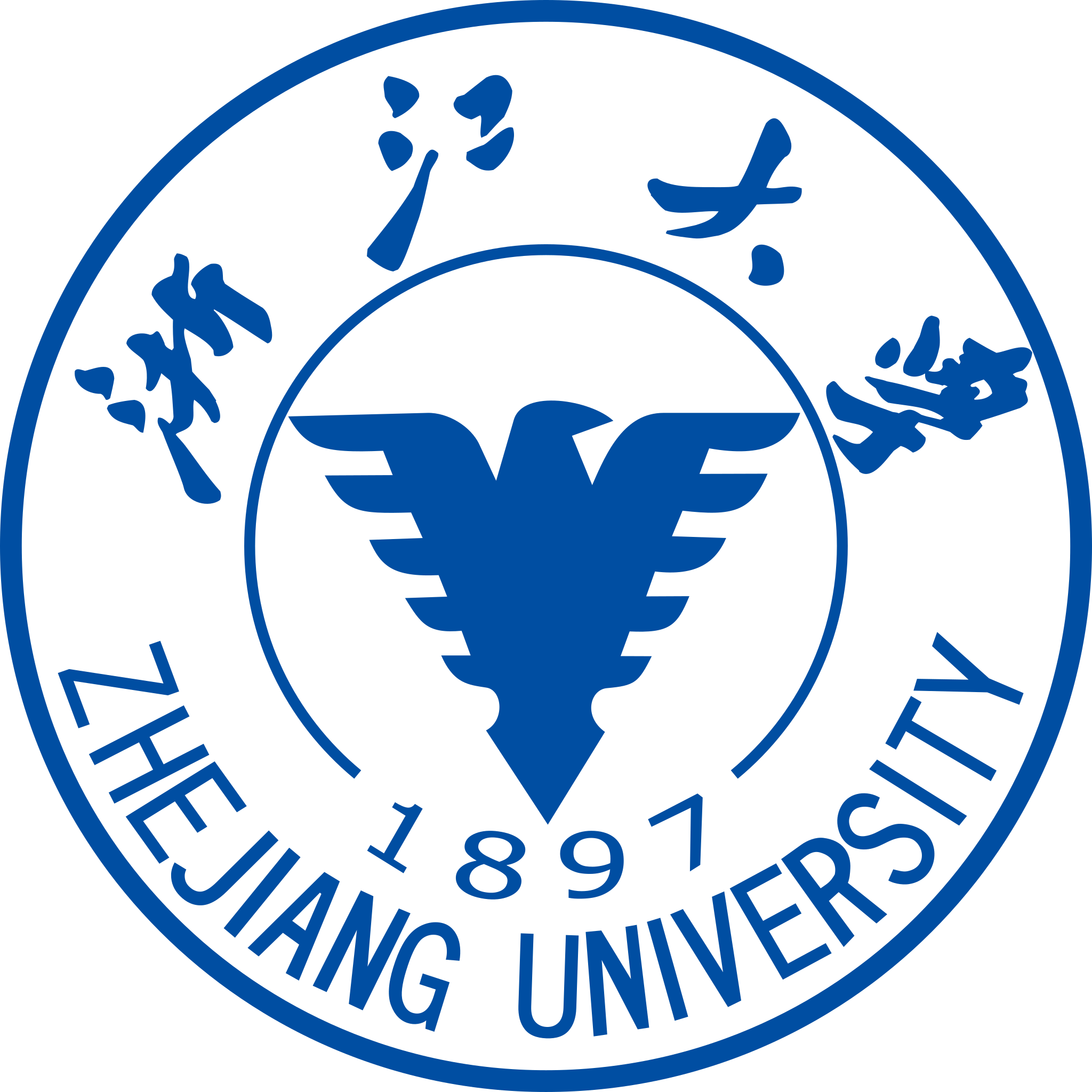}\hspace{2.5mm}%
                \includegraphics[height=12mm]{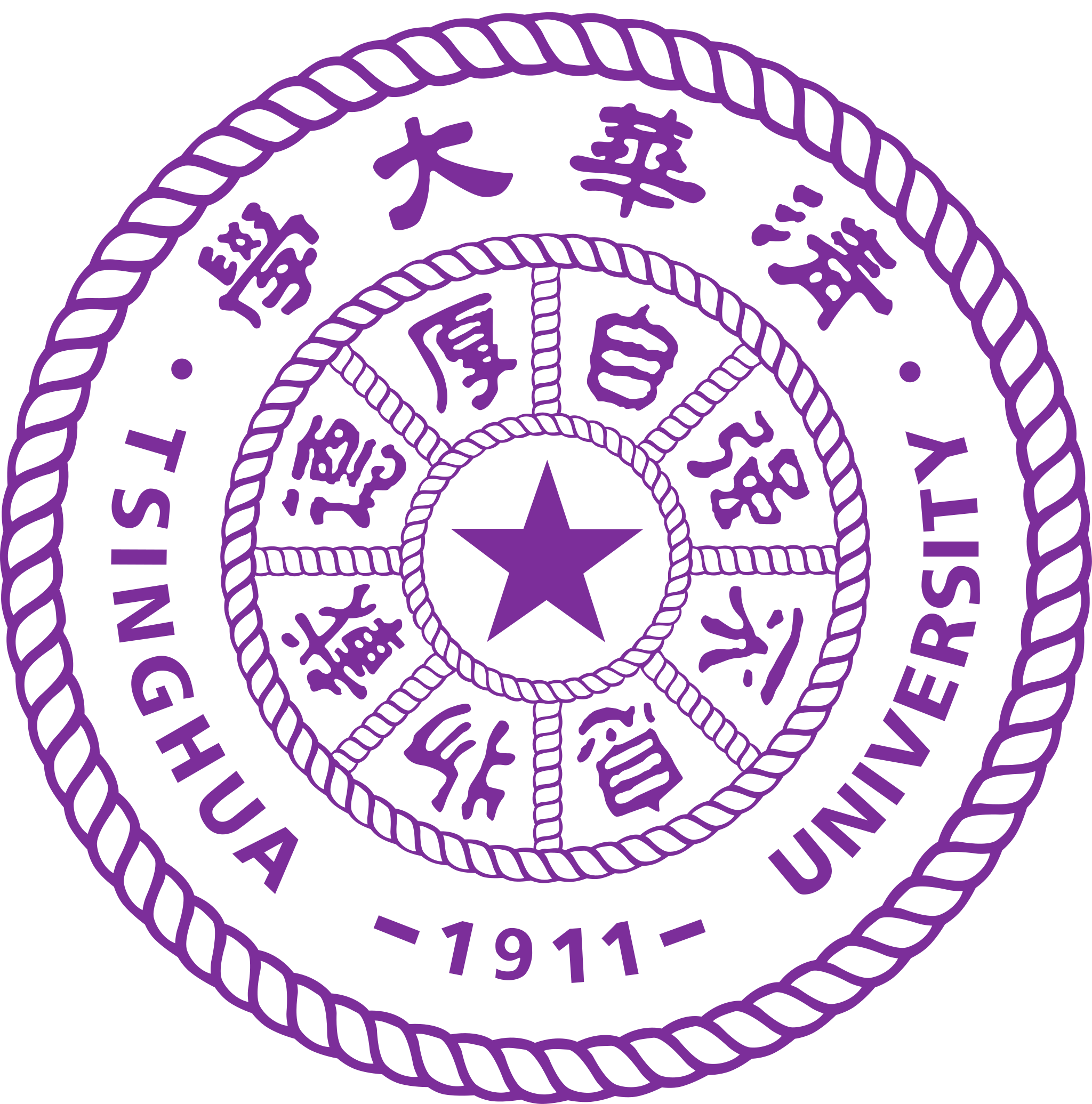}%
            }}%
        }%
    }
    \fancyhead[L]{
        \raisebox{-12.5mm}[0pt][0pt]{%
            \vbox{\vskip 3mm\hbox{\includegraphics[height=10mm]{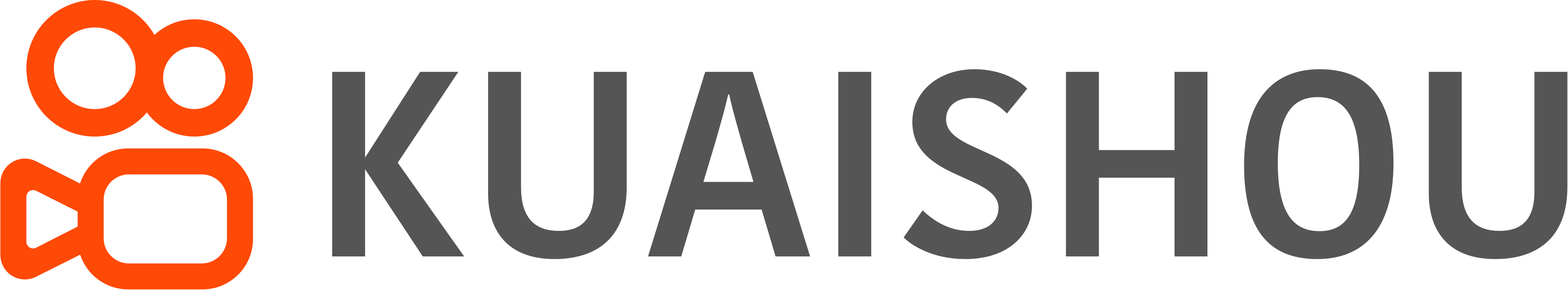}}}%
        }%
    }
    \fancyfoot[L]{\raisebox{1.15\baselineskip}[0pt][0pt]{\authorfootnotes}}
    \fancyfoot[C]{\thepage}
}
\renewcommand{\checkdatalist}{%
\begin{paperresources}
\paperresourceicon{\resourceprojecticon}{Project Page}{https://mobile-forge.github.io/}{mobile-forge.github.io}
\quad\quad
\paperresourceicon{\resourcehficon}{Datasets}{https://huggingface.co/collections/lgy0404/mobileforge-datasets}{lgy0404/mobileforge-datasets}\\
\paperresourceicon{\resourcegithubicon}{Code}{https://github.com/kwai/MobileForge}{github.com/kwai/MobileForge}
\quad\quad
\paperresourceicon{\resourcehficon}{Models}{https://huggingface.co/collections/lgy0404/mobileforge-models}{lgy0404/mobileforge-models}
\end{paperresources}}

\abstract{MLLM-based mobile GUI agents have made substantial progress in UI understanding and action execution, but adapting them to real target apps remains costly because mobile apps are numerous, frequently updated, and hard to cover with human-written tasks, demonstrations, or reward labels. Existing annotation-free GUI learning reduces manual supervision, yet lacks a unified substrate connecting target-app exploration, curriculum mining, rollout execution, and feedback, while policy optimization often relies on isolated rollouts and coarse rewards that are hard to convert into reliable improvement signals. We present \textbf{\ourmethod}, an annotation-free adaptation system for mobile GUI agents. \ourmethod consists of \textbf{\mobilegym}, which grounds task generation and rollout evaluation in real mobile app interaction, and \textbf{\underline{Hi}}erarchical \textbf{\underline{F}}eedback-Guided \textbf{\underline{P}}olicy \textbf{\underline{O}}ptimization (\textbf{\hifpo}), which turns trajectory outcomes, step-level process feedback, and corrective hints into hint-contextualized step-level GRPO updates. Using only automatically generated annotation-free adaptation data, \ourmethod adapts \llmname{Qwen3-VL-8B} to 67.2\% Pass@3 on AndroidWorld, close to the closed-data GUI-specialized \llmname{GUI-Owl-1.5-8B} base model at 69.0\%. The MobileForge-adapted \llmname{ForgeOwl-8B} further reaches 77.6\% Pass@3 on AndroidWorld and 41.0\% success on the out-of-domain MobileWorld GUI-only split, establishing the strongest open-data mobile GUI agent in our evaluation. Code, data, and trained models will be released at \url{https://mobile-forge.github.io/}.
}

\begin{document}
\maketitle
\enlargethispage{6\baselineskip}
\begingroup
\vspace*{-2.0em}
\captionsetup{font=footnotesize}
\noindent\makebox[\textwidth][c]{%
\begin{minipage}{0.99\textwidth}
    \centering
    \includegraphics[width=\paperfigwide]{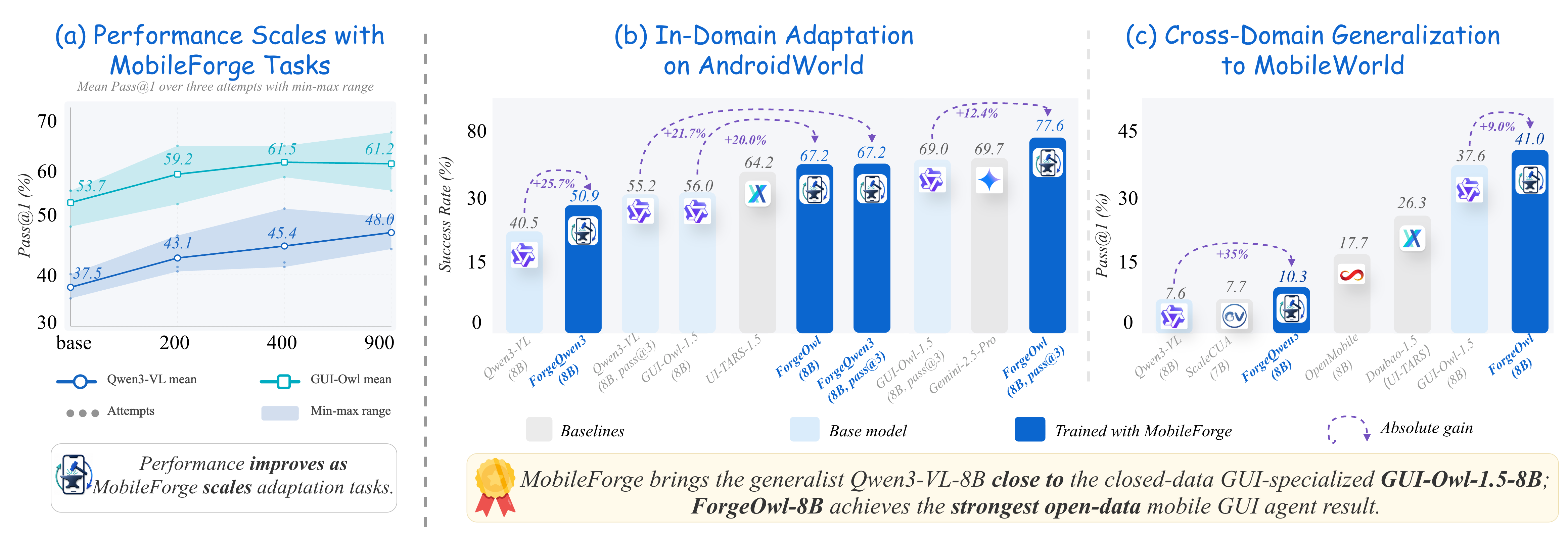}
    \vspace{-0.5em}
    \captionof{figure}{\textbf{Main performance of \ourmethod.}
    \ourmethod scales with generated AndroidWorld tasks, improves in-domain AndroidWorld performance, and transfers to the MobileWorld GUI-only split.}
    \label{fig:main-performance}
\end{minipage}}
\vspace{-0.8em}
\endgroup

\clearpage

\section{Introduction}
\label{sec:introduction}

MLLM-based mobile GUI agents have made substantial progress in UI understanding and action execution~\cite{gao2026ui,zhou2025mai,hong2024cogagent,xu2026mobile,yan2025step,liu2025llm,liu2026memguiagent}. However, real deployment requires adaptation beyond fixed benchmarks~\cite{xie2024osworld,chen2024spa,liu2026memguibench,sun2026agents,yuan2026osworld2}. Mobile apps are diverse and rapidly evolving~\cite{chai2026a3,wang2025fedmabench,sui2026androiddaily} . Consequently, human-authored tasks, expert demonstrations, and manually labeled rewards are costly to produce and quickly become outdated \citep{zhang2025tongui,wang2025mobilea3gent,sun2025genesis,yang2025zerogui}. This motivates annotation-free adaptation, in which agents explore target-app functionality, attempt grounded tasks, evaluate their own behavior, and learn from self-collected experience.

\begin{figure}[t]
    \centering
    \includegraphics[width=0.8\linewidth]{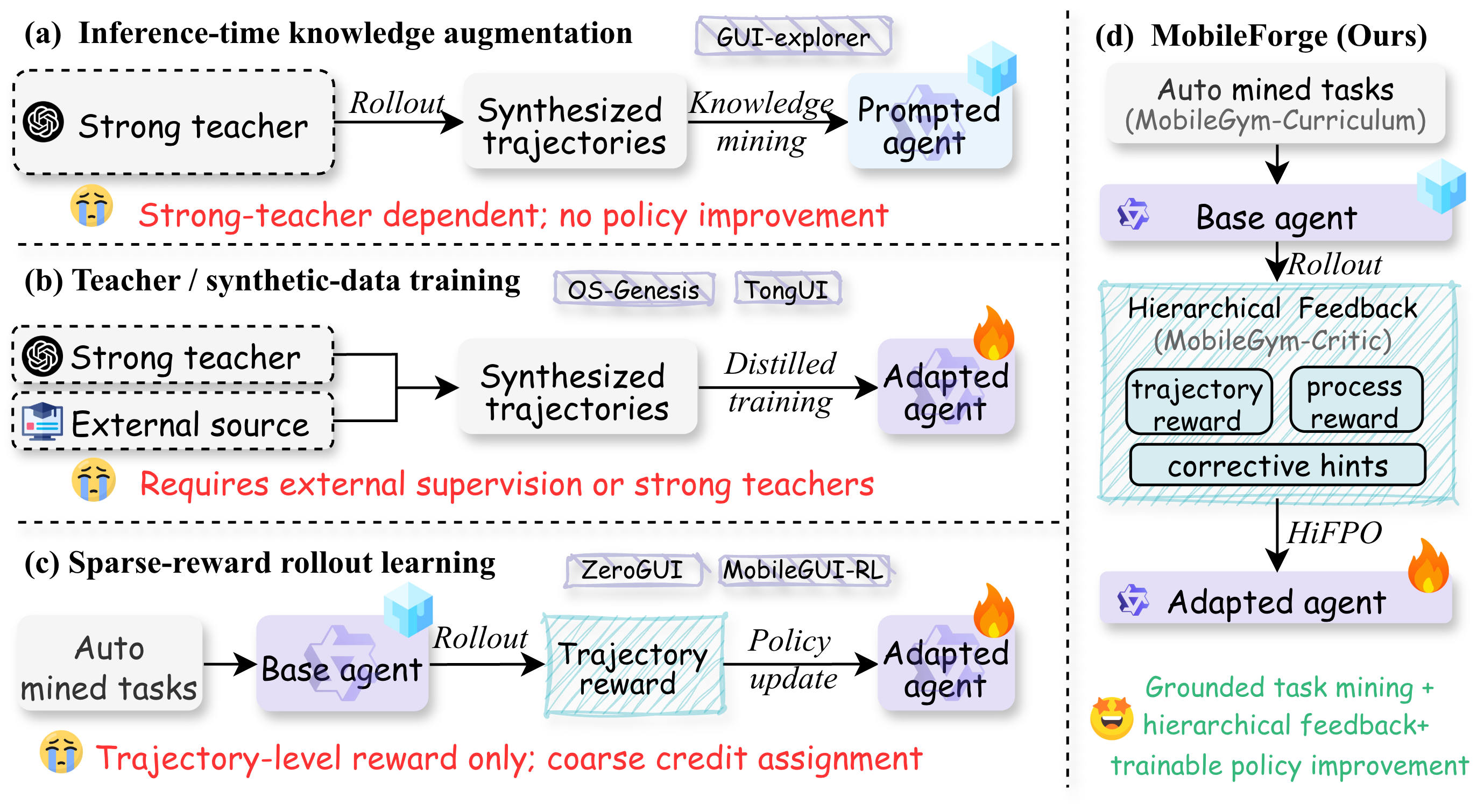}
    \vspace{-2mm}
    \caption{\textbf{Comparison of representative annotation-free GUI learning paradigms and \ourmethod.} Prior paradigms depend on external supervision, leave the policy unchanged, or rely on coarse trajectory-level rewards; \ourmethod combines grounded task mining, hierarchical feedback, and trainable policy optimization.}
    \label{fig:mobileforge-teaser}
\end{figure}

Recent annotation-free GUI agent work has reduced human supervision. TongUI\citep{zhang2025tongui} mines supervision from web tutorials, MobileA3gent\citep{wang2025mobilea3gent} uses decentralized user-phone trajectories, OS-Genesis\citep{sun2025genesis} uses GPT-4o\citep{hurst2024gpt} to execute synthesized tasks and treats its actions as SFT targets, effectively distilling a proprietary GPT-4o teacher, and GUI-explorer\citep{xie2025gui} mines transition-aware interaction knowledge through autonomous exploration. ZeroGUI\citep{yang2025zerogui} and MobileGUI-RL\citep{shi2025mobilegui} study automatic reward estimation and online reinforcement learning, while SEAgent \citep{sun2025seagent}, ACuRL \citep{xue2026acurl}, and UI-Oceanus \citep{wu2026uioceanus} explore self-evolving, continual, or synthetic-environment scaling for broader computer-use agents. Figure~\ref{fig:mobileforge-teaser} summarizes these paradigms and their limitations. \textit{Appendix~\ref{sec:related} provides a  detailed taxonomy of related work}.

Despite these advances, two bottlenecks still limit mobile target-app adaptation. \textbf{(I)} Existing methods \textbf{lack a unified mobile substrate} connecting target-app exploration, curriculum mining, rollout execution, and feedback, so generated tasks may be weakly grounded and evaluator feedback may remain detached from policy learning. \textbf{(II)} Policy optimization often treats rollouts as \textbf{isolated experiences with sparse reward}; even with step-level assessment, current loops rarely combine outcomes, process feedback, and corrective hints to accumulate reusable experience beyond the initial policy's capability boundary.

\obsbox{\textbf{Research question.} How can we build an annotation-free adaptation system for mobile GUI agents that grounds task generation in target-app interaction, generates fine-grained feedback, and converts self-collected experience into policy-improvement signals without human-written tasks, demonstrations, or reward labels?}

To address this question, we introduce \textbf{\ourmethod}, an annotation-free adaptation system with hierarchical feedback-guided policy optimization. \textbf{\emph{(i)}} \textbf{\mobilegym} is the interaction and evaluation substrate: \curriculum mines executable tasks from target-app traces, while \critic evaluates rollouts with outcome feedback, step-level feedback, and corrective hints. \textbf{\emph{(ii)}} \textbf{\underline{Hi}}erarchical \textbf{\underline{F}}eedback-Guided \textbf{\underline{P}}olicy \textbf{\underline{O}}ptimization (\textbf{\hifpo}) schedules hint-guided attempts, filters tasks and steps with hierarchical feedback, and trains the policy with hint-contextualized step-level GRPO.

We evaluate \ourmethod on AndroidWorld \citep{rawles2024androidworld} as the in-domain setting and MobileWorld GUI-only \citep{kong2026mobileworld} as the out-of-domain setting, with no MobileWorld rollout used for training. As shown in Figure~\ref{fig:main-performance}, annotation-free adaptation narrows the gap to closed-data GUI-specialized agents: \llmname{Qwen3-VL-8B}~\citep{bai2025qwen3} reaches 67.2\% Pass@3 on AndroidWorld using generated adaptation data, close to the \llmname{GUI-Owl-1.5-8B}~\citep{xu2026mobile} base model at 69.0\%. \ourmethod further improves GUI-Owl, yielding \llmname{ForgeOwl-8B} with 77.6\% Pass@3 on AndroidWorld and 41.0\% success on MobileWorld GUI-only, the strongest open-data mobile GUI agent.

\paragraph{Contributions.}
\textbf{(1)} We identify two bottlenecks in annotation-free mobile GUI adaptation: the lack of a unified target-app interaction and evaluation substrate, and isolated rollouts with coarse feedback for policy optimization.
\textbf{(2)} We propose \textbf{\mobilegym}, which grounds exploration, curriculum mining, rollout execution, and hierarchical evaluation in real mobile app interaction.
\textbf{(3)} We propose \textbf{\underline{Hi}}erarchical \textbf{\underline{F}}eedback-Guided \textbf{\underline{P}}olicy \textbf{\underline{O}}ptimization (\textbf{\hifpo}), which transforms multi-attempt feedback and corrective hints into hint-contextualized step-level GRPO updates.
\textbf{(4)} We show that \ourmethod improves generalist and GUI-specialized agents, transfers from AndroidWorld to MobileWorld GUI-only, and yields \llmname{ForgeOwl-8B}, the strongest open-data mobile GUI agent in our evaluation.

\section{\ourmethod}
\label{sec:method}

\begin{figure}[H]
    \centering
    \includegraphics[width=0.8\linewidth]{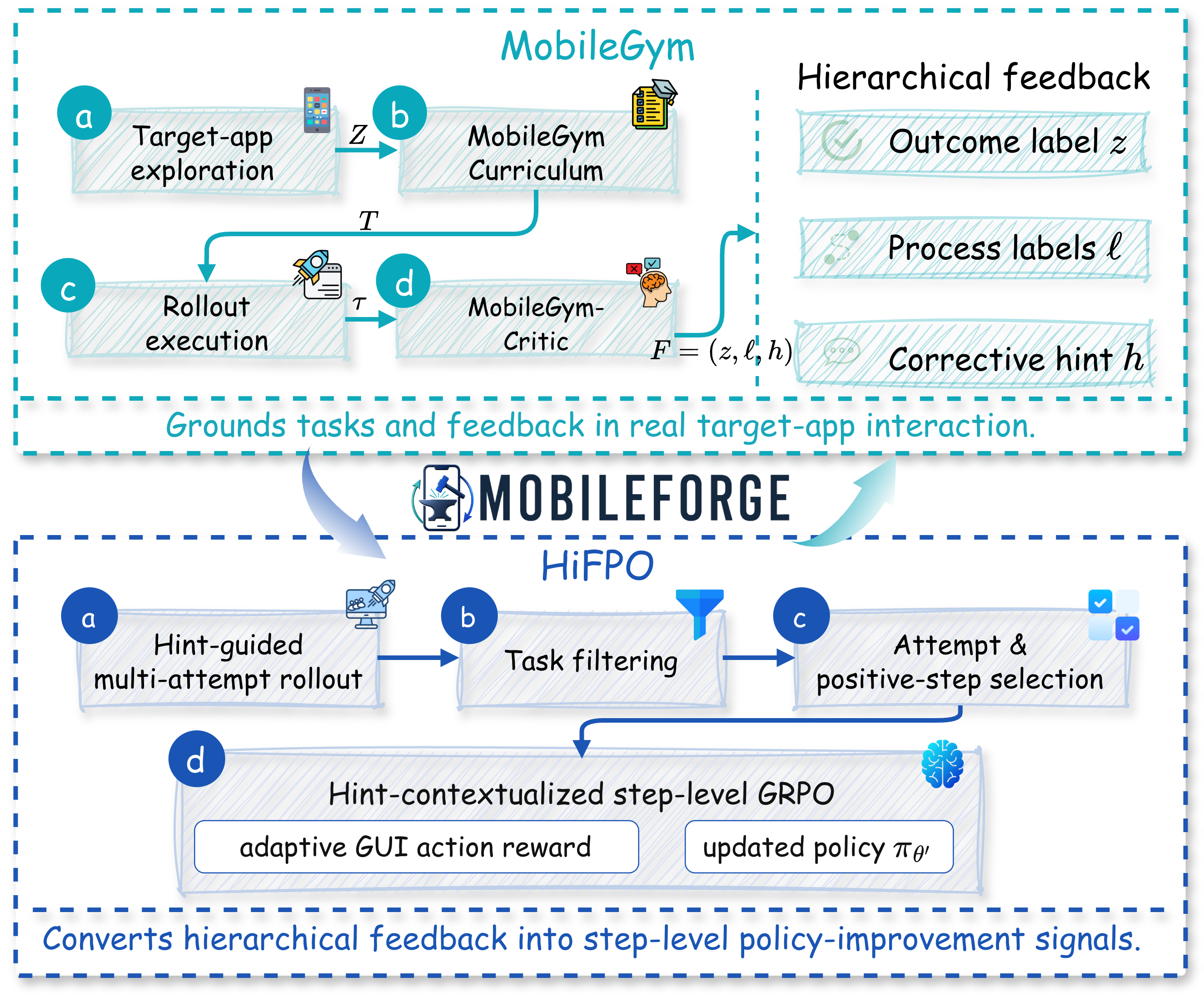}
    \caption{\textbf{Overview of \ourmethod.}
}
    \label{fig:mobileforge-overview}
\end{figure}

Given target apps and an initial policy, \ourmethod grounds task generation in real app interaction, evaluates self-collected rollouts, and turns hierarchical feedback into policy updates. Figure~\ref{fig:mobileforge-overview} illustrates the overall loop.

\subsection{Problem Setup}

We model mobile GUI control as sequential decision making. For generated task $x\in\mathcal{T}$, attempt $k$, and step $t$,
\begin{equation}
\label{eq:policy-interaction}
\begin{aligned}
s_k^{(t)}&=(x,I_k^{(t)},\mathcal{H}_k^{(t)},\eta_{<k}),\\
a_k^{(t)}&=(\alpha_k^{(t)},\psi_k^{(t)})
\sim\pi_\theta(\cdot\mid s_k^{(t)}),
\end{aligned}
\end{equation}
where $I$ is the screenshot, $\mathcal{H}$ the interaction history, $\eta_{<k}$ prior-attempt hints, and $(\alpha,\psi)$ the action type and arguments. Their sequence forms rollout $\tau_k$. \critic returns outcome and process labels $(z,\ell)$ plus hints; $R$ is reserved for the GRPO action reward.

\ourmethod contains two coupled components. \mobilegym supplies the interaction and evaluation substrate; \hifpo turns the resulting feedback into policy updates:
\begin{equation}
\label{eq:mobileforge-loop}
\begin{aligned}
\mathcal{Z} &\leftarrow \operatorname{Explore}(\mathcal{E}),\\
\mathcal{T} &\leftarrow \operatorname{Curriculum}(\mathcal{Z}),\\
\{\tau_k\}_{k=1}^{K} &\leftarrow
\operatorname{Rollout}(\pi_\theta,x,\eta_{<k}),\quad \forall x\in\mathcal{T},\\
\mathcal{F}_{k} &\leftarrow \operatorname{Critic}(x,\tau_k),\\
\theta' &\leftarrow \operatorname{HiFPO}(\theta,\mathcal{T},\tau,\mathcal{F}).
\end{aligned}
\end{equation}
Here $\mathcal{Z}$ is exploration evidence and $\mathcal{F}_k$ is hierarchical feedback. \textit{Appendix~\ref{sec:appendix_method_details} gives the full notation and pseudocode}.

\subsection{\mobilegym: Building Mobile GUI Agent Adaptation Substrate}
\label{sec:mobilegym}

\begin{figure}[H]
    \centering
    \includegraphics[width=0.99\linewidth]{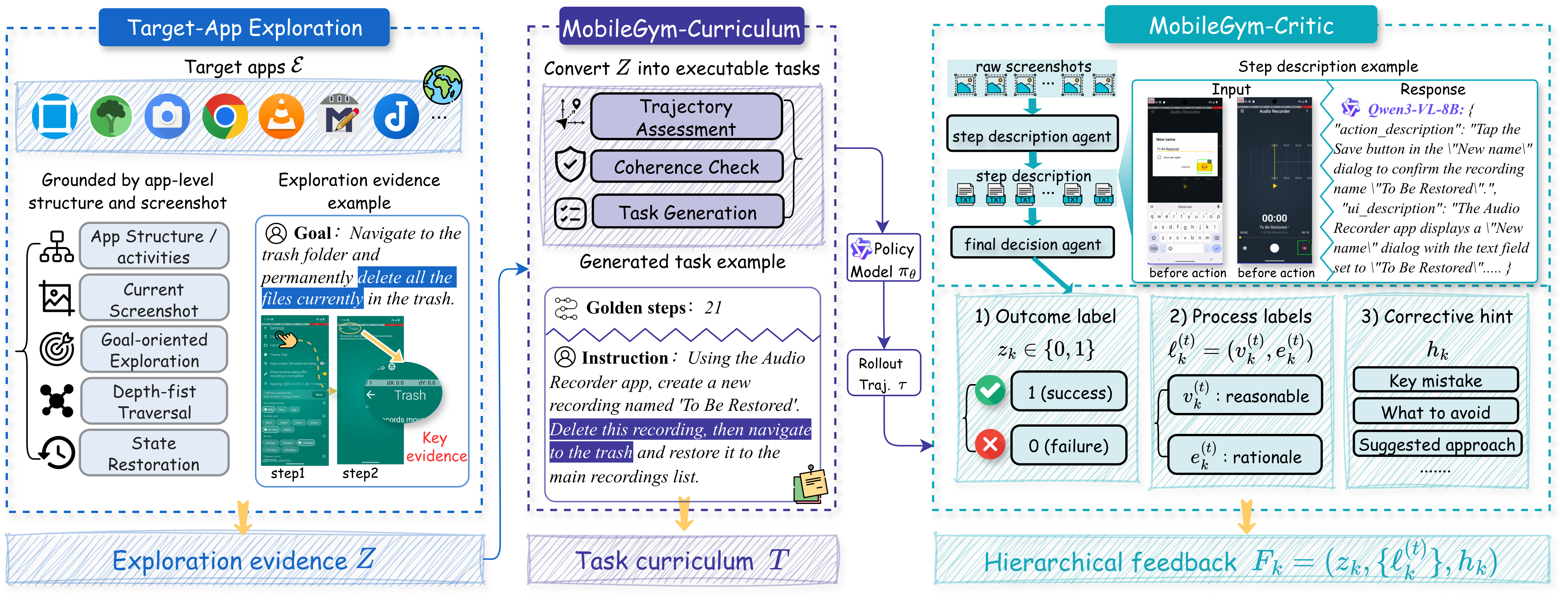}
    \caption{\textbf{\mobilegym as the annotation-free adaptation substrate.}
    \mobilegym grounds adaptation in real target-app interaction: it explores reachable GUI states, mines trajectory-grounded tasks through \curriculum, and evaluates completed attempts with \critic to produce trajectory-level outcomes, step-level process feedback, and corrective hints.}
    \label{fig:mobilegym}
\end{figure}

\mobilegym turns target-app interaction into exploration evidence, executable tasks, and hierarchical rollout feedback (Figure~\ref{fig:mobilegym}). It supplies the adaptation substrate, while \hifpo handles repeated-attempt optimization.

\subsubsection{Target-App Exploration.}

\ourmethod uses function-aware exploration inspired by GUI-explorer~\citep{xie2025gui}. App anchors and screenshots guide goal generation, while depth-first traversal records reachable screens, UI affordances, actions, and transitions in the evidence pool $\mathcal{Z}$. \textit{Appendix~\ref{sec:exploration_details} gives the full procedure}.



\subsubsection{\curriculum.}
\label{sec:curriculum}

\curriculum converts $\mathcal{Z}$ into executable tasks by checking trajectory coherence and completion, then generating variants grounded in the same reachable app states and functions. We write a task as $x=(\iota,B,c,v,p)$, where $\iota$ is the instruction, $B$ is an estimated step budget, $c$ is the core functionality, $v$ is the variation type, and $p$ describes prerequisites. \textit{Appendix~\ref{sec:curriculum_prompt} shows the core prompt}.

\subsubsection{Hierarchical Rollout Evaluation.}
\label{sec:evaluation}

For completed attempt $\tau_k$, \critic renders a visual action trace and returns
\begin{equation}
\label{eq:feedback-object}
\begin{aligned}
\mathcal{F}_{k}&=\left(z_{k},\{\ell_k^{(t)}\}_{t=1}^{T_k},h_k\right),\\
\ell_k^{(t)}&=(v_k^{(t)},e_k^{(t)}).
\end{aligned}
\end{equation}
Here $z_k$ marks task completion, $(v_k^{(t)},e_k^{(t)})$ provides the reasonable-step tag and rationale, and $h_k$ summarizes corrections. \textit{Appendix~\ref{sec:critic_prompts} gives the evaluator prompts}.

\subsection{\hifpo: Hierarchical Feedback-Guided Policy Optimization}
\label{sec:optimization}

\begin{figure}[H]
    \centering
    \includegraphics[width=0.99\linewidth]{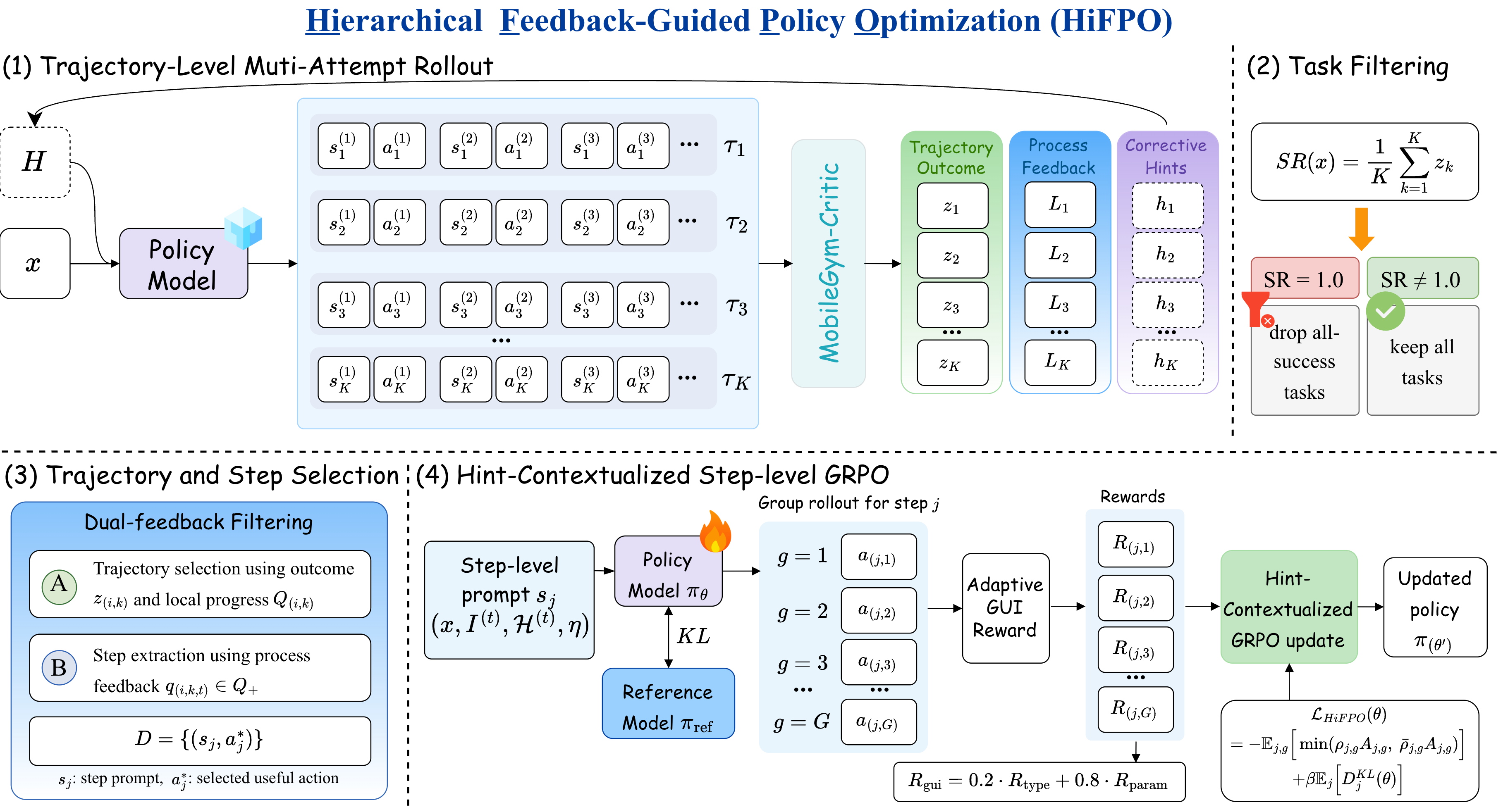}
    \caption{\textbf{\hifpo converts hierarchical feedback into policy updates.}
    \hifpo performs hint-guided multi-attempt rollout, removes mastered all-success tasks, retains difficult and partially solved tasks, extracts reasonable local steps from hierarchical feedback, and trains the policy with hint-contextualized step-level GRPO.}
    \label{fig:hifpo}
\end{figure}

\hifpo converts evaluated attempts into policy updates (Figure~\ref{fig:hifpo}). It reuses hints across attempts, removes mastered tasks, selects informative trajectories and steps, and optimizes the resulting local decisions. \textit{Appendix~\ref{sec:appendix_hifpo_formal} gives the complete filtering and optimization definitions}.

\subsubsection{Hint-Guided Multi-Attempt Rollout.}

\begin{figure}[t]
    \centering
    \includegraphics[width=\paperfigmain]{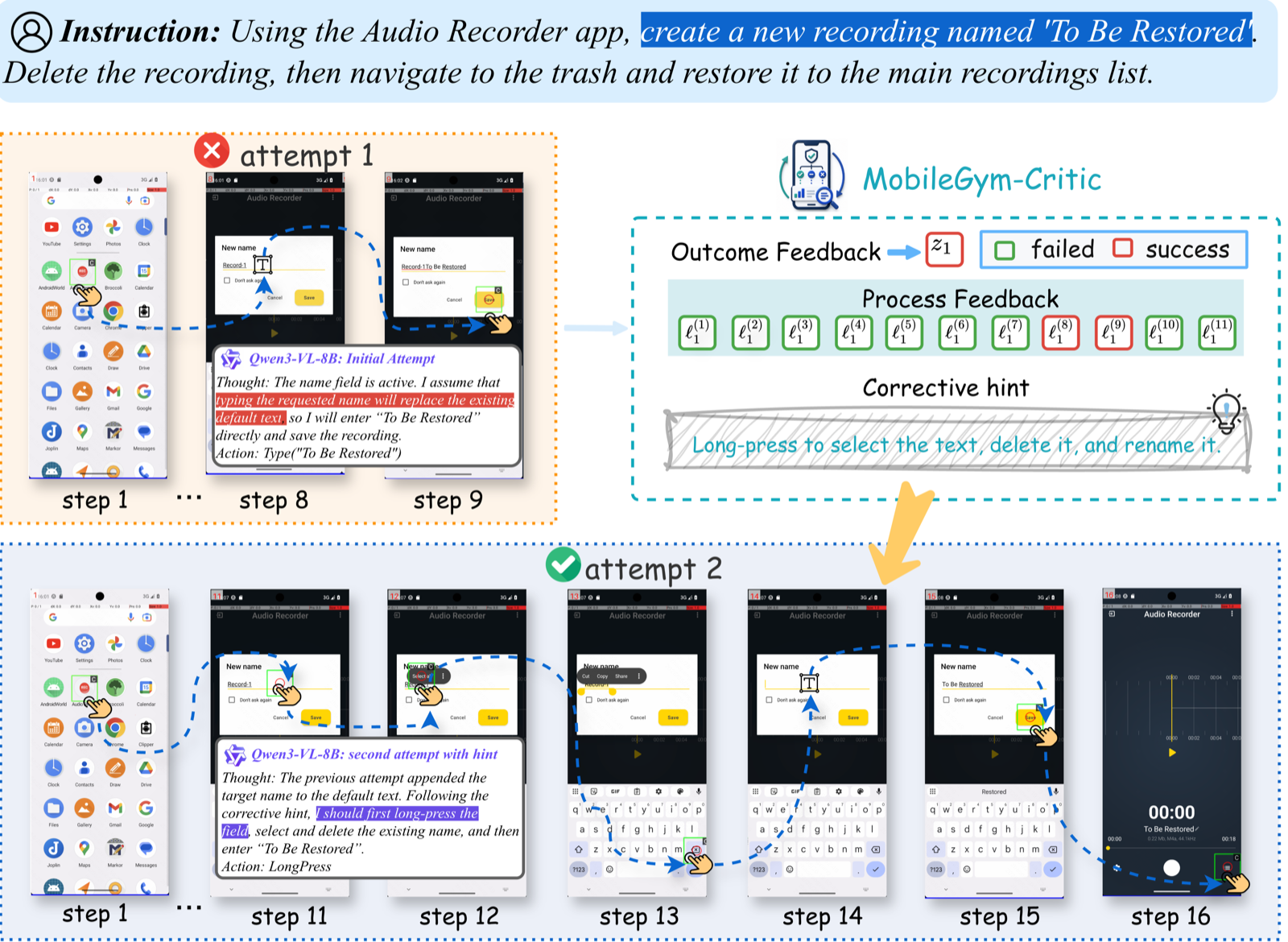}
    \caption{\textbf{Example of corrective hints improving rollout.}
    \critic summarizes the failure as a compact hint, and later attempts use the hint to avoid repeated mistakes and complete the task more reliably.}
    \label{fig:case-add-hint}
\end{figure}

For each task $x$, \hifpo runs $K$ serialized attempts with $\eta_{<k}=\operatorname{Aggregate}(h_1,\ldots,h_{k-1})$ and $\tau_k\sim\operatorname{Rollout}(\pi_\theta,x,\eta_{<k})$. Attempt $k$ therefore reuses only earlier hints, without leaking its own evaluation $h_k$. Figure~\ref{fig:case-add-hint} makes this correction concrete: the first attempt appends the target name to existing text and fails; \critic identifies the local error and advises clearing the field, enabling the second attempt to enter the correct name and complete the record--delete--restore workflow. \textit{Appendix~\ref{sec:rollout_prompts} gives the prompts}.

\subsubsection{Feedback-Based Step Extraction.}

\hifpo computes $\operatorname{SR}(x)=K^{-1}\sum_{k=1}^{K}z_k$ and removes only all-success tasks with $\operatorname{SR}(x)=1$. For each retained task, let $\chi_k^{(t)}=\mathbb{I}[v_k^{(t)}=1]$ and $Q_k=T_k^{-1}\sum_t\chi_k^{(t)}$. \hifpo selects the highest-$Q_k$ successful attempt when one exists, otherwise the highest-$Q_k$ attempt, and retains only its steps with $\chi_k^{(t)}=1$. The resulting set $\mathcal{D}=\{d_j=(s_j,a_j^\star)\}$ preserves useful local actions without reinforcing every action in a failed trajectory: $z$ guides task and trajectory selection, while $\ell$ determines which steps enter training.


\subsubsection{Hint-Contextualized Step-Level GRPO.}

GRPO avoids a learned value model by normalizing rewards within same-prompt groups~\citep{shao2024deepseekmath}; DeepSeek-R1 later scaled this recipe~\citep{guo2025deepseek}. GUI variants apply it to rule-scored actions or online trajectories~\citep{lu2025ui,luo2025gui,shi2025mobilegui}. \hifpo instead optimizes feedback-selected local decisions while conditioning every candidate group on prior-attempt hints $\eta_{<k}$.

For $d_j=(s_j,a_j^\star)$, we set $\tilde{s}_j=\operatorname{Prompt}(s_j)$, sample $\hat{o}_{j,g}\sim\pi_{\theta_{\rm old}}(\cdot\mid\tilde{s}_j)$, and parse each into $\hat{a}_{j,g}=(\hat{\alpha}_{j,g},\hat{\psi}_{j,g})$. With action-specific argument matcher $S_{\alpha_j^\star}$, the reward and group-relative advantage are
\begin{equation}
\label{eq:hint-step-advantage}
\begin{aligned}
R_{j,g}&=\lambda_{\rm type}\mathbb{I}[\hat{\alpha}_{j,g}=\alpha_j^\star]\\
&\quad+\lambda_{\rm arg}\mathbb{I}[\hat{\alpha}_{j,g}=\alpha_j^\star]
S_{\alpha_j^\star}(\hat{\psi}_{j,g},\psi_j^\star),\\
A_{j,g}&=\frac{R_{j,g}-\mu_j}{\sigma_j+\epsilon_{\rm std}}.
\end{aligned}
\end{equation}
Here $(\mu_j,\sigma_j)$ are the mean and standard deviation over the $G$ candidates. We use $(\lambda_{\rm type},\lambda_{\rm arg})=(0.2,0.8)$; unparseable responses receive zero reward.

Let $\rho_{j,g}$ denote the current-to-old policy ratio for candidate $g$ and $\bar{\rho}_{j,g}$ its clipped counterpart. The \hifpo objective is
\begin{equation}
\label{eq:hifpo-objective-main}
\begin{aligned}
\mathcal{L}_{\hifpo}(\theta)
&=-\mathbb{E}_{j,g}\!\left[\min(\rho_{j,g}A_{j,g},
\bar{\rho}_{j,g}A_{j,g})\right]\\
&\quad+\beta\mathbb{E}_{j}\!\left[D^{\rm KL}_{j}(\theta)\right].
\end{aligned}
\end{equation}
$D^{\rm KL}_{j}(\theta)$ regularizes against $\pi_{\rm ref}$. Appendix~\ref{sec:appendix_hifpo_formal} gives the complete optimization definitions. Thus, $z$ and $\ell$ select training decisions, $h$ enters through $\eta_{<k}$, and $R_{j,g}$ scores new actions. Each group therefore compares actions under the same feedback-aware state.


\section{Experiments}
\label{sec:experiments}

\subsection{Experimental Protocol}

We evaluate \ourmethod on AndroidWorld~\citep{rawles2024androidworld} as the in-domain setting and MobileWorld GUI-only~\citep{kong2026mobileworld} as the out-of-domain setting. Starting from \llmname{Qwen3-VL-8B} and \llmname{GUI-Owl-1.5-8B}~\citep{bai2025qwen3,xu2026mobile}, \ourmethod mines AndroidWorld-side annotation-free adaptation data and trains with 200-, 400-, and 900-task subsets. We report AndroidWorld Pass@1/2/3, MobileWorld success rate, and ablations over rollout hints, training objective, task filtering, evaluator choice, and curriculum grounding. \textit{Appendix~\ref{sec:appendix_experimental_protocol} gives the full protocol}. \textit{Appendix~\ref{sec:appendix_data_details} reports the generated data}, while \textit{Appendix~\ref{sec:appendix_training_details} reports the training details}.

\setcounter{insight}{0}

\begin{table*}[!t]
\centering
\paperCompactTable
\caption{AndroidWorld in-domain adaptation and scaling results. Pass@$k$ is computed over 116 tasks. Easy/Medium/Hard report single-attempt success rates by task difficulty. Overall Avg. is the mean of Pass@1/2/3 and Easy/Medium/Hard. Within each base-agent block, best values are bolded and second-best values are underlined. Relative-gain rows compare the 900-task model with the corresponding base agent.}
\label{tab:androidworld_main}
\begin{paperFit}[\textwidth]
\begin{tabular}{l c c c c c c c c}
\toprule
\multirow{2}{*}{\textbf{Base Agent}} & \multirow{2}{*}{\textbf{Tasks}} & \multicolumn{3}{c}{\textbf{Pass@$k$ Level}} & \multicolumn{3}{c}{\textbf{Task Difficulty Level}} & \multirow{2}{*}{\textbf{Overall Avg.}} \\
\cmidrule(lr){3-5}\cmidrule(lr){6-8}
& & \textbf{Pass@1} & \textbf{Pass@2} & \textbf{Pass@3} & \textbf{Easy} & \textbf{Medium} & \textbf{Hard} & \\
\midrule
\llmname{Qwen3-VL-8B} & 0 & 47/116 (40.5\%) & 57/116 (49.1\%) & 64/116 (55.2\%) & 44.8\% & 35.2\% & \textbf{19.3\%} & 40.7\% \\
\llmname{Qwen3-VL-8B} & 200 & 55/116 (47.4\%) & 64/116 (55.2\%) & 71/116 (61.2\%) & \underline{59.0\%} & 32.4\% & 12.3\% & 44.6\% \\
\llmname{Qwen3-VL-8B} & 400 & \textbf{61/116 (52.6\%)} & \underline{69/116 (59.5\%)} & \underline{73/116 (62.9\%)} & \underline{59.0\%} & \underline{38.9\%} & 14.0\% & \underline{47.8\%} \\
\rowcolor{oursrowcolor}
\llmname{Qwen3-VL-8B} & 900 & \underline{59/116 (50.9\%)} & \textbf{70/116 (60.3\%)} & \textbf{78/116 (67.2\%)} & \textbf{61.2\%} & \textbf{41.7\%} & \underline{17.5\%} & \textbf{49.8\%} \\
\rowcolor{baselinerowcolor}
\multicolumn{2}{r}{\footnotesize\emph{Rel. gain (900 vs. base)}} & \rise{+25.7\%} & \rise{+22.8\%} & \rise{+21.9\%} & \rise{+36.6\%} & \rise{+18.5\%} & \drop{-9.3\%} & \rise{+22.4\%} \\
\midrule
\llmname{GUI-Owl-1.5-8B} & 0 & 65/116 (56.0\%) & 79/116 (68.1\%) & 80/116 (69.0\%) & 66.7\% & 50.0\% & 19.3\% & 54.9\% \\
\llmname{GUI-Owl-1.5-8B} & 200 & \underline{75/116 (64.7\%)} & 85/116 (73.3\%) & \underline{86/116 (74.1\%)} & \underline{73.2\%} & 51.4\% & \underline{28.1\%} & 60.8\% \\
\llmname{GUI-Owl-1.5-8B} & 400 & \underline{75/116 (64.7\%)} & \underline{86/116 (74.1\%)} & \textbf{90/116 (77.6\%)} & \textbf{73.8\%} & \textbf{59.3\%} & 26.3\% & \underline{62.6\%} \\
\rowcolor{oursrowcolor}
\llmname{GUI-Owl-1.5-8B} & 900 & \textbf{78/116 (67.2\%)} & \textbf{87/116 (75.0\%)} & \textbf{90/116 (77.6\%)} & \underline{73.2\%} & \underline{57.4\%} & \textbf{29.8\%} & \textbf{63.4\%} \\
\rowcolor{baselinerowcolor}
\multicolumn{2}{r}{\footnotesize\emph{Rel. gain (900 vs. base)}} & \rise{+20.0\%} & \rise{+10.1\%} & \rise{+12.5\%} & \rise{+9.7\%} & \rise{+14.8\%} & \rise{+54.4\%} & \rise{+15.5\%} \\
\bottomrule
\end{tabular}
\end{paperFit}
\vspace{-1mm}
\end{table*}

\subsection{Overall Performance}
\label{sec:results}

\paragraph{In-Domain Adaptation on AndroidWorld.}
Table~\ref{tab:androidworld_main} summarizes the main AndroidWorld results and task-scaling study, while Figure~\ref{fig:main-performance}(a) visualizes the trend. Annotation-free adaptation brings the generalist \llmname{Qwen3-VL-8B} close to the closed-data GUI-specialized \llmname{GUI-Owl-1.5-8B}. \llmname{ForgeQwen3-8B} reaches 67.2\% Pass@3, versus 69.0\% for the GUI-Owl base. The same loop improves the stronger GUI-specialized agent: with 900 generated tasks, \llmname{ForgeOwl-8B} reaches 67.2\% Pass@1 and 77.6\% Pass@3. The gains span easy and medium tasks; \llmname{ForgeOwl-8B} also raises hard-task Pass@1 from 19.3\% to 29.8\%.

\paragraph{Cross-Domain Generalization to MobileWorld.}
Table~\ref{tab:mobileworld_main} reports out-of-domain MobileWorld GUI-only results. No MobileWorld rollout, task, or feedback is used for adaptation. \llmname{ForgeOwl-8B} reaches 41.0\% success on the 117-task split, surpassing existing open-data mobile GUI agents and approaching much larger or closed-data GUI-specialized systems. \llmname{ForgeQwen3-8B} transfers more modestly, suggesting that out-of-domain generalization still depends strongly on the base agent's mobile GUI competence.

\begin{table}[H]
\centering
\paperTable
\caption{Out-of-domain MobileWorld GUI-only success rate on the 117-task split. External baseline numbers follow the GUI-only reporting protocol; \ourmethod rows use AndroidWorld-derived adaptation data only. Within each method group, best values are bolded and second-best values are underlined. Relative-gain rows compare each adapted model with its corresponding 8B base agent.}
\label{tab:mobileworld_main}
\begin{paperFit}
\begin{tabular}{@{}l r@{}}
\toprule
\textbf{Agent} & \textbf{GUI-Only SR (\%)} \\
\midrule
\multicolumn{2}{l}{\textit{\textbf{Open-weight GUI / VLM agents}}} \\
\midrule
\llmname{GUI-Owl-1.5-32B}~\citep{xu2026mobile} & \textbf{43.9} \\
\llmname{MAI-UI-235B-A22B}~\citep{zhou2025mai} & \underline{39.7} \\
\llmname{GUI-Owl-1.5-8B}~\citep{xu2026mobile} & 37.6 \\
\llmname{MAI-UI-32B}~\citep{zhou2025mai} & 36.2 \\
\llmname{GUI-Owl-1.5-2B}~\citep{xu2026mobile} & 32.2 \\
\llmname{MAI-UI-8B}~\citep{zhou2025mai} & 27.5 \\
\llmname{Qwen3-VL-235B-Thinking}~\citep{bai2025qwen3} & 14.5 \\
\llmname{Qwen3-VL-235B}~\citep{bai2025qwen3} & 12.8 \\
\llmname{Qwen3-VL-32B}~\citep{bai2025qwen3} & 11.9 \\
\llmname{Qwen3-VL-8B}~\citep{bai2025qwen3} & 7.7 \\
\midrule
\multicolumn{2}{l}{\textit{\textbf{Open-data mobile GUI agents}}} \\
\midrule
\llmname{OpenMobile-8B}~\citep{OpenMobile2025} & \underline{17.7} \\
\llmname{ClawGUI-2B}~\citep{tang2026clawgui} & 17.1 \\
\llmname{OpenMobile-7B}~\citep{OpenMobile2025} & 14.8 \\
\llmname{ScaleCUA-7B}~\citep{ScaleCUA2025} & 7.7 \\
\rowcolor{oursrowcolor}
\llmname{ForgeQwen3-8B} \textit{(Ours)} & 10.3 \\
\rowcolor{baselinerowcolor}
\multicolumn{1}{r}{\footnotesize\emph{Rel. gain over \llmname{Qwen3-VL-8B}}} & \rise{+33.8\%} \\
\rowcolor{oursrowcolor}
\llmname{ForgeOwl-8B} \textit{(Ours)} & \textbf{41.0} \\
\rowcolor{baselinerowcolor}
\multicolumn{1}{r}{\footnotesize\emph{Rel. gain over \llmname{GUI-Owl-1.5-8B}}} & \rise{+9.0\%} \\
\bottomrule
\end{tabular}
\end{paperFit}
\end{table}

\insight{\ourmethod improves both generalist and GUI-specialized agents using only annotation-free AndroidWorld data. On MobileWorld GUI-only, \llmname{ForgeOwl-8B} leads all evaluated open-data mobile GUI agents.}
\FloatBarrier

\begin{figure}[H]
    \centering
    \includegraphics[width=\paperfigwide]{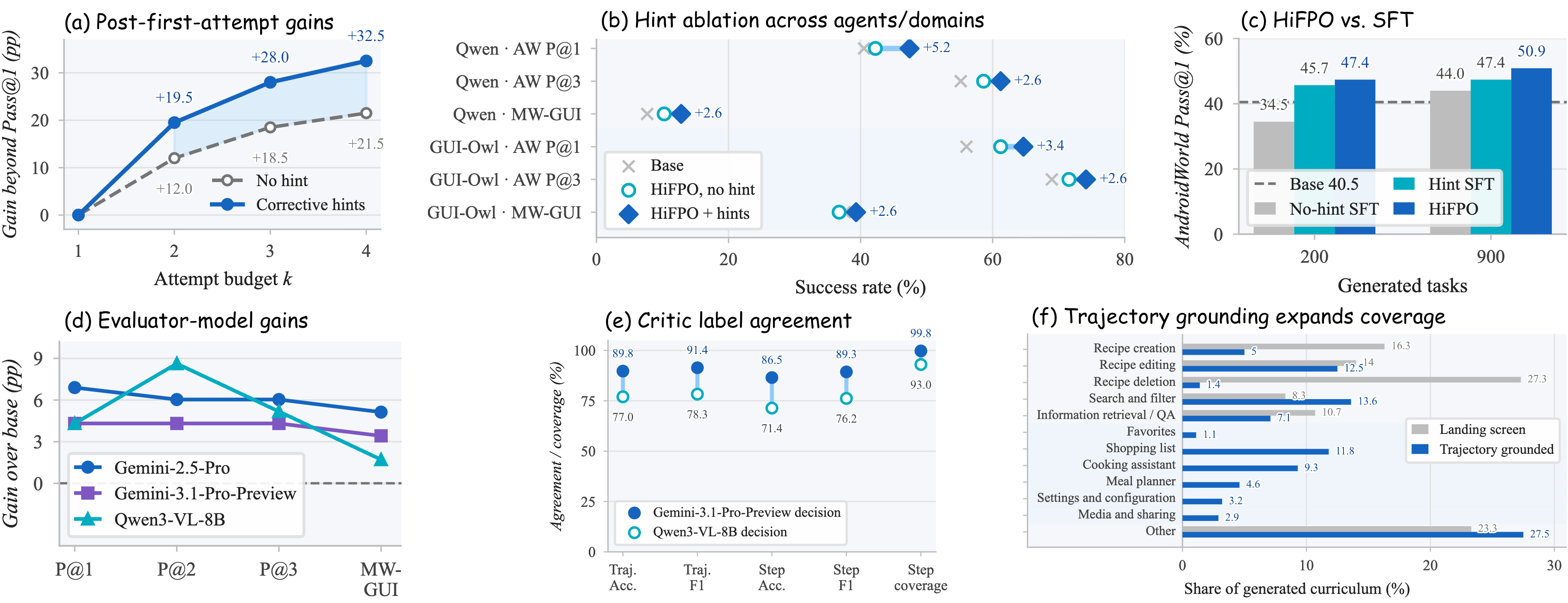}
    \caption{\textbf{Ablation analysis of \ourmethod.} (a--b) Effects of corrective hints on rollout and policy learning. (c) \hifpo versus SFT. (d--e) Final-decision evaluators and critic-label agreement. (f) Trajectory grounding and curriculum coverage. Appendix~\ref{sec:appendix_ablation_details} reports complete numerical results in Tables~\ref{tab:hint_rollout_ablation_main}--\ref{tab:curriculum_functional_coverage_main}, including raw counts, step-efficiency statistics, and precision/recall.}
    \label{fig:main_ablation}
\end{figure}

\subsection{Ablation and Analysis}
\label{sec:case_study_error_analysis}

\paragraph{Ablation on corrective rollout hints.}
Using the same 200 generated tasks, we ablate corrective hints in repeated rollouts and compare the resulting policies under matched \hifpo training. Figure~\ref{fig:main_ablation}(a) shows that hints become more beneficial as the attempt budget grows: overall rollout success rises from 52.0\% to 77.0\%, Pass@3 from 49.0\% to 72.5\%, while the average steps per attempt fall. This indicates that multi-attempt rollout benefits from accumulated feedback rather than additional sampling alone. Figure~\ref{fig:main_ablation}(b) confirms corresponding policy gains: hints raise AndroidWorld Pass@1/3 from 49/68 to 55/71 for \llmname{Qwen3-VL-8B} and from 71/83 to 75/86 for \llmname{GUI-Owl-1.5-8B}. The gains also transfer to MobileWorld GUI-only, where success increases from 12/117 to 15/117 and from 43/117 to 46/117, respectively, despite using only AndroidWorld-side adaptation data.
\insight{Corrective hints turn rollout feedback into reusable experience, improving rollout quality and policy learning across agents and benchmarks.}

\paragraph{Ablation on the training objective.}
We compare SFT and hint-contextualized GRPO on the same generated data, with and without corrective hints. Figure~\ref{fig:main_ablation}(c) shows that no-hint SFT is weak and can fall below the base model. Hint SFT is better, but hint-contextualized GRPO is strongest at both 200 and 900 tasks, reaching 50.9\% Pass@1 in the 900-task setting. This isolates the value of step-level group-relative optimization after the feedback-guided filtering stage.
\insight{Using the same annotation-free data, hint-contextualized GRPO outperforms direct SFT.}

\begin{table}[h]
\centering
\paperTable
\caption{Task-level success-rate filtering ablation. The final \ourmethod design removes mastered all-success tasks and retains all-fail plus mixed tasks. Best values in result columns are bolded and second-best values are underlined.}
\label{tab:task_filter_ablation_main}
\begin{tabular}{@{}l c r r r r@{}}
\toprule
\textbf{Filter} & \textbf{SR Range} & \textbf{Samples} & \textbf{Tasks} & \textbf{AW} & \textbf{MW-GUI} \\
\midrule
Medium only & $[0.1,0.9]$ & 1193 & 105 & \textbf{48.3\%} & \underline{12/117} \\
Medium + simple & $[0.1,1.0]$ & 1288 & 120 & \underline{46.6\%} & \textbf{15/117} \\
\rowcolor{oursrowcolor}
Medium + hard & $[0.0,0.9]$ & 1910 & 167 & \textbf{48.3\%} & \textbf{15/117} \\
All tasks & $[0.0,1.0]$ & 2137 & 200 & \textbf{48.3\%} & 10/117 \\
\bottomrule
\end{tabular}
\end{table}

\paragraph{Ablation on task filtering.}
We vary the task-level success-rate range used before step extraction. Table~\ref{tab:task_filter_ablation_main} shows that keeping all-fail and mixed tasks, corresponding to the $[0.0,0.9]$ range, gives the strongest combined AndroidWorld and MobileWorld result among the tested filters. Keeping all tasks admits mastered all-success tasks, while removing all-fail tasks discards useful local progress from difficult attempts.
\insight{The right filtering rule is not to remove failures; it is to remove mastered tasks and let step feedback recover useful local actions.}

\paragraph{Ablation on the evaluator model.}
We vary the final-decision model in \critic while fixing \llmname{Qwen3-VL-8B} for step descriptions. Figure~\ref{fig:main_ablation}(d) shows that \mbox{Gemini-2.5-Pro}~\citep{comanici2025gemini} yields the strongest 200-task policy result, while an all-Qwen critic still improves the base policy from 40.5\% to 44.8\% Pass@1 and from 55.2\% to 60.3\% Pass@3, showing that the gains do not require a proprietary evaluator. Because these labels drive task filtering, trajectory selection, and training targets, we also compare the \mbox{Gemini-3.1-Pro-Preview} and all-Qwen configurations against reference labels on 400 trajectories each. Figure~\ref{fig:main_ablation}(e) shows that the Gemini configuration reaches 89.75\% trajectory and 86.53\% step accuracy, with 91.40\% and 89.35\% F1, respectively; the all-Qwen critic reaches 77.00\% trajectory and 71.37\% step accuracy while covering 93.02\% of steps. Despite its lower agreement, the all-Qwen critic still yields 52/67/70 AndroidWorld Pass@1/2/3 and 11/117 MobileWorld GUI-only success.
\insight{Stronger evaluators improve label quality, but an open all-Qwen critic still yields policy gains.}

\paragraph{Ablation on curriculum grounding.}
Figure~\ref{fig:main_ablation}(f) compares the functional coverage of tasks generated from only the landing screen against trajectory-grounded \curriculum. The landing-screen baseline over-concentrates on recipe creation, editing, and deletion, with recipe deletion alone taking 27.3\% of tasks. In contrast, \curriculum covers broader functions such as shopping lists, cooking assistant flows, meal planning, settings, and media sharing.
\insight{Grounding task generation in exploration expands coverage beyond the landing screen.}

\FloatBarrier

\begin{figure}[H]
    \centering
    \includegraphics[width=0.99\linewidth]{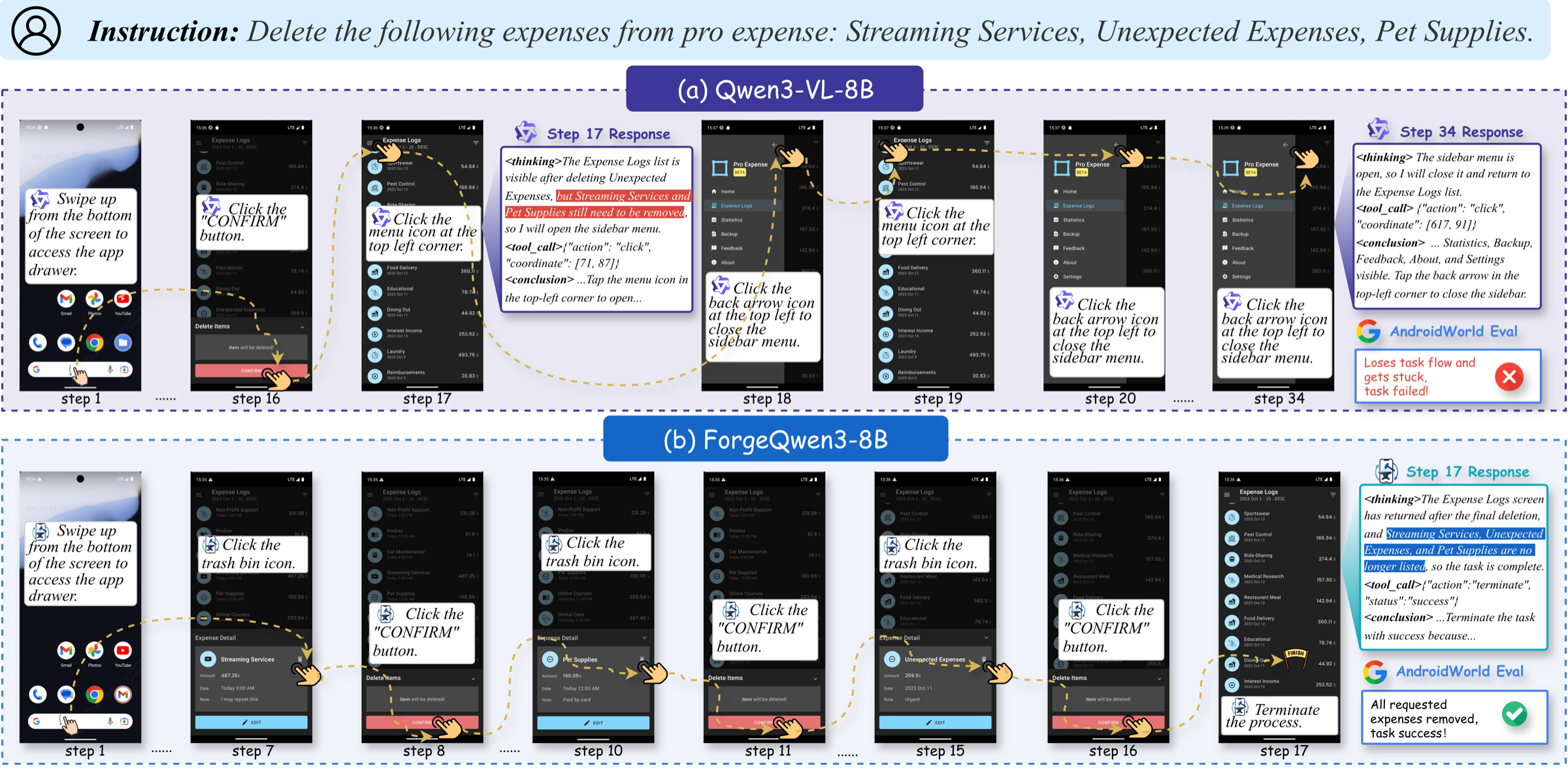}
    \caption{\textbf{Case study on AndroidWorld ExpenseDeleteMultiple2.}
    The base \llmname{Qwen3-VL-8B} loses the task flow after an early deletion, while \llmname{ForgeQwen3-8B} follows the repeated deletion workflow and completes the task.}
    \label{fig:case-study-good}
\end{figure}

\paragraph{Case Study.}
Figure~\ref{fig:case-study-good} compares \llmname{Qwen3-VL-8B} and the adapted \llmname{ForgeQwen3-8B} on AndroidWorld. The base model reaches a deletion confirmation but then loses the task flow, repeatedly opening and closing the sidebar instead of continuing through the remaining requested expenses. After \ourmethod adaptation, \llmname{ForgeQwen3-8B} follows the app-specific deletion pattern across multiple items and completes the requested removals. \textit{Appendix~\ref{sec:appendix_track_completion_cases} provides additional paired trajectory comparisons}, and Figures~\ref{fig:track_completion_androidworld_qwen}, \ref{fig:track_completion_mobileworld_qwen}, and~\ref{fig:track_completion_mobileworld_guiowl} show the same pattern across AndroidWorld and MobileWorld tasks.

\begin{figure}[H]
    \centering
    \includegraphics[width=0.7\linewidth]{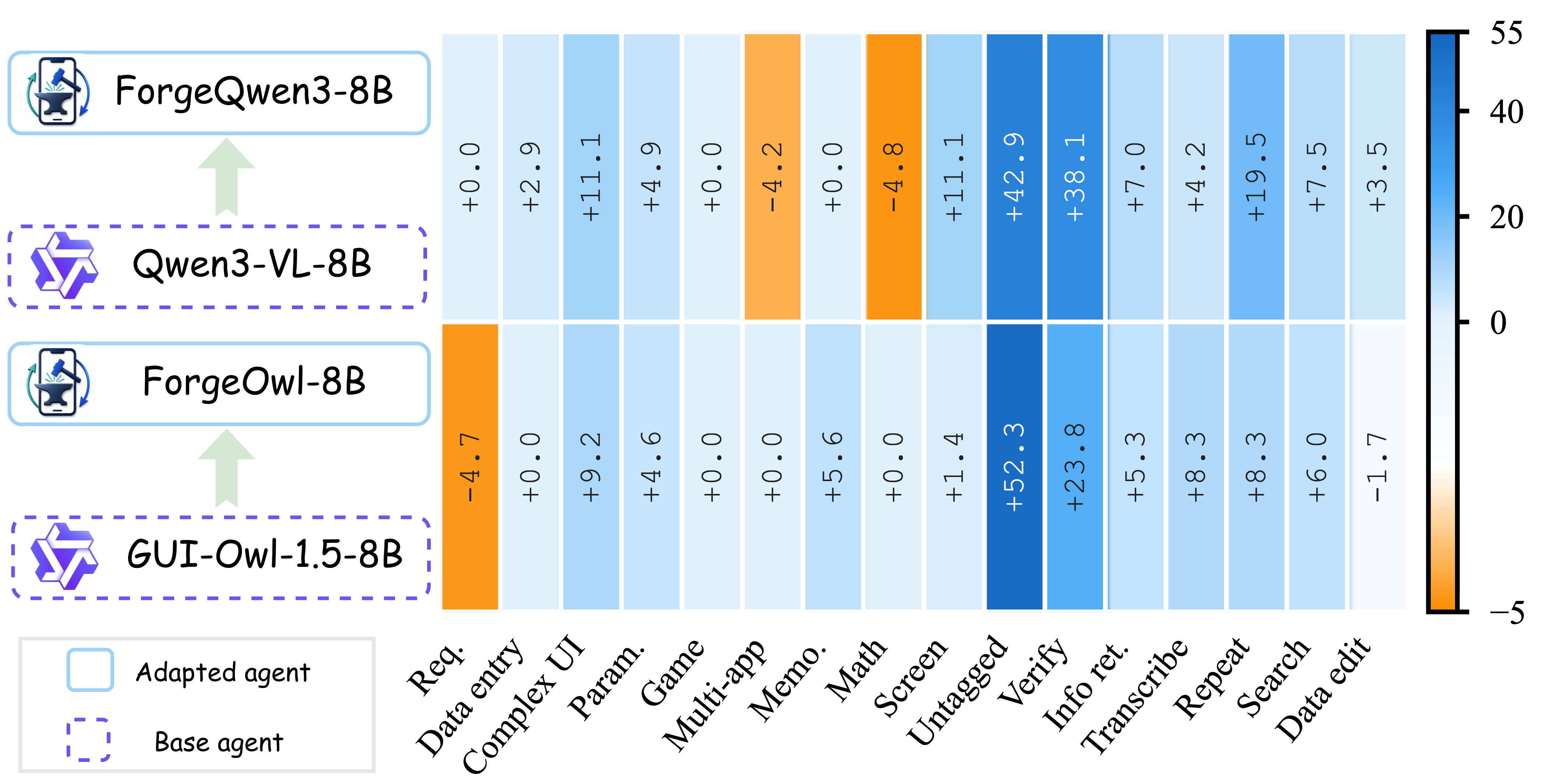}
    \caption{\textbf{AndroidWorld tag-wise failure-rate reduction.}
    Each cell reports the reduction in attempt-level failure rate after \ourmethod adaptation relative to the corresponding base agent; higher is better. Blue cells indicate fewer failures after adaptation, while orange cells indicate regressions.}
    \label{fig:tag_failure_heatmap}
\end{figure}

\paragraph{Error Analysis.}
Figure~\ref{fig:tag_failure_heatmap} shows that gains concentrate on app-grounded UI skills such as verification, search, screen reading, repetition, and information retrieval. Remaining failures cluster around games, multi-app tasks, and memorization or counting, reflecting long-horizon and cross-app challenges rather than basic UI grounding.

\section{Conclusion}

We present \ourmethod for annotation-free mobile GUI agent adaptation. \mobilegym grounds task generation and rollout evaluation in target apps; \hifpo converts step feedback and corrective hints into hint-contextualized GRPO updates. Across AndroidWorld and MobileWorld, \ourmethod improves generalist and GUI-specialized models; \llmname{ForgeOwl-8B} is the strongest open-data mobile GUI agent evaluated.

\paragraph{Limitations.}
\ourmethod is still bounded by the app ecosystem. Training remains limited to AndroidWorld-side apps, leaving broader coverage, long multi-app workflows, persistent user state, and unusual task rules challenging.



\bibliographystyle{plainnat}
\bibliography{main}

\begin{thebibliography}{59}
\providecommand{\natexlab}[1]{#1}
\providecommand{\url}[1]{\texttt{#1}}
\expandafter\ifx\csname urlstyle\endcsname\relax
  \providecommand{\doi}[1]{doi: #1}\else
  \providecommand{\doi}{doi: \begingroup \urlstyle{rm}\Url}\fi

\bibitem[Bai et~al.(2025)Bai, Cai, Chen, Chen, Chen, Cheng, Deng, Ding, Gao,
  Ge, et~al.]{bai2025qwen3}
Shuai Bai, Yuxuan Cai, Ruizhe Chen, Keqin Chen, Xionghui Chen, Zesen Cheng,
  Lianghao Deng, Wei Ding, Chang Gao, Chunjiang Ge, et~al.
\newblock Qwen3-vl technical report.
\newblock \emph{arXiv preprint arXiv:2511.21631}, 2025.

\bibitem[Chai et~al.(2026)Chai, Tang, Xiao, Lin, Li, Zhang, Liu, Zhao, Liu,
  Wang, et~al.]{chai2026a3}
Yuxiang Chai, Shunye Tang, Han Xiao, Weifeng Lin, Hanhao Li, Jiayu Zhang, Liang
  Liu, Pengxiang Zhao, Guangyi Liu, Guozhi Wang, et~al.
\newblock A3: Android agent arena for mobile gui agents with essential-state
  procedural evaluation.
\newblock In \emph{Findings of the Association for Computational Linguistics:
  ACL 2026}, pages 3774--3789, 2026.

\bibitem[Chen et~al.(2024)Chen, Yuen, Xie, Yang, Chen, Wu, Yixing, Zhou, Liu,
  Wang, et~al.]{chen2024spa}
Jingxuan Chen, Derek Yuen, Bin Xie, Yuhao Yang, Gongwei Chen, Zhihao Wu,
  Li~Yixing, Xurui Zhou, Weiwen Liu, Shuai Wang, et~al.
\newblock Spa-bench: A comprehensive benchmark for smartphone agent evaluation.
\newblock In \emph{NeurIPS 2024 Workshop on Open-World Agents}, 2024.

\bibitem[Cheng et~al.(2024)Cheng, Sun, Chu, Xu, Li, Zhang, and
  Wu]{cheng2024seeclick}
Kanzhi Cheng, Qiushi Sun, Yougang Chu, Fangzhi Xu, Yantao Li, Jianbing Zhang,
  and Zhiyong Wu.
\newblock Seeclick: Harnessing gui grounding for advanced visual gui agents.
\newblock \emph{arXiv preprint arXiv:2401.10935}, 2024.

\bibitem[Cheng et~al.(2026)Cheng, Li, Ma, Chen, Cao, Sun, Ding, Xu, Yan, Chen,
  et~al.]{OpenMobile2025}
Kanzhi Cheng, Zehao Li, Zheng Ma, Nuo Chen, Jialin Cao, Qiushi Sun, Zichen
  Ding, Fangzhi Xu, Hang Yan, Jiajun Chen, et~al.
\newblock {OpenMobile}: Building open mobile agents with task and trajectory
  synthesis.
\newblock \emph{arXiv preprint arXiv:2604.15093}, 2026.

\bibitem[Comanici et~al.(2025)Comanici, Bieber, Schaekermann, Pasupat,
  Sachdeva, Dhillon, Blistein, Ram, Zhang, Rosen, et~al.]{comanici2025gemini}
Gheorghe Comanici, Eric Bieber, Mike Schaekermann, Ice Pasupat, Noveen
  Sachdeva, Inderjit Dhillon, Marcel Blistein, Ori Ram, Dan Zhang, Evan Rosen,
  et~al.
\newblock Gemini 2.5: Pushing the frontier with advanced reasoning,
  multimodality, long context, and next generation agentic capabilities.
\newblock \emph{arXiv preprint arXiv:2507.06261}, 2025.

\bibitem[Deng et~al.(2023)Deng, Gu, Zheng, Chen, Stevens, Wang, Sun, and
  Su]{deng2023mind2web}
Xiang Deng, Yu~Gu, Boyuan Zheng, Shijie Chen, Sam Stevens, Boshi Wang, Huan
  Sun, and Yu~Su.
\newblock {Mind2Web}: Towards a generalist agent for the web.
\newblock In \emph{Advances in Neural Information Processing Systems},
  volume~36, 2023.

\bibitem[Gao et~al.(2026)Gao, Gu, Liu, Qiu, Shen, Wen, Xia, Xu, Zeng, Zhou,
  et~al.]{gao2026ui}
Changlong Gao, Zhangxuan Gu, Yulin Liu, Xinyu Qiu, Shuheng Shen, Yue Wen,
  Tianyu Xia, Zhenyu Xu, Zhengwen Zeng, Beitong Zhou, et~al.
\newblock Ui-venus-1.5 technical report.
\newblock \emph{arXiv e-prints}, pages arXiv--2602, 2026.

\bibitem[Gou et~al.(2025)Gou, Wang, Zheng, Xie, Chang, Shu, Sun, and
  Su]{gou2025uground}
Boyu Gou, Ruohan Wang, Boyuan Zheng, Yanan Xie, Cheng Chang, Yiheng Shu, Huan
  Sun, and Yu~Su.
\newblock Navigating the digital world as humans do: Universal visual grounding
  for {GUI} agents.
\newblock In \emph{The Thirteenth International Conference on Learning
  Representations}, 2025.

\bibitem[Guo et~al.(2025)Guo, Yang, Zhang, Song, Zhang, Xu, Zhu, Ma, Wang, Bi,
  et~al.]{guo2025deepseek}
Daya Guo, Dejian Yang, Haowei Zhang, Junxiao Song, Ruoyu Zhang, Runxin Xu,
  Qihao Zhu, Shirong Ma, Peiyi Wang, Xiao Bi, et~al.
\newblock Deepseek-r1: Incentivizing reasoning capability in llms via
  reinforcement learning.
\newblock \emph{arXiv preprint arXiv:2501.12948}, 2025.

\bibitem[He et~al.(2024)He, Yao, Ma, Yu, Dai, Zhang, Lan, and
  Yu]{he2024webvoyager}
Hongliang He, Wenlin Yao, Kaixin Ma, Wenhao Yu, Yong Dai, Hongming Zhang,
  Zhenzhong Lan, and Dong Yu.
\newblock Webvoyager: Building an end-to-end web agent with large multimodal
  models.
\newblock \emph{arXiv preprint arXiv:2401.13919}, 2024.

\bibitem[Hong et~al.(2024)Hong, Wang, Lv, Xu, Yu, Ji, Wang, Wang, Dong, Ding,
  et~al.]{hong2024cogagent}
Wenyi Hong, Weihan Wang, Qingsong Lv, Jiazheng Xu, Wenmeng Yu, Junhui Ji, Yan
  Wang, Zihan Wang, Yuxiao Dong, Ming Ding, et~al.
\newblock Cogagent: A visual language model for gui agents.
\newblock In \emph{Proceedings of the IEEE/CVF conference on computer vision
  and pattern recognition}, pages 14281--14290, 2024.

\bibitem[Hurst et~al.(2024)Hurst, Lerer, Goucher, Perelman, Ramesh, Clark,
  Ostrow, Welihinda, Hayes, Radford, et~al.]{hurst2024gpt}
Aaron Hurst, Adam Lerer, Adam~P Goucher, Adam Perelman, Aditya Ramesh, Aidan
  Clark, AJ~Ostrow, Akila Welihinda, Alan Hayes, Alec Radford, et~al.
\newblock Gpt-4o system card.
\newblock \emph{arXiv preprint arXiv:2410.21276}, 2024.

\bibitem[Kong et~al.(2026)Kong, Zhang, Yang, Gao, Liu, Tong, Cai, Zhou, Zhang,
  Chen, et~al.]{kong2026mobileworld}
Quyu Kong, Xu~Zhang, Zhenyu Yang, Nolan Gao, Chen Liu, Panrong Tong, Chenglin
  Cai, Hanzhang Zhou, Jianan Zhang, Liangyu Chen, et~al.
\newblock Mobileworld: Benchmarking autonomous mobile agents in agent-user
  interactive and mcp-augmented environments.
\newblock \emph{arXiv preprint}, 2026.

\bibitem[Li et~al.(2025)Li, Meng, Lin, Luo, Tian, Ma, Huang, and
  Chua]{li2025screenspotpro}
Kaixin Li, Ziyang Meng, Hongzhan Lin, Ziyang Luo, Yuchen Tian, Jing Ma, Zhiyong
  Huang, and Tat-Seng Chua.
\newblock {ScreenSpot-Pro}: {GUI} grounding for professional high-resolution
  computer use.
\newblock \emph{arXiv preprint arXiv:2504.07981}, 2025.

\bibitem[Lin et~al.(2025)Lin, Li, Gao, Yang, Wu, Bai, Lei, Wang, and
  Shou]{lin2025showui}
Kevin~Qinghong Lin, Linjie Li, Difei Gao, Zhengyuan Yang, Shiwei Wu, Zechen
  Bai, Stan~Weixian Lei, Lijuan Wang, and Mike~Zheng Shou.
\newblock {ShowUI}: One vision-language-action model for {GUI} visual agent.
\newblock In \emph{Proceedings of the IEEE/CVF Conference on Computer Vision
  and Pattern Recognition}, pages 19498--19508, 2025.

\bibitem[Liu et~al.(2025{\natexlab{a}})Liu, Zhao, Liang, Liu, Guo, Xiao, Lin,
  Chai, Han, Ren, et~al.]{liu2025llm}
Guangyi Liu, Pengxiang Zhao, Yaozhen Liang, Liang Liu, Yaxuan Guo, Han Xiao,
  Weifeng Lin, Yuxiang Chai, Yue Han, Shuai Ren, et~al.
\newblock Llm-powered gui agents in phone automation: Surveying progress and
  prospects.
\newblock \emph{arXiv preprint arXiv:2504.19838}, 2025{\natexlab{a}}.

\bibitem[Liu et~al.(2025{\natexlab{b}})Liu, Zhao, Liu, Chen, Chai, Ren, Wang,
  He, and Meng]{liu2025learnact}
Guangyi Liu, Pengxiang Zhao, Liang Liu, Zhiming Chen, Yuxiang Chai, Shuai Ren,
  Hao Wang, Shibo He, and Wenchao Meng.
\newblock Learnact: Few-shot mobile gui agent with a unified demonstration
  benchmark.
\newblock \emph{arXiv preprint arXiv:2504.13805}, 2025{\natexlab{b}}.

\bibitem[Liu et~al.(2026{\natexlab{a}})Liu, Wu, Liu, Zhao, Liu, Li, Zhang,
  Wang, Guo, and Liu]{liu2026memguiagent}
Guangyi Liu, Gao Wu, Congxiao Liu, Pengxiang Zhao, Liang Liu, Mading Li,
  Qi~Zhang, Mengyan Wang, Liang Guo, and Yong Liu.
\newblock Memgui-agent: An end-to-end long-horizon mobile gui agent with
  proactive context management.
\newblock \emph{arXiv preprint arXiv:2606.19926}, 2026{\natexlab{a}}.

\bibitem[Liu et~al.(2026{\natexlab{b}})Liu, Zhao, Liang, Luo, Tang, Chai, Lin,
  Xiao, Wang, Chen, et~al.]{liu2026memguibench}
Guangyi Liu, Pengxiang Zhao, Yaozhen Liang, Qinyi Luo, Shunye Tang, Yuxiang
  Chai, Weifeng Lin, Han Xiao, WenHao Wang, Siheng Chen, et~al.
\newblock Memgui-bench: Benchmarking memory of mobile gui agents in dynamic
  environments.
\newblock \emph{arXiv preprint arXiv:2602.06075}, 2026{\natexlab{b}}.

\bibitem[Liu et~al.(2024)Liu, Yu, Zhang, Xu, Lei, Lai, Gu, Ding, Men, Yang,
  Zhang, Deng, Zeng, Du, Zhang, Shen, Zhang, Su, Sun, Huang, Dong, and
  Tang]{liu2024agentbench}
Xiao Liu, Hao Yu, Hanchen Zhang, Yifan Xu, Xuanyu Lei, Hanyu Lai, Yu~Gu,
  Hangliang Ding, Kaiwen Men, Kejuan Yang, Shudan Zhang, Xiang Deng, Aohan
  Zeng, Zhengxiao Du, Chenhui Zhang, Sheng Shen, Tianjun Zhang, Yu~Su, Huan
  Sun, Minlie Huang, Yuxiao Dong, and Jie Tang.
\newblock {AgentBench}: Evaluating {LLM}s as agents.
\newblock In \emph{The Twelfth International Conference on Learning
  Representations}, 2024.

\bibitem[Liu et~al.(2025{\natexlab{c}})Liu, Xie, Ding, Li, Yang, Wu, Wang, Sun,
  Liu, Wang, et~al.]{ScaleCUA2025}
Zhaoyang Liu, JingJing Xie, Zichen Ding, Zehao Li, Bowen Yang, Zhenyu Wu,
  Xuehui Wang, Qiushi Sun, Shi Liu, Weiyun Wang, et~al.
\newblock Scalecua: Scaling open-source computer use agents with cross-platform
  data.
\newblock \emph{arXiv preprint arXiv:2509.15221}, 2025{\natexlab{c}}.

\bibitem[Lu et~al.(2024)Lu, Yang, Shen, and Awadallah]{lu2024omniparser}
Yadong Lu, Jianwei Yang, Yelong Shen, and Ahmed Awadallah.
\newblock {OmniParser} for pure vision based {GUI} agent.
\newblock \emph{arXiv preprint arXiv:2408.00203}, 2024.

\bibitem[Lu et~al.(2025)Lu, Chai, Guo, Yin, Liu, Wang, Xiao, Ren, Xiong, and
  Li]{lu2025ui}
Zhengxi Lu, Yuxiang Chai, Yaxuan Guo, Xi~Yin, Liang Liu, Hao Wang, Han Xiao,
  Shuai Ren, Guanjing Xiong, and Hongsheng Li.
\newblock Ui-r1: Enhancing efficient action prediction of gui agents by
  reinforcement learning.
\newblock \emph{arXiv preprint arXiv:2503.21620}, 2025.

\bibitem[Luo et~al.(2025)Luo, Wang, He, Chen, Li, and Xia]{luo2025gui}
Run Luo, Lu~Wang, Wanwei He, Longze Chen, Jiaming Li, and Xiaobo Xia.
\newblock Gui-r1: A generalist r1-style vision-language action model for gui
  agents.
\newblock \emph{arXiv preprint arXiv:2504.10458}, 2025.

\bibitem[Qin et~al.(2025)Qin, Ye, Fang, Wang, Liang, Tian, Zhang, Li, Li,
  Huang, et~al.]{qin2025ui}
Yujia Qin, Yining Ye, Junjie Fang, Haoming Wang, Shihao Liang, Shizuo Tian,
  Junda Zhang, Jiahao Li, Yunxin Li, Shijue Huang, et~al.
\newblock Ui-tars: Pioneering automated gui interaction with native agents.
\newblock \emph{arXiv preprint arXiv:2501.12326}, 2025.

\bibitem[Rawles et~al.(2023)Rawles, Li, Rodriguez, Riva, and
  Lillicrap]{rawles2023androidwild}
Christopher Rawles, Alice Li, Daniel Rodriguez, Oriana Riva, and Timothy
  Lillicrap.
\newblock {Android in the Wild}: A large-scale dataset for {Android} device
  control.
\newblock In \emph{Advances in Neural Information Processing Systems},
  volume~36, 2023.

\bibitem[Rawles et~al.(2024)Rawles, Clinckemaillie, Chang, Waltz, Lau, Fair,
  Li, Bishop, Li, Campbell-Ajala, et~al.]{rawles2024androidworld}
Christopher Rawles, Sarah Clinckemaillie, Yifan Chang, Jonathan Waltz,
  Gabrielle Lau, Marybeth Fair, Alice Li, William Bishop, Wei Li, Folawiyo
  Campbell-Ajala, et~al.
\newblock Androidworld: A dynamic benchmarking environment for autonomous
  agents.
\newblock \emph{arXiv preprint arXiv:2405.14573}, 2024.

\bibitem[Shao et~al.(2024)Shao, Wang, Zhu, Xu, Song, Bi, Zhang, Zhang, Li, Wu,
  et~al.]{shao2024deepseekmath}
Zhihong Shao, Peiyi Wang, Qihao Zhu, Runxin Xu, Junxiao Song, Xiao Bi, Haowei
  Zhang, Mingchuan Zhang, YK~Li, Yang Wu, et~al.
\newblock Deepseekmath: Pushing the limits of mathematical reasoning in open
  language models.
\newblock \emph{arXiv preprint arXiv:2402.03300}, 2024.

\bibitem[Shi et~al.(2025)Shi, Yu, Li, Wang, Zhang, Liu, Mi, and
  Yu]{shi2025mobilegui}
Yucheng Shi, Wenhao Yu, Zaitang Li, Yonglin Wang, Hongming Zhang, Ninghao Liu,
  Haitao Mi, and Dong Yu.
\newblock Mobilegui-rl: Advancing mobile gui agent through reinforcement
  learning in online environment.
\newblock \emph{arXiv preprint arXiv:2507.05720}, 2025.

\bibitem[Sui et~al.(2026)Sui, Huang, Li, Xu, Lv, Yan, Shen, Liu, Fan, Meng,
  et~al.]{sui2026androiddaily}
Yifan Sui, Xin Huang, Hongbing Li, Fang Xu, Jiahe Lv, Haolong Yan, Yeqing Shen,
  Litao Liu, Zhimin Fan, Ziyang Meng, et~al.
\newblock Androiddaily: A verifiable benchmark for mobile gui agents on
  real-world closed-source applications.
\newblock \emph{arXiv preprint arXiv:2605.27761}, 2026.

\bibitem[Sun et~al.(2025{\natexlab{a}})Sun, Cheng, Ding, Jin, Wang, Xu, Wu,
  Jia, Chen, Liu, et~al.]{sun2025genesis}
Qiushi Sun, Kanzhi Cheng, Zichen Ding, Chuanyang Jin, Yian Wang, Fangzhi Xu,
  Zhenyu Wu, Chengyou Jia, Liheng Chen, Zhoumianze Liu, et~al.
\newblock Os-genesis: Automating gui agent trajectory construction via reverse
  task synthesis.
\newblock In \emph{Proceedings of the 63rd Annual Meeting of the Association
  for Computational Linguistics (Volume 1: Long Papers)}, pages 5555--5579,
  2025{\natexlab{a}}.

\bibitem[Sun et~al.(2026)Sun, Han, Zhang, Pang, Wang, Cao, Huang, Duroiu,
  Zhang, Lin, et~al.]{sun2026agents}
Yiyou Sun, Xinyang Han, Weichen Zhang, Yuanbo Pang, Tianyu Wang, Yuhan Cao,
  Yixiao Huang, Chris Duroiu, Haoyun Zhang, Jeffrey Lin, et~al.
\newblock Agents' last exam.
\newblock \emph{arXiv preprint arXiv:2606.05405}, 2026.

\bibitem[Sun et~al.(2025{\natexlab{b}})Sun, Zhao, Yu, Wen, Va, Xu, Li, and
  Zhang]{sun2025gui}
Yuchen Sun, Shanhui Zhao, Tao Yu, Hao Wen, Samith Va, Mengwei Xu, Yuanchun Li,
  and Chongyang Zhang.
\newblock Gui-xplore: Empowering generalizable gui agents with one exploration.
\newblock In \emph{Proceedings of the Computer Vision and Pattern Recognition
  Conference}, pages 19477--19486, 2025{\natexlab{b}}.

\bibitem[Sun et~al.(2025{\natexlab{c}})Sun, Liu, Zang, Cao, Dong, Wu, Lin, and
  Wang]{sun2025seagent}
Zeyi Sun, Ziyu Liu, Yuhang Zang, Yuhang Cao, Xiaoyi Dong, Tong Wu, Dahua Lin,
  and Jiaqi Wang.
\newblock Seagent: Self-evolving computer use agent with autonomous learning
  from experience.
\newblock \emph{arXiv preprint arXiv:2508.04700}, 2025{\natexlab{c}}.

\bibitem[Tang et~al.(2026)Tang, Lu, Zhang, Lu, Xiao, Zhuang, and
  Shen]{tang2026clawgui}
Fei Tang, Zhiqiong Lu, Boxuan Zhang, Weiming Lu, Jun Xiao, Yueting Zhuang, and
  Yongliang Shen.
\newblock {ClawGUI}: A unified framework for training, evaluating, and
  deploying gui agents.
\newblock \emph{arXiv preprint arXiv:2604.11784}, 2026.

\bibitem[Wang et~al.(2024{\natexlab{a}})Wang, Xu, Jia, Zhang, Yan, Shen, Zhang,
  Huang, and Sang]{wang2024mobileagentv2}
Junyang Wang, Haiyang Xu, Haitao Jia, Xi~Zhang, Ming Yan, Weizhou Shen,
  Ji~Zhang, Fei Huang, and Jitao Sang.
\newblock {Mobile-Agent-v2}: Mobile device operation assistant with effective
  navigation via multi-agent collaboration.
\newblock \emph{arXiv preprint arXiv:2406.01014}, 2024{\natexlab{a}}.

\bibitem[Wang et~al.(2024{\natexlab{b}})Wang, Xu, Ye, Yan, Shen, Zhang, Huang,
  and Sang]{wang2024mobileagent}
Junyang Wang, Haiyang Xu, Jiabo Ye, Ming Yan, Weizhou Shen, Ji~Zhang, Fei
  Huang, and Jitao Sang.
\newblock {Mobile-Agent}: Autonomous multi-modal mobile device agent with
  visual perception.
\newblock \emph{arXiv preprint arXiv:2401.16158}, 2024{\natexlab{b}}.
\newblock ICLR 2024 Workshop on Large Language Model Agents.

\bibitem[Wang et~al.(2025{\natexlab{a}})Wang, Yu, Ye, Zhang, Liu, Liu, Chen,
  and Wang]{wang2025fedmabench}
Wenhao Wang, Zijie Yu, Rui Ye, Jianqing Zhang, Guangyi Liu, Liang Liu, Siheng
  Chen, and Yanfeng Wang.
\newblock Fedmabench: Benchmarking mobile gui agents on decentralized
  heterogeneous user data.
\newblock In \emph{Proceedings of the 2025 Conference on Empirical Methods in
  Natural Language Processing}, pages 26398--26419, 2025{\natexlab{a}}.

\bibitem[Wang et~al.(2025{\natexlab{b}})Wang, Yuan, Yu, Liu, Ye, Jin, Chen, and
  Wang]{wang2025mobilea3gent}
Wenhao Wang, Mengying Yuan, Zijie Yu, Guangyi Liu, Rui Ye, Tian Jin, Siheng
  Chen, and Yanfeng Wang.
\newblock Mobilea3gent: Training mobile gui agents using decentralized
  self-sourced data from diverse users.
\newblock \emph{arXiv preprint arXiv:2502.02982}, 2025{\natexlab{b}}.

\bibitem[Wen et~al.(2024)Wen, Li, Liu, Zhao, Yu, Li, Jiang, Liu, Zhang, and
  Liu]{wen2024autodroid}
Hao Wen, Yuanchun Li, Guohong Liu, Shanhui Zhao, Tao Yu, Toby Jia-Jun Li, Shiqi
  Jiang, Yunhao Liu, Yaqin Zhang, and Yunxin Liu.
\newblock {AutoDroid}: {LLM}-powered task automation in {Android}.
\newblock In \emph{Proceedings of the 30th Annual International Conference on
  Mobile Computing and Networking}, pages 543--557. ACM, 2024.
\newblock \doi{10.1145/3636534.3649379}.

\bibitem[Wu et~al.(2026)Wu, Guo, Cao, Lu, Zhu, Qu, Chen, Qin, Wang, Zhang,
  et~al.]{wu2026uioceanus}
Mengzhou Wu, Yuzhe Guo, Yuan Cao, Haochuan Lu, Songhe Zhu, Pingzhe Qu, Xin
  Chen, Kang Qin, Zhongpu Wang, Xiaode Zhang, et~al.
\newblock Ui-oceanus: Scaling gui agents with synthetic environmental dynamics.
\newblock \emph{arXiv preprint arXiv:2604.02345}, 2026.

\bibitem[Wu et~al.(2025)Wu, Wu, Xu, Wang, Sun, Jia, Cheng, Ding, Chen, Liang,
  and Qiao]{wu2025osatlas}
Zhiyong Wu, Zhenyu Wu, Fangzhi Xu, Yian Wang, Qiushi Sun, Chengyou Jia, Kanzhi
  Cheng, Zichen Ding, Liheng Chen, Paul~Pu Liang, and Yu~Qiao.
\newblock {OS-ATLAS}: A foundation action model for generalist {GUI} agents.
\newblock In \emph{The Thirteenth International Conference on Learning
  Representations}, 2025.

\bibitem[Xie et~al.(2025)Xie, Shao, Chen, Zhou, Li, Liu, Zhang, and
  Nie]{xie2025gui}
Bin Xie, Rui Shao, Gongwei Chen, Kaiwen Zhou, Yinchuan Li, Jie Liu, Min Zhang,
  and Liqiang Nie.
\newblock Gui-explorer: Autonomous exploration and mining of transition-aware
  knowledge for gui agent.
\newblock \emph{arXiv preprint arXiv:2505.16827}, 2025.

\bibitem[Xie et~al.(2024)Xie, Zhang, Chen, Li, Zhao, Cao, Hua, Cheng, Shin,
  Lei, et~al.]{xie2024osworld}
Tianbao Xie, Danyang Zhang, Jixuan Chen, Xiaochuan Li, Siheng Zhao, Ruisheng
  Cao, Toh~J Hua, Zhoujun Cheng, Dongchan Shin, Fangyu Lei, et~al.
\newblock Osworld: Benchmarking multimodal agents for open-ended tasks in real
  computer environments.
\newblock \emph{Advances in Neural Information Processing Systems},
  37:\penalty0 52040--52094, 2024.

\bibitem[Xu et~al.(2026)Xu, Zhang, Liu, Wang, Zhu, Zhou, Hu, Gao, Cao, Wang,
  et~al.]{xu2026mobile}
Haiyang Xu, Xi~Zhang, Haowei Liu, Junyang Wang, Zhaozai Zhu, Shengjie Zhou,
  Xuhao Hu, Feiyu Gao, Junjie Cao, Zihua Wang, et~al.
\newblock Mobile-agent-v3.5: Multi-platform fundamental gui agents.
\newblock \emph{arXiv preprint arXiv:2602.16855}, 2026.

\bibitem[Xu et~al.(2025)Xu, Liu, Sun, Cheng, Yu, Lai, Zhang, Zhang, Tang, and
  Dong]{xu2025androidlab}
Yifan Xu, Xiao Liu, Xueqiao Sun, Siyi Cheng, Hao Yu, Hanyu Lai, Shudan Zhang,
  Dan Zhang, Jie Tang, and Yuxiao Dong.
\newblock Androidlab: Training and systematic benchmarking of android
  autonomous agents.
\newblock In \emph{Proceedings of the 63rd Annual Meeting of the Association
  for Computational Linguistics (Volume 1: Long Papers)}, pages 2144--2166,
  2025.

\bibitem[Xu et~al.(2024)Xu, Wang, Wang, Lu, Xie, Saha, Sahoo, Yu, and
  Xiong]{xu2024aguvis}
Yiheng Xu, Zekun Wang, Junli Wang, Dunjie Lu, Tianbao Xie, Amrita Saha, Doyen
  Sahoo, Tao Yu, and Caiming Xiong.
\newblock Aguvis: Unified pure vision agents for autonomous gui interaction.
\newblock \emph{arXiv preprint arXiv:2412.04454}, 2024.

\bibitem[Xue et~al.(2026)Xue, Liao, Shi, Wang, Zhang, Song, Su, and
  Sun]{xue2026acurl}
Tianci Xue, Zeyi Liao, Tianneng Shi, Zilu Wang, Kai Zhang, Dawn Song, Yu~Su,
  and Huan Sun.
\newblock Autonomous continual learning of computer-use agents for environment
  adaptation.
\newblock \emph{arXiv preprint arXiv:2602.10356}, 2026.

\bibitem[Yan et~al.(2023)Yan, Yang, Zhu, Lin, Li, Wang, Yang, Zhong, McAuley,
  Gao, et~al.]{yan2023gpt}
An~Yan, Zhengyuan Yang, Wanrong Zhu, Kevin Lin, Linjie Li, Jianfeng Wang,
  Jianwei Yang, Yiwu Zhong, Julian McAuley, Jianfeng Gao, et~al.
\newblock Gpt-4v in wonderland: Large multimodal models for zero-shot
  smartphone gui navigation.
\newblock \emph{arXiv preprint arXiv:2311.07562}, 2023.

\bibitem[Yan et~al.(2025)Yan, Wang, Huang, Shen, Meng, Fan, Tan, Gao, Shi,
  Yang, et~al.]{yan2025step}
Haolong Yan, Jia Wang, Xin Huang, Yeqing Shen, Ziyang Meng, Zhimin Fan, Kaijun
  Tan, Jin Gao, Lieyu Shi, Mi~Yang, et~al.
\newblock Step-gui technical report.
\newblock \emph{arXiv preprint arXiv:2512.15431}, 2025.

\bibitem[Yang et~al.(2025)Yang, Su, Liu, Dong, Yu, Su, Wang, Liu, Zhu, Li,
  et~al.]{yang2025zerogui}
Chenyu Yang, Shiqian Su, Shi Liu, Xuan Dong, Yue Yu, Weijie Su, Xuehui Wang,
  Zhaoyang Liu, Jinguo Zhu, Hao Li, et~al.
\newblock Zerogui: Automating online gui learning at zero human cost.
\newblock \emph{arXiv preprint arXiv:2505.23762}, 2025.

\bibitem[You et~al.(2025)You, Zhang, Schoop, Weers, Swearngin, Nichols, Yang,
  and Gan]{you2025ferretui}
Keen You, Haotian Zhang, Eldon Schoop, Floris Weers, Amanda Swearngin, Jeffrey
  Nichols, Yinfei Yang, and Zhe Gan.
\newblock {Ferret-UI}: Grounded mobile {UI} understanding with multimodal
  {LLM}s.
\newblock In \emph{Computer Vision -- ECCV 2024}, volume 15122 of \emph{Lecture
  Notes in Computer Science}, pages 240--255. Springer Nature Switzerland,
  2025.
\newblock \doi{10.1007/978-3-031-73039-9_14}.

\bibitem[Yuan et~al.(2026)Yuan, Zhou, Xiong, Wu, Sun, Song, Cui, Wang, Wu, Li,
  et~al.]{yuan2026osworld2}
Mengqi Yuan, Zilong Zhou, Xinzhuang Xiong, Weiming Wu, Jiayang Sun, Jiamin
  Song, Kaiqian Cui, Bowen Wang, Haoyuan Wu, Yitong Li, et~al.
\newblock Osworld2. 0: Benchmarking computer use agents on long-horizon
  real-world tasks.
\newblock \emph{arXiv preprint arXiv:2606.29537}, 2026.

\bibitem[Zhang et~al.(2025{\natexlab{a}})Zhang, Shang, Gao, Zhang, Xie, Ma,
  Yuan, Wu, Zhu, and Li]{zhang2025tongui}
Bofei Zhang, Zirui Shang, Zhi Gao, Wang Zhang, Rui Xie, Xiaojian Ma, Tao Yuan,
  Xinxiao Wu, Song-Chun Zhu, and Qing Li.
\newblock Tongui: Building generalized gui agents by learning from multimodal
  web tutorials.
\newblock \emph{arXiv e-prints}, pages arXiv--2504, 2025{\natexlab{a}}.

\bibitem[Zhang et~al.(2025{\natexlab{b}})Zhang, Yang, Liu, Li, Han, Chen,
  Huang, Fu, and Yu]{zhang2025appagent}
Chi Zhang, Zhao Yang, Jiaxuan Liu, Yanda Li, Yucheng Han, Xin Chen, Zebiao
  Huang, Bin Fu, and Gang Yu.
\newblock {AppAgent}: Multimodal agents as smartphone users.
\newblock In \emph{Proceedings of the 2025 CHI Conference on Human Factors in
  Computing Systems}. ACM, 2025{\natexlab{b}}.
\newblock \doi{10.1145/3706598.3713600}.

\bibitem[Zhao et~al.(2025)Zhao, Liu, Liang, He, Lu, Huang, Guo, Zhang, Wang,
  Liu, et~al.]{zhao2025mas}
Pengxiang Zhao, Guangyi Liu, Yaozhen Liang, Weiqing He, Zhengxi Lu, Yuehao
  Huang, Yaxuan Guo, Kexin Zhang, Hao Wang, Liang Liu, et~al.
\newblock Mas-bench: A unified benchmark for shortcut-augmented hybrid mobile
  gui agents.
\newblock \emph{arXiv preprint arXiv:2509.06477}, 2025.

\bibitem[Zhou et~al.(2025)Zhou, Zhang, Tong, Zhang, Chen, Kong, Cai, Liu, Wang,
  Zhou, et~al.]{zhou2025mai}
Hanzhang Zhou, Xu~Zhang, Panrong Tong, Jianan Zhang, Liangyu Chen, Quyu Kong,
  Chenglin Cai, Chen Liu, Yue Wang, Jingren Zhou, et~al.
\newblock Mai-ui technical report: Real-world centric foundation gui agents.
\newblock \emph{arXiv preprint arXiv:2512.22047}, 2025.

\bibitem[Zhou et~al.(2024)Zhou, Xu, Zhu, Zhou, Lo, Sridhar, Cheng, Ou, Bisk,
  Fried, Alon, and Neubig]{zhou2024webarena}
Shuyan Zhou, Frank~F. Xu, Hao Zhu, Xuhui Zhou, Robert Lo, Abishek Sridhar,
  Xianyi Cheng, Tianyue Ou, Yonatan Bisk, Daniel Fried, Uri Alon, and Graham
  Neubig.
\newblock {WebArena}: A realistic web environment for building autonomous
  agents.
\newblock In \emph{The Twelfth International Conference on Learning
  Representations}, 2024.

\end{thebibliography}

\clearpage
\appendix
\clearpage
\section{Detailed Related Work}
\label{sec:related}

Figure~\ref{fig:mobileforge-teaser} organizes annotation-free GUI learning into three representative paradigms according to how experience is obtained and whether it updates the policy. These methods build on increasingly capable GUI grounding and action models \citep{cheng2024seeclick,you2025ferretui,wu2025osatlas,xu2024aguvis}, but differ in their reliance on external supervision, use of target-app interaction, and granularity of learning signals. Grounding models provide the initial ability to understand screenshots, locate interface elements, and emit structured actions; the paradigms below determine how that capability is adapted after deployment. Together, these dimensions explain why reducing manual annotation alone does not necessarily yield target-grounded experience, fine-grained credit assignment, and reusable policy improvement.

\paragraph{Inference-Time Knowledge Augmentation.}
This paradigm explores an interface, summarizes the collected experience, and retrieves the resulting knowledge to guide a fixed policy at inference time. GUI-explorer autonomously traverses target apps, mines transition-aware knowledge from observation--action--outcome triples, and injects retrieved app-specific knowledge into later prompts \citep{xie2025gui}. GUI-Xplore similarly uses compact exploration to expose functions beyond the landing screen and improve generalization across GUI tasks \citep{sun2025gui}. Related mobile agents also augment execution with app knowledge: AppAgent learns operating knowledge through autonomous exploration, while AutoDroid uses automatically acquired app knowledge to guide LLM-based task execution \citep{zhang2025appagent,wen2024autodroid}. Mobile-Agent and Mobile-Agent-v2 complement stored knowledge with inference-time planning, reflection, and multi-agent coordination \citep{wang2024mobileagent,wang2024mobileagentv2}. These methods show that direct exploration can reveal app affordances absent from static model knowledge and can refresh guidance without model training. This separation is useful for API-only agents or rapidly changing apps because knowledge can be replaced independently of model weights. However, the collected experience remains prompt-side context: the underlying policy is unchanged, and its weaknesses can recur when retrieval is incomplete, the task changes, or a relevant failure has not yet been summarized.

\paragraph{Teacher / Synthetic-Data Training.}
A second paradigm converts external or synthesized experience into offline supervision. TongUI transforms multimodal web tutorials into generalized GUI trajectories across platforms and applications \citep{zhang2025tongui}. MobileA3gent instead collects decentralized user-phone trajectories, automatically annotates their intents, and applies federated training \citep{wang2025mobilea3gent}. OS-Genesis reverses the task-first pipeline: it explores GUI transitions, synthesizes tasks from observed state changes, filters the resulting trajectories, and uses GPT-4o to execute the tasks; treating these actions as SFT targets effectively distills a proprietary teacher \citep{sun2025genesis,hurst2024gpt}. Human interaction corpora and few-shot demonstration-based adaptation provide related sources of mobile supervision \citep{rawles2023androidwild,liu2025learnact}, while UI-Oceanus broadens coverage by synthesizing environmental dynamics and training on transition-oriented objectives \citep{wu2026uioceanus}. Unlike inference-time augmentation, these methods produce trainable policy improvements and reduce the need for manually authored GUI trajectories. Their supervision nevertheless originates from tutorials, user traces, strong teachers, demonstrations, or synthetic environments. Scaling such data improves coverage, but does not ensure that its state-action distribution matches the policy currently being adapted. It may therefore be stale or mismatched with the current state and reachable functions of a target app. Even when data are collected from the same interface, the selected paths primarily capture teacher behavior rather than diagnosing the adapting policy's own failures, limiting cross-attempt correction and fine-grained learning targets.

\paragraph{Sparse-Reward Rollout Learning.}
The third paradigm closes the interaction-to-training loop by rolling out the current policy and optimizing it with automatically estimated rewards. ZeroGUI automatically generates tasks, evaluates trajectory-level success with a VLM, and updates the GUI policy through online reinforcement learning \citep{yang2025zerogui}. MobileGUI-RL further combines self-exploration, task filtering, trajectory-aware advantages, and composite rewards for online mobile GUI optimization \citep{shi2025mobilegui}. Broader self-evolving computer-use systems pursue a similar goal: SEAgent combines curriculum generation, step-wise assessment, experience accumulation, and policy learning, while ACuRL formulates environment adaptation as continual interaction, automatic evaluation, and repeated training \citep{sun2025seagent,xue2026acurl}. Their continual-learning perspective is complementary, but these systems target desktop or general computer-use environments rather than a unified mobile interaction and evaluation substrate. Collectively, these methods demonstrate that agents can improve from self-collected interaction, but their principal optimization signals remain trajectory-level success or composite scalar rewards. A single score can conflate task completion, efficiency, and action quality, making it difficult to decide which local behavior should be retained. Such signals provide coarse credit assignment for long-horizon mobile tasks: a failed rollout may contain correct local decisions, while a successful rollout may contain redundant or accidental actions. Independent attempts also lack an explicit mechanism for carrying a diagnosed mistake into the next rollout.

\ourmethod addresses these limitations within the rollout-learning paradigm. \mobilegym grounds task mining and evaluation in reachable target-app interaction and returns trajectory outcomes, step-level process feedback, and corrective hints. \hifpo reuses those hints across serialized attempts, filters mastered tasks, preserves informative trajectories and local decisions, and performs hint-contextualized step-level policy optimization. Unlike static retrieval or offline distillation, the same feedback both guides the next attempt and determines reusable policy updates. Thus, self-collected experience supports immediate correction and trainable improvement without human-written tasks, demonstrations, or reward labels.

\section{Method Notation and Algorithm}
\label{sec:appendix_method_details}

Table~\ref{tab:method_notation} lists the main symbols used in Section~\ref{sec:method}. Algorithm~\ref{alg:mobileforge} gives a compact procedural view of the annotation-free adaptation loop.

\begin{table}[t]
\centering
\paperTable
\caption{Key notation used in the method.}
\label{tab:method_notation}
\begin{tabular}{L{0.23\linewidth} L{0.66\linewidth}}
\toprule
\textbf{Symbol} & \textbf{Meaning} \\
\midrule
$\mathcal{E}$ & Target mobile app environment. \\
$\mathcal{Z}$ & Exploration evidence collected from target apps. \\
$\mathcal{T}$ & Generated adaptation curriculum. \\
$x$ & A generated task in curriculum $\mathcal{T}$. \\
$\tau_k$ & The $k$-th rollout attempt for task $x$. \\
$\eta_{<k}$ & Corrective hint context from earlier attempts of task $x$. \\
$\mathcal{F}_{k}$ & Hierarchical feedback for attempt $\tau_k$. \\
$I_k^{(t)}$ & Screenshot observation at attempt $k$ and step $t$. \\
$s_k^{(t)}$ & Decision state at attempt $k$ and step $t$. \\
$a_k^{(t)}=(\alpha,\psi)$ & Structured GUI action with type $\alpha$ and arguments $\psi$. \\
$z_k$ & Trajectory outcome label; $1$ means task success and $0$ means failure. \\
$\ell_k^{(t)}$ & Step-level process label. \\
$v_k^{(t)}$ & Binary reasonableness label for step $t$ in attempt $k$. \\
$e_k^{(t)}$ & Natural-language rationale for the step-level label. \\
$\chi_k^{(t)}$ & Indicator that a step is marked reasonable by \critic. \\
$Q_k$ & Local quality score of attempt $\tau_k$. \\
$h_k$ & Corrective hint generated after an attempt. \\
$d_j=(s_j,a_j^\star)$ & Step-level training sample and selected action. \\
$\tilde{s}_j$ & Hint-contextualized prompt used by step-level GRPO. \\
$G$ & Number of candidate responses sampled for a step-level prompt. \\
$\hat{o}_{j,g},\hat{a}_{j,g}$ & Candidate response and its parsed GUI action. \\
$R_{j,g}$ & Adaptive GUI action reward for candidate $g$ in a GRPO group. \\
$A_{j,g}$ & Group-relative advantage of candidate $g$ for step $j$. \\
$\rho_{j,g}$ & Policy ratio for candidate response $g$. \\
\bottomrule
\end{tabular}
\end{table}

\begin{algorithm}[t]
\caption{\ourmethod annotation-free adaptation loop}
\label{alg:mobileforge}
\KwIn{Target apps $\mathcal{E}$, policy $\pi_\theta$, attempts per task $K$}
\KwOut{Adapted policy $\pi_{\theta'}$}
Explore target apps and record reachable GUI transitions\;
Generate trajectory-grounded tasks from the explored transitions\;
\ForEach{task $x\in\mathcal{T}$}{
    initialize hint context $\eta_{<1}\leftarrow\emptyset$\;
    \For{$k=1$ \KwTo $K$}{
        run attempt $\tau_k$ with policy $\pi_\theta$ and hints $\eta_{<k}$\;
        evaluate the attempt to obtain outcome label $z$, step labels $\ell$, and hint $h$\;
        update the hint context for later attempts\;
    }
}
Remove mastered tasks, select informative attempts, and extract useful local steps\;
Train the policy with hint-contextualized step-level GRPO\;
\Return{$\pi_{\theta'}$}\;
\end{algorithm}

\section{Pipeline Details}
\label{sec:appendix_pipeline}

Table~\ref{tab:appendix_pipeline} summarizes the main system stages and their inputs and outputs.

\begin{table}[t]
\centering
\paperTable
\caption{Pipeline stages used by \ourmethod.}
\label{tab:appendix_pipeline}
\begin{paperFit}
\begin{tabular}{@{}l l l@{}}
\toprule
\textbf{Stage} & \textbf{Input} & \textbf{Output} \\
\midrule
Target exploration & Target apps & Exploration evidence \\
Curriculum generation & Exploration evidence & Executable task curriculum \\
HiFPO hint-guided rollout & Tasks and current policy & Multi-attempt trajectories \\
MobileGym hierarchical evaluation & Completed trajectories & Outcome feedback, process feedback, hints \\
Step-level filtering & Tasks, trajectories, feedback & Filtered step-level training samples \\
Policy optimization & Step-level samples & Adapted policy \\
\bottomrule
\end{tabular}
\end{paperFit}
\end{table}

\paragraph{Filtering order.}
The task-level success-rate filter is applied before best-trajectory selection. This matters because the task success rate must be computed from the original multi-attempt task group. The main setting uses $\operatorname{SR}_{\min}=0.0$ and $\operatorname{SR}_{\max}<1.0$, which removes all-success mastered tasks and keeps all-fail and partially solved tasks. Best-trajectory selection and reasonable-step filtering are then applied to extract useful local decisions.

\section{Formal HiFPO Details}
\label{sec:appendix_hifpo_formal}

This section expands the concise description of \hifpo in Section~\ref{sec:optimization}. For a task $x$, corrective hints from previous attempts are aggregated as
\begin{equation}
\label{eq:hint-context}
    \eta_{<k}=\operatorname{Aggregate}(h_1,\ldots,h_{k-1}),
\end{equation}
and the next attempt is sampled by
\begin{equation}
\label{eq:hifpo-rollout}
    \tau_k\sim \operatorname{Rollout}(\pi_\theta,x,\eta_{<k}).
\end{equation}
The empirical multi-attempt success rate is
\begin{equation}
\label{eq:success-rate}
    \operatorname{SR}(x)=\frac{1}{K}\sum_{k=1}^{K} z_k.
\end{equation}
Tasks with $\operatorname{SR}(x)=1$ are removed; tasks with $\operatorname{SR}(x)=0$ or $0<\operatorname{SR}(x)<1$ are retained for step extraction.

For every step, the process label induces a reasonable-step indicator
\begin{equation}
\label{eq:reasonable-step}
    \chi_k^{(t)}=\mathbb{I}[v_k^{(t)}=1],
\end{equation}
and each attempt receives a local quality score
\begin{equation}
\label{eq:quality-score}
    Q_k=\frac{1}{T_k}\sum_{t=1}^{T_k}\chi_k^{(t)}.
\end{equation}
For a retained task, \hifpo selects one informative attempt:
\begin{equation}
\label{eq:attempt-selection}
k^\star(x) =
\begin{cases}
\arg\max_{k:z_k=1} Q_k, & \exists k,\ z_k=1,\\
\arg\max_k Q_k, & \text{otherwise}.
\end{cases}
\end{equation}
The filtered step-level training set is
\begin{equation}
\label{eq:training-set}
    \mathcal{D}=
    \bigcup_{\substack{x\in\mathcal{T}\\ \operatorname{SR}(x)<1}}
    \{(s_{k^\star(x)}^{(t)},a_{k^\star(x)}^{(t)})\mid
    \chi_{k^\star(x)}^{(t)}=1\}.
\end{equation}
Within each set in the union, $s_k^{(t)}$, $a_k^{(t)}$, and $\chi_k^{(t)}$ refer to the rollouts of the current task $x$.

After filtering, each training example is a selected local decision $d_j=(s_j,a_j^\star)\in\mathcal{D}$, where $a_j^\star=(\alpha_j^\star,\psi_j^\star)$. The decision state is rendered as a hint-contextualized prompt
\begin{equation}
\label{eq:hint-prompt}
    \tilde{s}_j
    =\operatorname{Prompt}(s_j).
\end{equation}
For each prompt, the old policy samples a response group:
\begin{equation}
\label{eq:hint-grpo-sampling}
    \hat{o}_{j,g}\sim \pi_{\theta_{\rm old}}(\cdot\mid\tilde{s}_j),
    \quad g=1,\ldots,G.
\end{equation}
Each response is parsed into $\hat a_{j,g}=\operatorname{Parse}(\hat{o}_{j,g})=(\hat{\alpha}_{j,g},\hat{\psi}_{j,g})$. Unparseable responses receive zero action reward. For parseable responses, the adaptive GUI action reward is
\begin{equation}
\label{eq:gui-reward}
\begin{aligned}
    r^{\rm type}_{j,g}
    &=\mathbb{I}[\hat{\alpha}_{j,g}=\alpha_j^\star],\\
    r^{\rm arg}_{j,g}
    &=r^{\rm type}_{j,g}\,
      S_{\alpha_j^\star}(\hat{\psi}_{j,g},\psi_j^\star),\\
    R_{j,g}
    &=\lambda_{\rm type}r^{\rm type}_{j,g}
      +\lambda_{\rm arg}r^{\rm arg}_{j,g}.
\end{aligned}
\end{equation}
Here $S_{\alpha_j^\star}$ is gated by type correctness and uses action-specific matching; \textit{Appendix~\ref{sec:appendix_reward} gives the concrete rule table}.

Rewards are normalized within the response group:
\begin{equation}
\label{eq:group-stats}
\begin{aligned}
    \mu_j&=\frac{1}{G}\sum_{g=1}^{G}R_{j,g},\\
    \sigma_j&=\operatorname{Std}_{g=1,\ldots,G}(R_{j,g}),
\end{aligned}
\end{equation}
with group-relative advantage
\begin{equation}
\label{eq:grpo-adv}
    A_{j,g}=
    \frac{R_{j,g}-\mu_j}{\sigma_j+\epsilon_{\rm std}}.
\end{equation}
The policy ratio and its clipped counterpart are
\begin{equation}
\label{eq:grpo-ratio}
\begin{aligned}
    \rho_{j,g}(\theta)=
    \frac{\pi_\theta(\hat{o}_{j,g}\mid \tilde{s}_j)}
         {\pi_{\theta_{\rm old}}(\hat{o}_{j,g}\mid \tilde{s}_j)},\\
    \bar{\rho}_{j,g}(\theta)=
    \operatorname{clip}\!\left(
    \rho_{j,g}(\theta),
    1-\epsilon_{\rm low},
    1+\epsilon_{\rm high}
    \right).
\end{aligned}
\end{equation}
\hifpo minimizes
\begin{equation}
\label{eq:hifpo-objective}
\begin{aligned}
\mathcal{L}_{\hifpo}(\theta)
=&-\mathbb{E}_{j,g}
\left[\min\left(\rho_{j,g}A_{j,g},
\bar{\rho}_{j,g}A_{j,g}\right)\right]\\
&+\beta\,\mathbb{E}_{j}
\left[D^{\rm KL}_{j}(\theta)\right],
\end{aligned}
\end{equation}
where $D^{\rm KL}_{j}(\theta)$ regularizes the policy against $\pi_{\rm ref}$
for step prompt $\tilde{s}_j$.

\section{Experimental Protocol Details}
\label{sec:appendix_experimental_protocol}

\paragraph{Benchmarks.}
AndroidWorld~\citep{rawles2024androidworld} is the in-domain setting: \ourmethod explores the AndroidWorld app ecosystem, mines adaptation tasks, collects \hifpo rollouts, and evaluates on 116 AndroidWorld tasks with Pass@1, Pass@2, and Pass@3. MobileWorld GUI-only~\citep{kong2026mobileworld} is the out-of-domain setting: we evaluate on its 117-task split and use no MobileWorld rollout, task, or feedback for adaptation.
This mobile protocol complements broader web and cross-environment agent benchmarks \citep{deng2023mind2web,zhou2024webarena,he2024webvoyager,liu2024agentbench} and systematic mobile training and evaluation suites \citep{xu2025androidlab,zhao2025mas}.

\paragraph{Base agents and adaptation scale.}
We use two 8B-scale instruct base agents: the open generalist \llmname{Qwen3-VL-8B-Instruct}, abbreviated as \llmname{Qwen3-VL-8B}, and the GUI-specialized \llmname{GUI-Owl-1.5-8B-Instruct}, abbreviated as \llmname{GUI-Owl-1.5-8B}~\citep{bai2025qwen3,xu2026mobile}. \ourmethod generates 3,249 AndroidWorld-side candidate tasks grounded in 527 source trajectory identifiers from 20 apps. To study scaling under realistic compute constraints, we train with 200-, 400-, and 900-task subsets. The main 900-task 8B runs use eight 80GB GPUs and take roughly 80 hours.

\paragraph{Ablation scope.}
The ablations isolate corrective hints during rollout, hint-contextualized GRPO versus SFT, task-level success-rate filtering, final-decision evaluator choice, and trajectory-grounded curriculum coverage.

\section{Detailed Ablation Results}
\label{sec:appendix_ablation_details}

Figure~\ref{fig:main_ablation} summarizes the principal ablation trends in the main paper. The tables below preserve the complete numerical results, including the raw success counts and percentages, rollout-efficiency statistics, critic precision and recall, and curriculum category counts.

\begin{table}[h]
\centering
\paperTable
\caption{Trajectory-level rollout ablation on 200 generated tasks with \llmname{Qwen3-VL-8B}. Corrective hints are generated from previous attempts of the same task.}
\label{tab:hint_rollout_ablation_main}
\begin{tabular}{@{}l c c r@{}}
\toprule
\textbf{Metric} & \textbf{No Hint Context} & \textbf{With Corrective Hints} & \textbf{Gain} \\
\midrule
Overall success & 52.0\% & \textbf{77.0\%} & \rise{+25.0 pp} \\
Pass@1 & 30.5\% & \textbf{44.5\%} & \rise{+14.0 pp} \\
Pass@2 & 42.5\% & \textbf{64.0\%} & \rise{+21.5 pp} \\
Pass@3 & 49.0\% & \textbf{72.5\%} & \rise{+23.5 pp} \\
Pass@4 & 52.0\% & \textbf{77.0\%} & \rise{+25.0 pp} \\
Avg. steps / attempt & 18.4 & \textbf{17.2} & \rise{-1.2} \\
Total steps (success only) & 2,593 & \textbf{4,711} & \rise{+2,118} \\
\bottomrule
\end{tabular}
\end{table}

\begin{table*}[!t]
\centering
\paperCompactTable
\caption{Downstream corrective-hint ablation under matched 200-task \hifpo training. All adaptation data are from AndroidWorld; no MobileWorld task, rollout, or feedback is used for adaptation.}
\label{tab:downstream_hint_ablation_main}
\begin{paperFit}[\textwidth]
\begin{tabular}{@{}l l c c c@{}}
\toprule
\textbf{Base Agent} &
\textbf{Training Setting} &
\textbf{AndroidWorld Pass@1} &
\textbf{AndroidWorld Pass@3} &
\textbf{MobileWorld GUI-Only} \\
\midrule
\llmname{Qwen3-VL-8B}
& Base, no training
& 47/116 (40.5\%)
& 64/116 (55.2\%)
& 9/117 (7.7\%) \\
\llmname{Qwen3-VL-8B}
& \hifpo without corrective hints
& 49/116 (42.2\%)
& 68/116 (58.6\%)
& 12/117 (10.3\%) \\
\rowcolor{oursrowcolor}
\llmname{Qwen3-VL-8B}
& \textbf{\hifpo with corrective hints}
& \textbf{55/116 (47.4\%)}
& \textbf{71/116 (61.2\%)}
& \textbf{15/117 (12.8\%)} \\
\midrule
\llmname{GUI-Owl-1.5-8B}
& Base, no training
& 65/116 (56.0\%)
& 80/116 (69.0\%)
& 44/117 (37.6\%) \\
\llmname{GUI-Owl-1.5-8B}
& \hifpo without corrective hints
& 71/116 (61.2\%)
& 83/116 (71.6\%)
& 43/117 (36.8\%) \\
\rowcolor{oursrowcolor}
\llmname{GUI-Owl-1.5-8B}
& \textbf{\hifpo with corrective hints}
& \textbf{75/116 (64.7\%)}
& \textbf{86/116 (74.1\%)}
& \textbf{46/117 (39.3\%)} \\
\bottomrule
\end{tabular}
\end{paperFit}
\end{table*}

\begin{table}[h]
\centering
\paperTable
\caption{Training objective ablation with \llmname{Qwen3-VL-8B}. AndroidWorld numbers report Pass@1 over 116 tasks. The best result is bolded and second-best results are underlined.}
\label{tab:training_objective_ablation_main}
\begin{tabular}{@{}l c r@{}}
\toprule
\textbf{Method} & \textbf{Tasks} & \textbf{AndroidWorld Pass@1} \\
\midrule
Base & 0 & 47/116 (40.5\%) \\
No-hint SFT & 200 & 40/116 (34.5\%) \\
Hint SFT & 200 & 53/116 (45.7\%) \\
Hint-contextualized GRPO & 200 & \underline{55/116 (47.4\%)} \\
No-hint SFT & 900 & 51/116 (44.0\%) \\
Hint SFT & 900 & \underline{55/116 (47.4\%)} \\
\rowcolor{oursrowcolor}
Hint-contextualized GRPO & 900 & \textbf{59/116 (50.9\%)} \\
\bottomrule
\end{tabular}
\end{table}

\begin{table}[h]
\centering
\paperTable
\caption{Final-decision model ablation for \llmname{Qwen3-VL-8B} in the 200-task hint-contextualized GRPO setting. The step-description model is kept fixed. Best values are bolded and second-best values are underlined.}
\label{tab:evaluator_model_ablation_main}
\begin{tabular}{@{}l c c c r@{}}
\toprule
\textbf{Decision Model} & \textbf{Pass@1} & \textbf{Pass@2} & \textbf{Pass@3} & \textbf{MW-GUI} \\
\midrule
Base, no training & 47/116 & 57/116 & 64/116 & 9/117 \\
\rowcolor{oursrowcolor}
Gemini-2.5-Pro & \textbf{55/116} & \underline{64/116} & \textbf{71/116} & \textbf{15/117} \\
Gemini-3.1-Pro-Preview & \underline{52/116} & 62/116 & 69/116 & \underline{13/117} \\
Qwen3-VL-8B & \underline{52/116} & \textbf{67/116} & \underline{70/116} & 11/117 \\
\bottomrule
\end{tabular}
\end{table}

\begin{table}[h]
\centering
\paperCompactTable
\caption{\critic agreement with reference labels and downstream policy results. Each configuration contains 400 trajectories. AndroidWorld reports Pass@1/2/3, and MobileWorld reports GUI-only success count.}
\label{tab:critic_reference_agreement_main}
\begin{paperFit}
\begin{tabular}{@{}l c c@{}}
\toprule
\textbf{Step-Description Model}
& \llmname{Qwen3-VL-8B}
& \llmname{Qwen3-VL-8B} \\
\textbf{Final-Decision Model}
& Gemini-3.1-Pro-Preview
& \llmname{Qwen3-VL-8B} \\
\midrule
\textbf{Trajectory Acc./P/R/F1 (\%)}
& 89.75 / 90.08 / 92.77 / 91.40
& 77.00 / 72.81 / 84.69 / 78.30 \\
\textbf{Step Acc./P/R/F1 (\%)}
& 86.53 / 84.81 / 94.41 / 89.35
& 71.37 / 65.05 / 91.94 / 76.19 \\
\textbf{Step Coverage}
& 99.77\%
& 93.02\% \\
\textbf{AW Pass@1/2/3}
& 52/62/69
& 52/67/70 \\
\textbf{MW-GUI}
& 13/117
& 11/117 \\
\bottomrule
\end{tabular}
\end{paperFit}
\end{table}

\begin{table}[h]
\centering
\paperCompactTable
\caption{Functional coverage of Broccoli tasks. Percentages are relative to each generated curriculum.}
\label{tab:curriculum_functional_coverage_main}
\begin{tabular}{@{}l r r r r@{}}
\toprule
\textbf{Functionality} & \multicolumn{2}{c}{\textbf{Landing-Screen Baseline}} & \multicolumn{2}{c}{\textbf{\ourmethod Curriculum}} \\
\cmidrule(lr){2-3}\cmidrule(lr){4-5}
& \textbf{Count} & \textbf{\%} & \textbf{Count} & \textbf{\%} \\
\midrule
Recipe creation & 49 & 16.3 & 14 & 5.0 \\
Recipe editing & 42 & 14.0 & 35 & 12.5 \\
Recipe deletion & 82 & 27.3 & 4 & 1.4 \\
Search and filter & 25 & 8.3 & 38 & 13.6 \\
Information retrieval / QA & 32 & 10.7 & 20 & 7.1 \\
Favorites & 0 & 0.0 & 3 & 1.1 \\
Shopping list & 0 & 0.0 & 33 & 11.8 \\
Cooking assistant & 0 & 0.0 & 26 & 9.3 \\
Meal planner & 0 & 0.0 & 13 & 4.6 \\
Settings and configuration & 0 & 0.0 & 9 & 3.2 \\
Media and sharing & 0 & 0.0 & 8 & 2.9 \\
Other & 70 & 23.3 & 77 & 27.5 \\
\midrule
Total & 300 & 100.0 & 280 & 100.0 \\
\bottomrule
\end{tabular}
\end{table}

\section{Annotation-Free Adaptation Data Details}
\label{sec:appendix_data_details}

Table~\ref{tab:generated_task_corpus} reports the generated AndroidWorld-side adaptation task pool by source app. The full pool contains 3,249 candidate tasks grounded in 527 source trajectory identifiers from 20 apps.

\begin{table}[h]
\centering
\paperTable
\caption{Generated AndroidWorld-side adaptation tasks by source app. The full pool contains 3,249 candidate tasks grounded in 527 source trajectory identifiers from 20 apps.}
\label{tab:generated_task_corpus}
\begin{tabular}{@{}l r@{\qquad\qquad}l r@{}}
\toprule
\textbf{App} & \textbf{\#Tasks} & \textbf{App} & \textbf{\#Tasks} \\
\midrule
Files & 258 & Tasks & 150 \\
Pro Expense & 247 & Simple Draw Pro & 145 \\
Broccoli Recipe & 241 & OsmAnd & 136 \\
Simple SMS Messenger & 240 & Joplin & 130 \\
Markor & 233 & Simple Calendar Pro & 107 \\
Clock & 215 & OpenTracks & 104 \\
Retro Music & 189 & Settings & 102 \\
Contacts & 166 & Camera & 94 \\
VLC & 161 & Audio Recorder & 91 \\
Chrome & 157 & Simple Gallery Pro & 83 \\
\bottomrule
\end{tabular}
\end{table}

\section{Exploration Phase Details}
\label{sec:exploration_details}

\mobilegym begins by collecting raw interaction evidence from each target app. We adopt a function-aware exploration strategy inspired by GUI-explorer~\citep{xie2025gui}, replacing unguided random walks with app-grounded goal generation and systematic traversal.
This stage relies on screen parsing, visual grounding, and structured action generation studied by prior mobile and cross-platform GUI agents \citep{yan2023gpt,lu2024omniparser,gou2025uground,lin2025showui,qin2025ui,li2025screenspotpro}.

\paragraph{Exploration anchors.}
For an Android app, the explorer extracts structural anchors from app metadata, especially the activity list declared by the APK. These anchors provide a compact description of reachable app functions and screens. They do not serve as demonstrations; they only ground exploration goals in functions that the app plausibly exposes.

\paragraph{Function-aware goal generation.}
At each exploration state, an MLLM receives the current screenshot together with the app name, package name, and available activity anchors. It generates concrete user goals that start from the current screen, are expected to be executable within a bounded number of steps, and cover diverse interaction patterns such as viewing, editing, searching, sharing, configuration, and information lookup. This design makes the explored trajectories more likely to touch real app functions than generic task templates.

\paragraph{Depth-first trajectory collection.}
The explorer pursues generated goals with depth-first traversal. When branching to a new goal from an earlier state, it restores the parent state by relaunching the app and replaying the parent action prefix, then continues exploration from that state. For each transition, it records the task goal, before/after screenshots, selected action, target element, execution metadata, and a short summary. The output is a rich but unstructured evidence pool $\mathcal{Z}$ used by \curriculum to mine executable adaptation tasks.

\section{Prompt Templates}
\label{sec:appendix_prompts}

\subsection{Curriculum Generation Prompt}
\label{sec:curriculum_prompt}

We represent a task as $x=(\iota,B,c,v,p)$, where $\iota$ is the instruction, $B$ the estimated step budget, $c$ the core functionality, $v$ the variation type, and $p$ the prerequisites. The \curriculum prompt jointly performs trajectory assessment and task generation. It observes the exploration goal, visualized trajectory screenshots, few-shot examples, generation principles, and existing tasks for the app. The core template is shown below.

\begin{promptlistingbox}[MobileGym-Curriculum core prompt]
You are an expert Curriculum Generator - a teacher designing
comprehensive learning tasks for GUI agents. Your core purpose is to
create a complete curriculum covering all app functionalities with
progressive difficulty levels to systematically teach GUI agents how
to use the target app.

Your job is to:
1. EVALUATE the original task for reasonableness and completion.
2. GENERATE new diverse curriculum tasks that comprehensively cover
   the app's functionality.

## App Information
App Name: {app_name}
Original Task Goal: {original_goal}

## Few-shot Examples
{fewshot_examples}

## Task Generation Principles
{task_principles}

## Already Generated Tasks for {app_name}
{existing_tasks}
IMPORTANT: Do not generate tasks that are too similar to the above.

## Curriculum Design Instructions

### Step 1: Task Evaluation
1. Reasonableness Assessment:
   - Is this a reasonable task that a user might actually want to
     perform in this app?
   - Are the requirements clear and achievable?
   - Does the task make sense in the context of the app?

2. Step-by-Step Quality Analysis:
   - Analyze representative visible steps in the screenshot sequence.
   - A reasonable step logically progresses toward task completion.
   - An unreasonable step is unnecessary, wrong, counterproductive,
     stuck in a loop, or moves backward unnecessarily.
   - Failed trajectories may contain reasonable steps.
   - Successful trajectories may contain detours or inefficiencies.
   - Compute trajectory_quality_score = reasonable_steps / total_steps.

3. Overall Completion Assessment:
   - Did the agent complete the stated task?
   - Were the required steps performed correctly?
   - Did the agent reach the intended goal state?

### Step 2: Curriculum Task Generation
Generate 3-8 new learning tasks that:
- cover different core functionalities of {app_name};
- vary in length from 1 to 40 steps;
- are pedagogically useful for teaching GUI agents;
- avoid redundancy with existing tasks;
- focus on core functionality variations, not only parameter changes;
- limit each functionality to at most 3 parameter variations.

## Output Format
Return JSON:
{
  "evaluation": {
    "task_reasonable": true/false,
    "task_completed": true/false,
    "reasonableness_explanation": "...",
    "completion_explanation": "...",
    "confidence_score": 0.0-1.0,
    "step_quality_analysis": {
      "total_steps": 10,
      "reasonable_steps": 8,
      "trajectory_quality_score": 0.8,
      "quality_summary": "..."
    }
  },
  "generated_tasks": [
    {
      "task_id": "task_1",
      "instruction": "Self-contained learning task",
      "estimated_steps": 5,
      "core_functionality": "Main functionality being taught",
      "variation_type": "simplification/parameter_change/"
                        "scenario_application/step_progression",
      "prerequisites": "Only non-obvious prerequisites"
    }
  ]
}
\end{promptlistingbox}

\subsection{MobileGym-Critic Prompts}
\label{sec:critic_prompts}

\critic uses a hierarchical prompting procedure. The first prompt converts each visualized action into a compact step description. The second prompt makes the final trajectory decision and step-level process assessment. The third prompt turns failures or inefficient steps into corrective hints for later attempts.

\begin{promptlistingbox}[MobileGym-Critic step description prompt]
System:
You are an expert mobile device assistant. Analyze a two-panel image
showing the Before Action and After Action state of a user's workflow.
Focus only on the Before Action panel. Output JSON.

User:
The overall task is: {task_description}

The image shows a before/after action state. The user action is
visualized with markers, and the raw action log is:
- Action Type: {log_action}
- Action Detail: {log_detail}

Return JSON:
{
  "action_description": "A crisp description of the action performed",
  "ui_description": "Task-relevant UI elements visible before the action"
}
\end{promptlistingbox}

\begin{promptlistingbox}[MobileGym-Critic final decision prompt]
System:
You are an expert in evaluating mobile UI automation tasks.
Use the evaluation guidelines:
- The final UI state must satisfy all task requirements.
- Existing conditions can satisfy requirements without repetition.
- A logically correct action sequence can imply success even if the
  last screen alone is insufficient.
- The agent may make and correct mistakes.
- If a task is inherently infeasible, the agent can still succeed by
  correctly identifying the impossibility and giving appropriate feedback.

User:
Task Description: {task_description}

Here is a step-by-step breakdown of the agent's actions, including raw
action logs and VLM-generated descriptions:
{step_descriptions_with_raw_logs}

You are also provided with a composite image of the last screenshots.
Synthesize the visual evidence with the full step list.

Required assessment:
1. Decide whether the attempt completed the task.
2. Assess whether the task itself is feasible.
3. If failed, identify the failure_step.
4. Analyze every step for reasonableness and provide a concise rationale.

Return JSON:
{
  "decision": 1 or 0,
  "reason": "Explanation of the final judgment",
  "failure_step": 4,
  "task_feasible": true/false,
  "task_feasible_reason": "Why the task is feasible or infeasible",
  "task_barriers": [],
  "reasonable_steps": [1, 2, 4],
  "unreasonable_steps": [3],
  "step_analysis": {
    "1": {
      "reasonableness": "reasonable",
      "explanation": "..."
    },
    "3": {
      "reasonableness": "unreasonable",
      "explanation": "..."
    }
  }
}
\end{promptlistingbox}

\begin{promptlistingbox}[MobileGym-Critic corrective hint prompt]
System:
You are an intelligent agent performing self-reflection on your task
execution. Analyze your own actions, identify mistakes and
inefficiencies, and extract lessons for future attempts. Use the
step-by-step reasonableness analysis. Output JSON.

User:
## Task
{task_description}

## My Steps (with Reasonableness Analysis)
{step_descriptions_with_step_analysis}

## Evaluation Result
{failure_or_inefficiency_reason}

Reflect on the execution:
1. Identify key mistakes, especially unreasonable steps.
2. Specify what to avoid in future attempts.
3. Propose concrete alternative approaches.
4. Extract important task insights.

Return JSON:
{
  "key_mistake": "Concise summary of the main mistake",
  "what_to_avoid": ["..."],
  "suggested_approach": ["..."],
  "important_insights": ["..."],
  "hint_summary": "Brief self-reminder for the next attempt"
}
\end{promptlistingbox}

\subsection{Hint-Guided Rollout Prompts}
\label{sec:rollout_prompts}

During \hifpo rollout, the hint context $\eta_{<k}$ is appended to the task instruction before attempt $k$. The same mechanism is used for both base agents; the difference lies in each agent's native step prompt. The shared hint block has the following structure.

\begin{promptlistingbox}[Corrective hint context block]
EVALUATION HINTS FROM PREVIOUS ATTEMPTS
The following insights come from analysis of previous failed attempts.
Please review these carefully to avoid repeating the same mistakes:

--- Hint from Attempt #{attempt_id} ---
Summary: {hint_summary}
Key Mistake: {key_mistake}
What to Avoid:
  - {avoid_item_1}
  - {avoid_item_2}
Suggested Approach:
  - {approach_item_1}
  - {approach_item_2}
Important Insights:
  - {insight_item_1}

END OF EVALUATION HINTS
\end{promptlistingbox}

For Qwen3-VL, the structured hint block is inserted into the query field, while the screenshot remains the visual observation for the current step.

\begin{promptlistingbox}[Qwen3-VL hint-contextualized rollout prompt]
System:
You are a helpful assistant that can help with tasks on a mobile device.
Use the mobile_use tool with actions such as click, long_press, swipe,
type, answer, system_button, wait, and terminate.
Return exactly:
  <thinking>...</thinking>
  <tool_call>{"name": "mobile_use", "arguments": {...}}</tool_call>
  <conclusion>UI observation: ... Intended action: ...</conclusion>
If the task is completed, terminate with status="success"; if it is
infeasible, terminate with status="failure". Trust the current UI state
over action history. If hints are present in the task description, use
them to avoid repeating previous mistakes.

User:
The user query: {task_instruction}
{EVALUATION_HINTS_FROM_PREVIOUS_ATTEMPTS}

Task progress (You have done the following operation on the current
device):
{previous_step_conclusions}
<image>
\end{promptlistingbox}

For GUI-Owl, the hint block is likewise appended to the instruction, while the prompt preserves GUI-Owl's concise action-plus-tool-call format.

\begin{promptlistingbox}[GUI-Owl hint-contextualized rollout prompt]
System:
You may use the mobile_use tool with actions such as key, click,
long_press, swipe, type, system_button, open, wait, answer, and
terminate. For each step, output:
  Action: {short imperative}
  <tool_call>{"name": "mobile_use", "arguments": {...}}</tool_call>

User:
Please generate the next move according to the UI screenshot,
instruction and previous actions.

Instruction: {task_instruction}
{EVALUATION_HINTS_FROM_PREVIOUS_ATTEMPTS}

Previous actions:
{action_history_or_no_previous_action}
<image>
\end{promptlistingbox}

\section{Adaptive GUI Action Reward}
\label{sec:appendix_reward}

The step-level GRPO stage uses a rule-based GUI action reward. For each hint-contextualized prompt, the policy samples a group of responses. Each response is first parsed into a structured action $\hat a=(\hat\alpha,\hat\psi)$ and compared against the selected action $a^\star=(\alpha^\star,\psi^\star)$ extracted from hierarchical feedback. The parser supports the output templates used by both base agents and maps action aliases into a canonical mobile action space before scoring.

The optimized scalar reward is
\begin{equation}
\label{eq:appendix_reward}
R(\hat a,a^\star)
=\lambda_{\rm type}r_{\rm type}
+\lambda_{\rm arg}r_{\rm arg},
\end{equation}
where $\lambda_{\rm type}$ and $\lambda_{\rm arg}$ are configurable weights. The implementation also logs a binary format score for malformed tool calls, but this format score is not included in $R$. The type score is
\begin{equation}
\label{eq:appendix_type_reward}
r_{\rm type}=\mathbb{I}[\hat\alpha=\alpha^\star].
\end{equation}
The argument score is gated by the type score:
\begin{equation}
\label{eq:appendix_arg_reward}
r_{\rm arg}=
\begin{cases}
S_{\alpha^\star}(\hat\psi,\psi^\star), & r_{\rm type}=1,\\
0, & r_{\rm type}=0.
\end{cases}
\end{equation}
Thus a response with the wrong action type receives no parameter credit. The main experiments use the reward weights reported in Section~\ref{sec:experiments}; the same reward implementation exposes these weights as hyperparameters.

\begin{table}[t]
\centering
\paperTable
\caption{Rule-based argument score $S_{\alpha}$ used by the adaptive GUI action reward.}
\label{tab:adaptive_gui_reward}
\begin{tabular}{p{0.20\linewidth} p{0.70\linewidth}}
\toprule
\textbf{Action} & \textbf{Argument score} \\
\midrule
click & If the target is a box, reward is $1$ when the predicted point is inside the box; otherwise it decays with distance to the box center normalized by the box diagonal. If the target is a point, reward is $\max(0,1-d/50)$ in the normalized 0--1000 coordinate space. \\
long press & Same coordinate score as click. \\
swipe & If a target direction is provided, reward is $1$ only when the predicted direction matches. If no target direction is provided, a valid swipe parameter receives full credit. \\
type & Token-level F1 similarity between predicted text and target text. \\
answer & Token-level F1 similarity when a target answer is provided; otherwise full credit after the action type is correct. \\
system button & Exact match of the system button identity, such as Back, Home, Menu, or Enter. \\
wait & Full credit after the action type is correct; no additional argument is required by the training target. \\
terminate & Exact match of the termination status when one is provided; otherwise full credit after the action type is correct. \\
open & Exact app-name match receives full credit; containment-based partial app-name match receives partial credit. If no target app name is provided, the argument score is treated as satisfied. \\
key & Exact key-name match when one is provided; otherwise full credit after the action type is correct. \\
\bottomrule
\end{tabular}
\end{table}

\section{Training Details}
\label{sec:appendix_training_details}

\begin{figure*}[!t]
\centering
\begin{subfigure}{0.32\textwidth}
  \includegraphics[width=\linewidth]{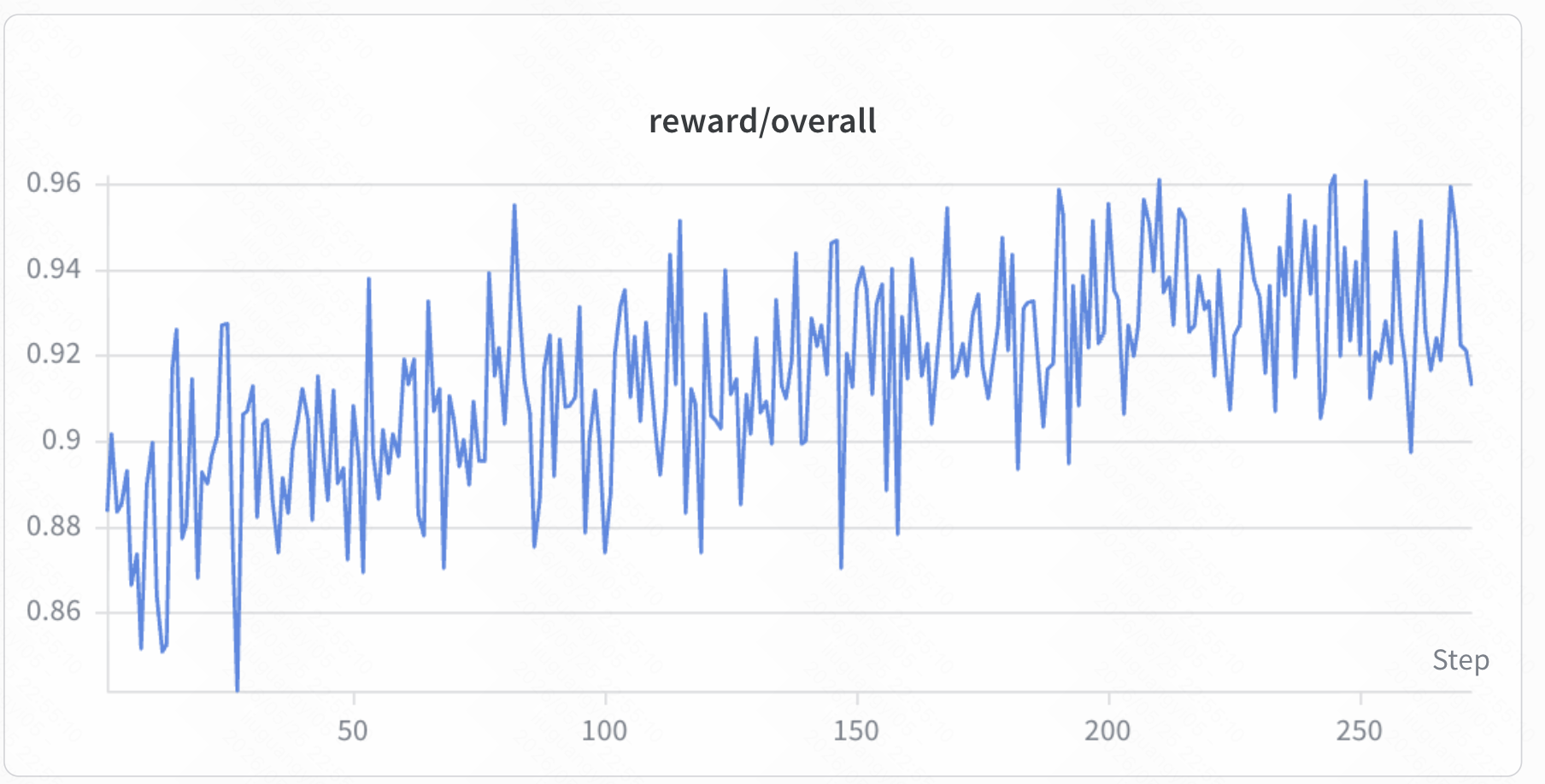}
  \caption{\llmname{Qwen3-VL-8B}: overall}
\end{subfigure}
\hfill
\begin{subfigure}{0.32\textwidth}
  \includegraphics[width=\linewidth]{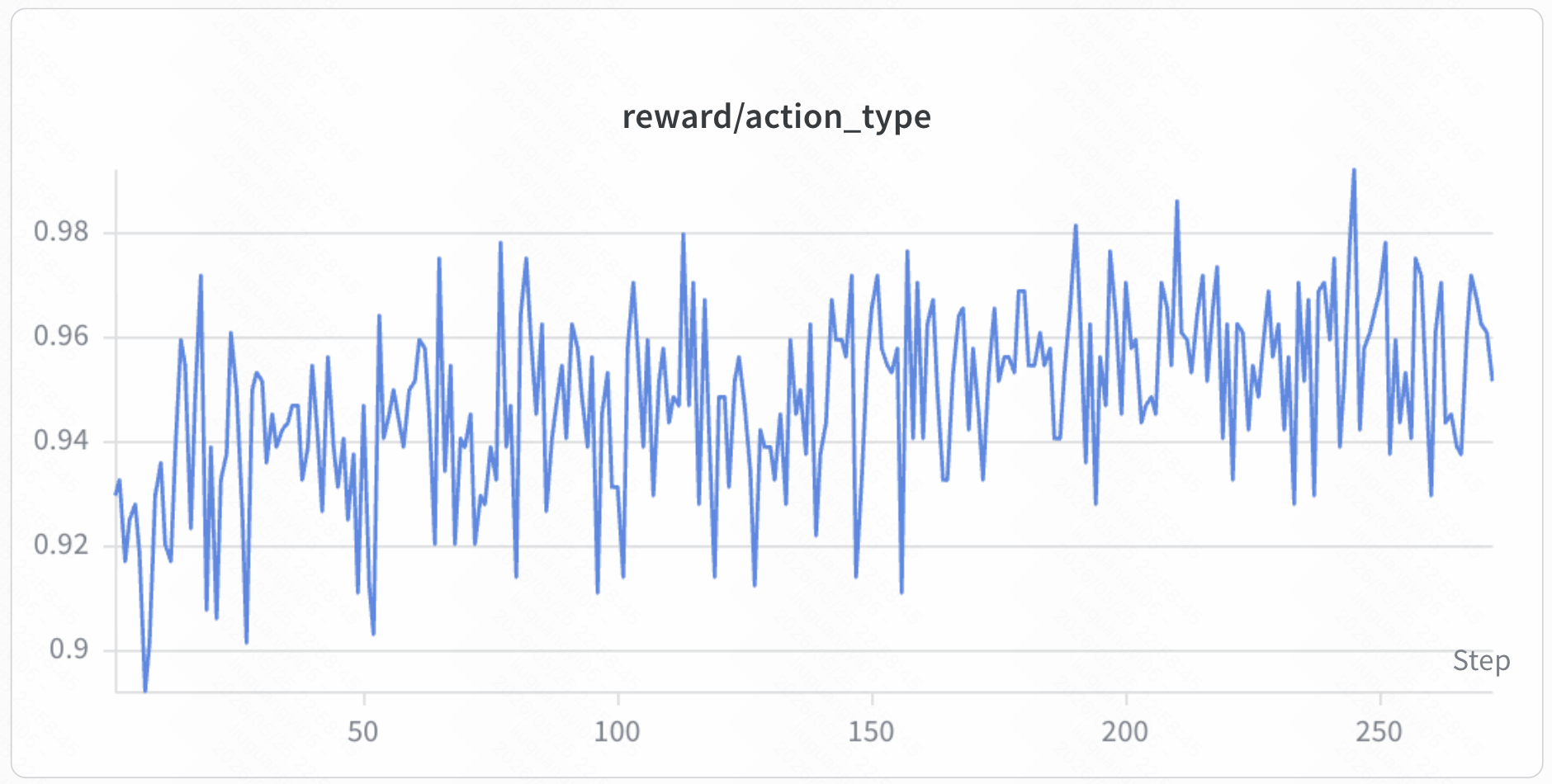}
  \caption{\llmname{Qwen3-VL-8B}: action type}
\end{subfigure}
\hfill
\begin{subfigure}{0.32\textwidth}
  \includegraphics[width=\linewidth]{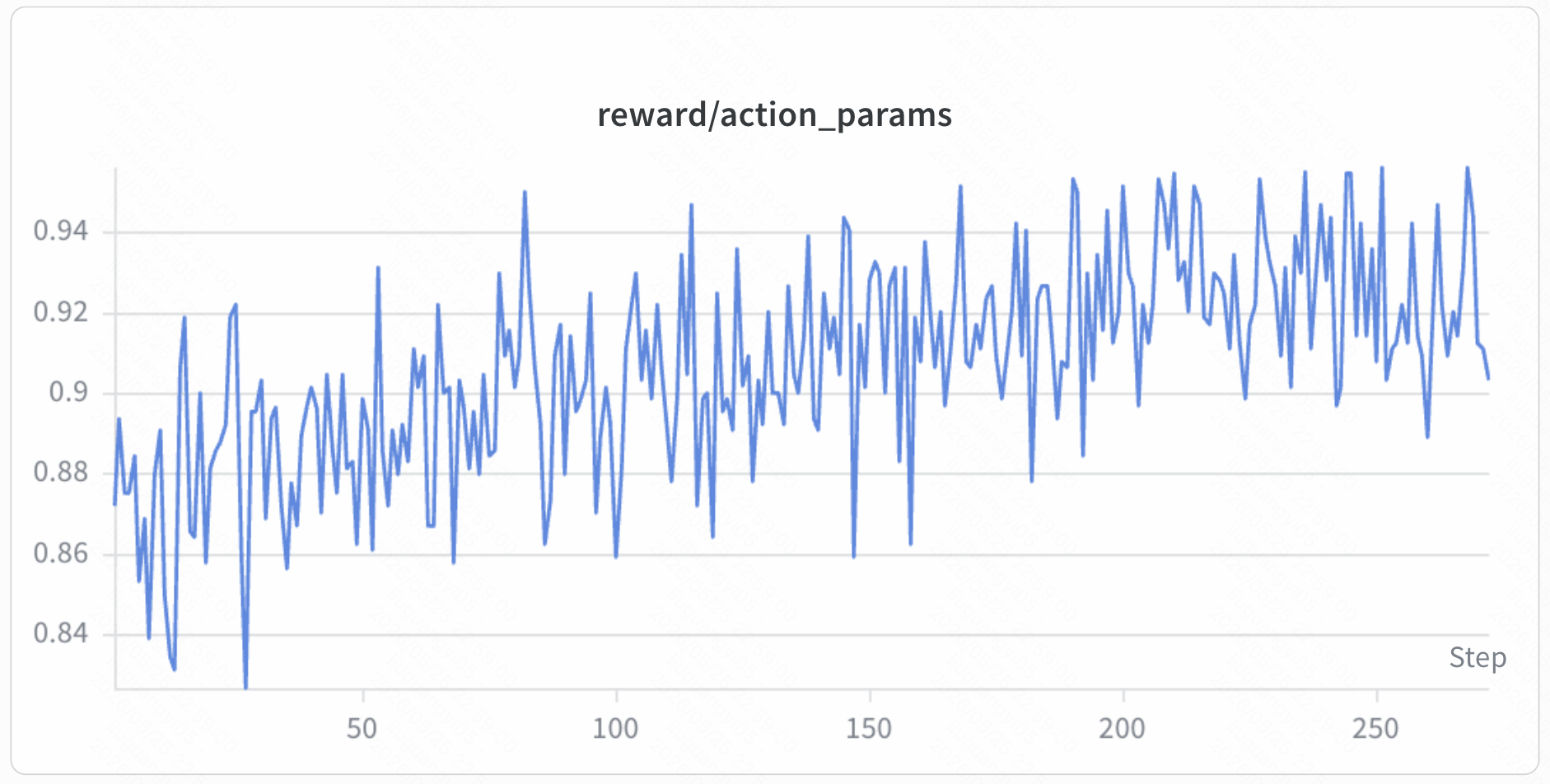}
  \caption{\llmname{Qwen3-VL-8B}: arguments}
\end{subfigure}

\vspace{2mm}
\begin{subfigure}{0.32\textwidth}
  \includegraphics[width=\linewidth]{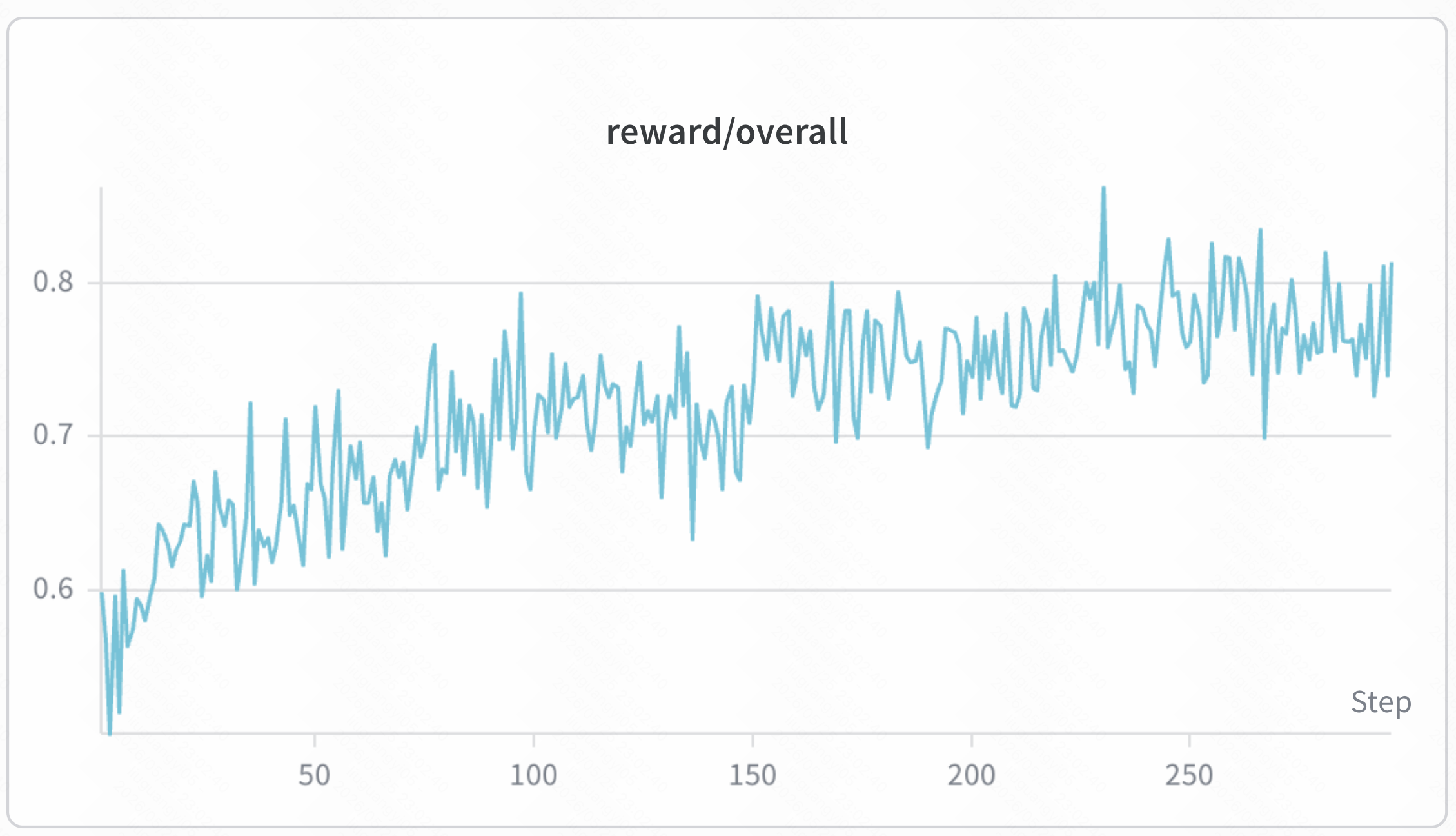}
  \caption{\llmname{GUI-Owl-1.5-8B}: overall}
\end{subfigure}
\hfill
\begin{subfigure}{0.32\textwidth}
  \includegraphics[width=\linewidth]{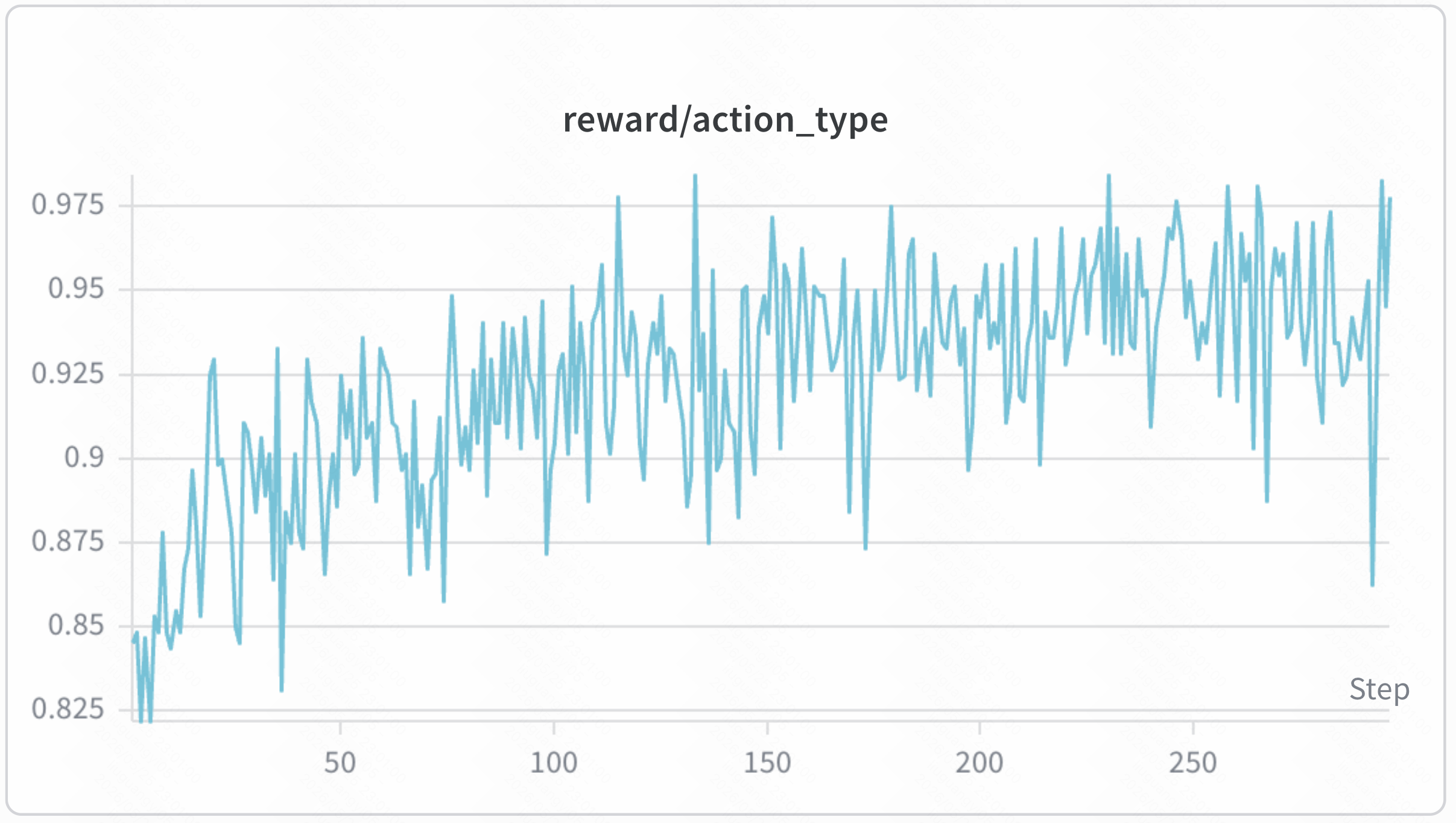}
  \caption{\llmname{GUI-Owl-1.5-8B}: action type}
\end{subfigure}
\hfill
\begin{subfigure}{0.32\textwidth}
  \includegraphics[width=\linewidth]{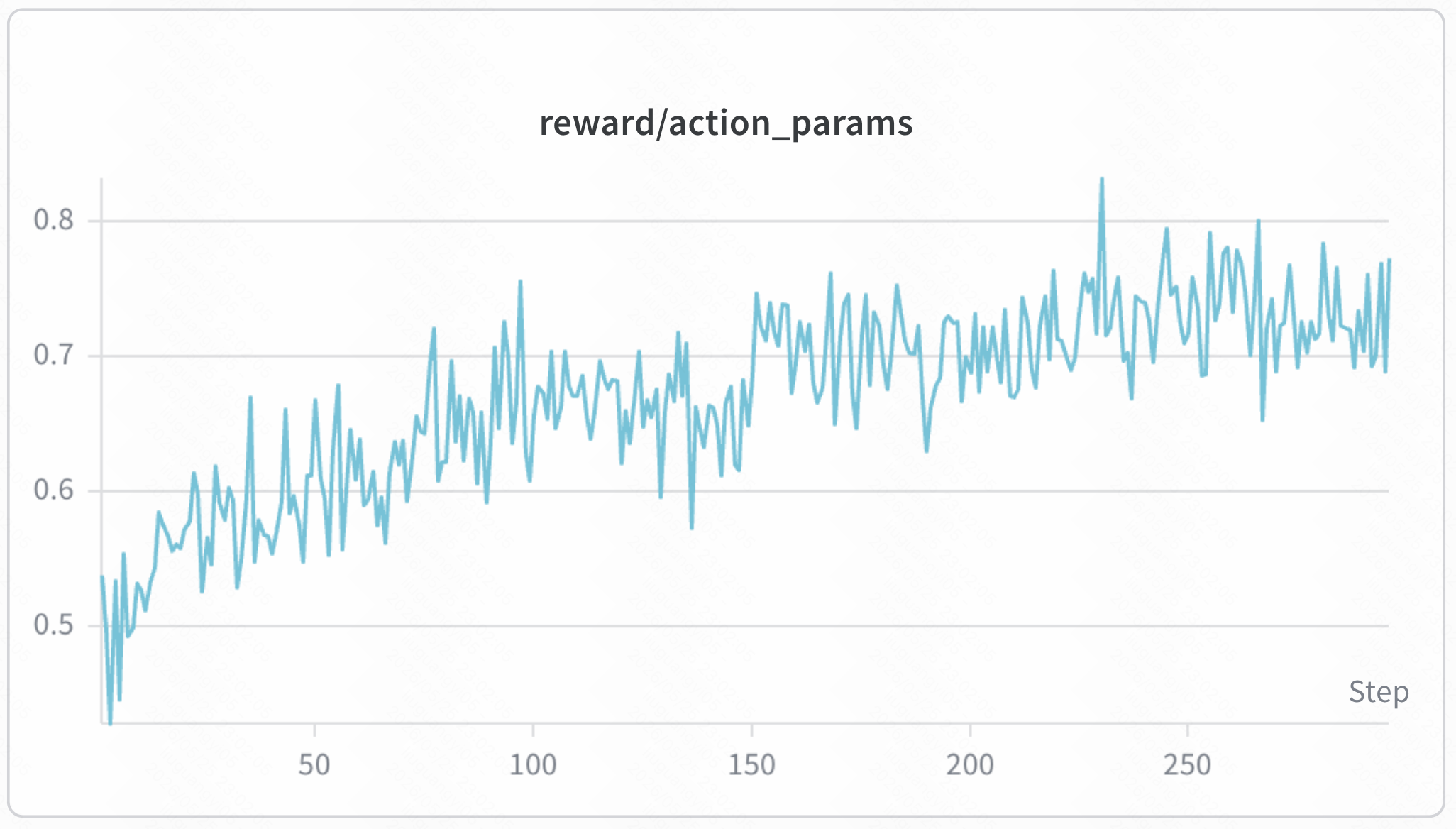}
  \caption{\llmname{GUI-Owl-1.5-8B}: arguments}
\end{subfigure}
\caption{Training reward curves for the 900-task \hifpo runs. The first row reports \llmname{Qwen3-VL-8B} and the second reports \llmname{GUI-Owl-1.5-8B}; columns show overall, action-type, and action-argument rewards. The overall reward combines the latter two with weights 0.2 and 0.8.}
\label{fig:training_curves}
\end{figure*}

Table~\ref{tab:training_details} lists the main hyperparameters and hardware setting used for the 900-task 8B runs. Figure~\ref{fig:training_curves} reports the corresponding reward curves. The curves show that both models learn the action-argument component throughout training; \llmname{GUI-Owl-1.5-8B} starts from a lower overall reward but improves steadily, while \llmname{Qwen3-VL-8B} starts high and receives smaller but still positive reward gains.

\begin{table*}[!t]
\centering
\paperTable
\caption{Main \hifpo training configuration for the 900-task 8B runs. The same configuration is used for both base agents unless noted otherwise.}
\label{tab:training_details}
\begin{tabular}{L{0.19\textwidth} L{0.73\textwidth}}
\toprule
\textbf{Category} & \textbf{Setting} \\
\midrule
Hardware and runtime & 8 x 80GB GPUs; approximately 80 hours for a 900-task 8B run. \\
Training data & 900 generated AndroidWorld-side tasks for the main runs; held-out validation data is not used for MobileWorld adaptation. \\
Sequence length & Maximum prompt length 2048 tokens; maximum response length 2048 tokens; overlong prompts are filtered. \\
Filtering & Success-rate range $[0.0,0.9]$; best-trajectory filtering enabled; mastered tasks removed; corrective hints retained. \\
GRPO rollout & 5 sampled responses per step-level prompt; temperature 1.0; top-$p$ 1.0; tensor parallel size 2. \\
Optimization & 4 epochs; global batch size 128; rollout batch size 512; AdamW with learning rate $1.0\times10^{-6}$, weight decay $1.0\times10^{-2}$, and max gradient norm 1.0. \\
KL regularization & GRPO advantage estimator; KL loss enabled with low-variance KL penalty and coefficient $1.0\times10^{-2}$. \\
Model training & Vision tower is not frozen; gradient checkpointing and FSDP full-shard training are enabled; parameters and optimizer states are offloaded. \\
Reward weights & Action-type reward weight 0.2; action-argument reward weight 0.8. \\
Validation & Validation every 50 steps; greedy validation decoding with temperature 0 and one response per prompt. \\
\bottomrule
\end{tabular}
\end{table*}

\paragraph{Training dynamics.}
The component curves clarify where the gains arise. For both agents, action-type reward remains comparatively high, whereas action-argument reward provides most of the sustained improvement; with an argument weight of 0.8, the overall curve therefore tracks parameter quality closely. The pronounced step-to-step variation reflects the heterogeneous states and action schemas represented by the filtered training prompts rather than a single repeated behavior. Despite this variation, all three rewards maintain an upward trend and show no late-stage collapse. This pattern is consistent with stable optimization under the selected filtering and KL settings, and it indicates that \hifpo supplies useful learning signals to both a generalist VLM and a GUI-specialized base.

\paragraph{Reproducibility protocol.}
Both 8B base agents use the same configuration. Each main run trains for four epochs on 900 generated AndroidWorld-side tasks; no MobileWorld data enters adaptation. Prompts and responses are capped at 2,048 tokens, and overlong prompts are filtered. Training retains corrective hints and best trajectories while excluding mastered tasks. Each retained step-level prompt yields five GRPO responses at temperature 1.0 and top-$p$ 1.0; action-type and action-argument rewards use weights 0.2 and 0.8. Updates use AdamW and a low-variance KL penalty with coefficient $1.0\times10^{-2}$. The vision tower remains trainable under gradient checkpointing and FSDP full sharding.

\paragraph{Controlled comparison.}
The two main runs share the same generated-task budget, filtering rules, rollout group size, optimizer, reward weights, and validation cadence. This holds the adaptation pipeline fixed while changing only the initial policy, so the curves reflect how the generalist and GUI-specialized bases respond to the same \hifpo signals. Validation runs every 50 steps with greedy decoding and one response per prompt. MobileWorld is fully held out from adaptation and checkpoint selection.

\clearpage

\subsection{Track-Completion Cases}
\label{sec:appendix_track_completion_cases}

\begin{figure*}[!p]
\centering
\begin{subfigure}[t]{0.49\textwidth}
  \centering
  \includegraphics[width=\linewidth,height=0.68\textheight,keepaspectratio]{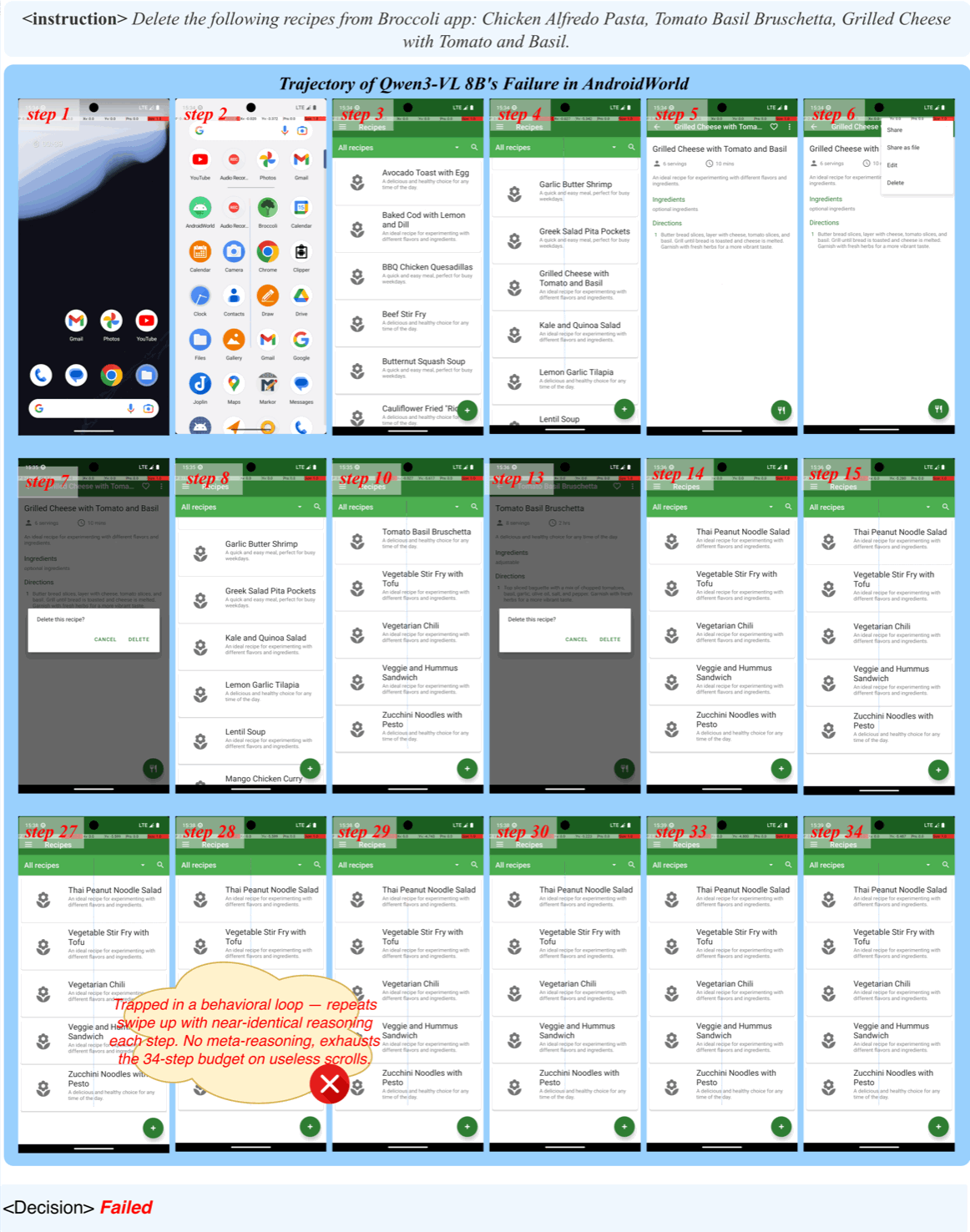}
  \caption{\llmname{Qwen3-VL-8B} base: failed}
\end{subfigure}
\hfill
\begin{subfigure}[t]{0.49\textwidth}
  \centering
  \includegraphics[width=\linewidth,height=0.68\textheight,keepaspectratio]{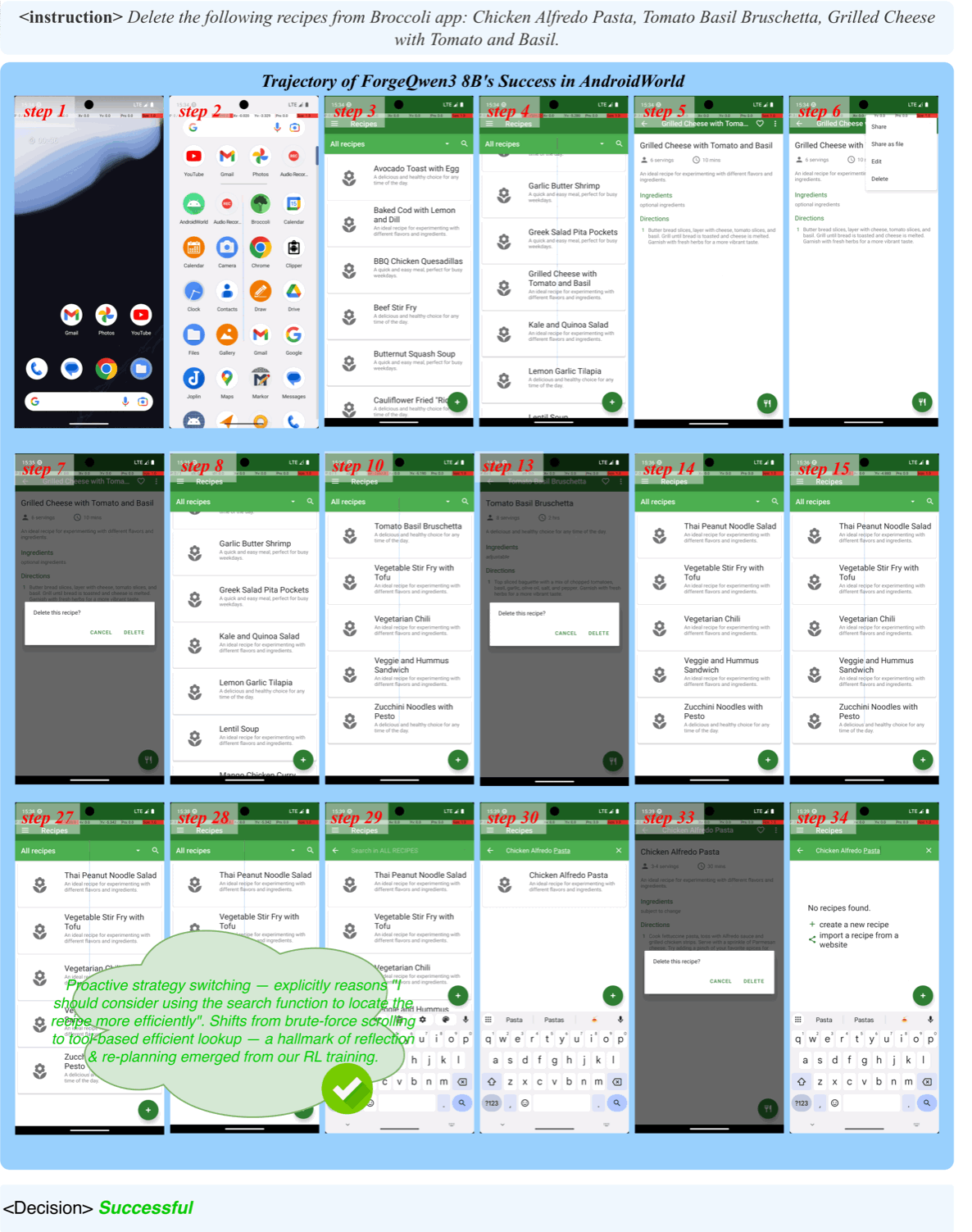}
  \caption{\llmname{ForgeQwen3-8B}: successful}
\end{subfigure}
\caption{\textbf{AndroidWorld track-completion comparison for \llmname{Qwen3-VL-8B}.}
On the same Broccoli recipe-deletion task, the base model falls into repeated scrolling after partial progress, while \ourmethod adaptation enables \llmname{ForgeQwen3-8B} to switch to a targeted search strategy and complete the remaining deletion.}
\label{fig:track_completion_androidworld_qwen}
\end{figure*}

\begin{figure*}[!p]
\centering
\begin{subfigure}[t]{0.49\textwidth}
  \centering
  \includegraphics[width=\linewidth,height=0.68\textheight,keepaspectratio]{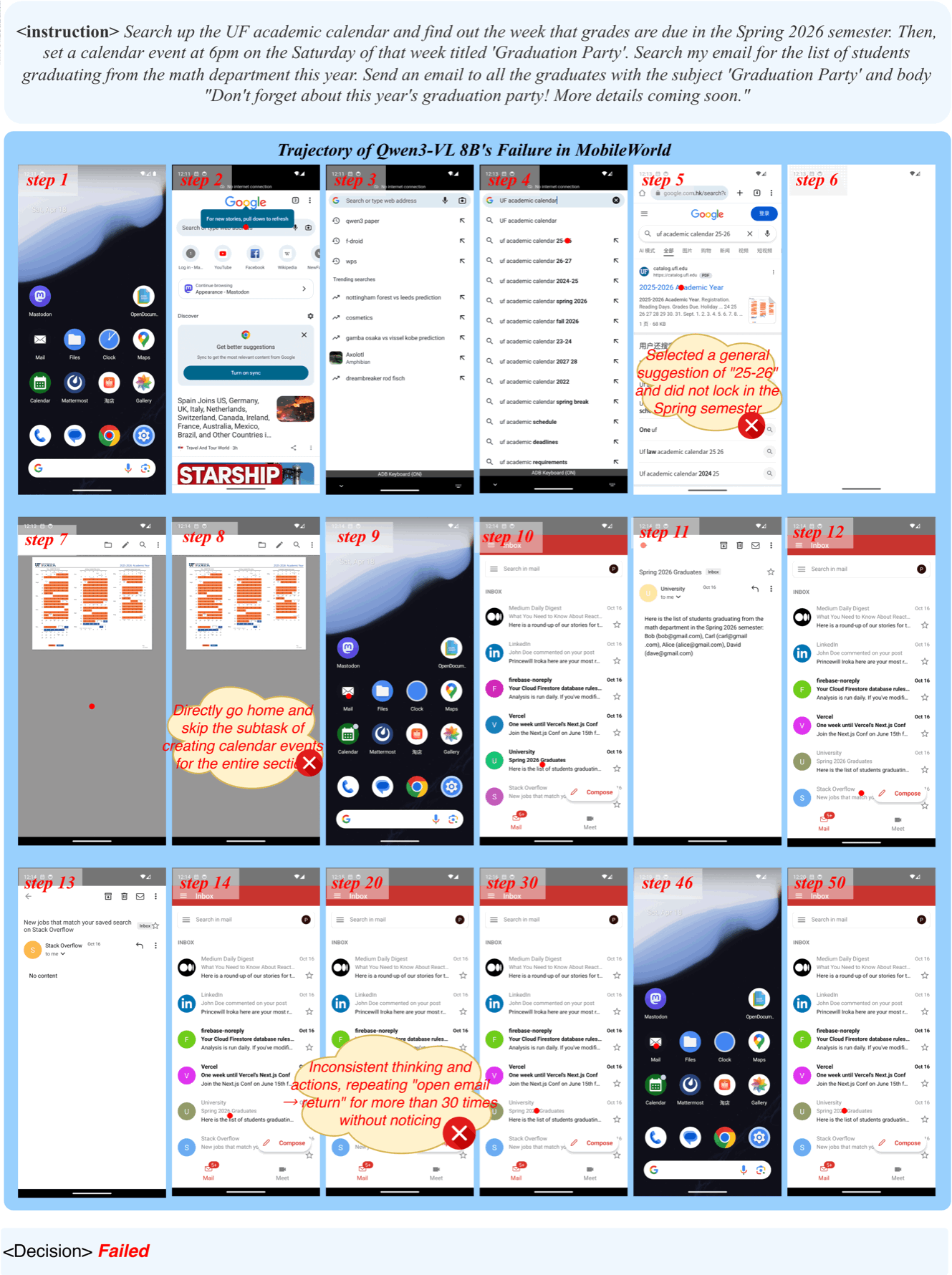}
  \caption{\llmname{Qwen3-VL-8B} base: failed}
\end{subfigure}
\hfill
\begin{subfigure}[t]{0.49\textwidth}
  \centering
  \includegraphics[width=\linewidth,height=0.68\textheight,keepaspectratio]{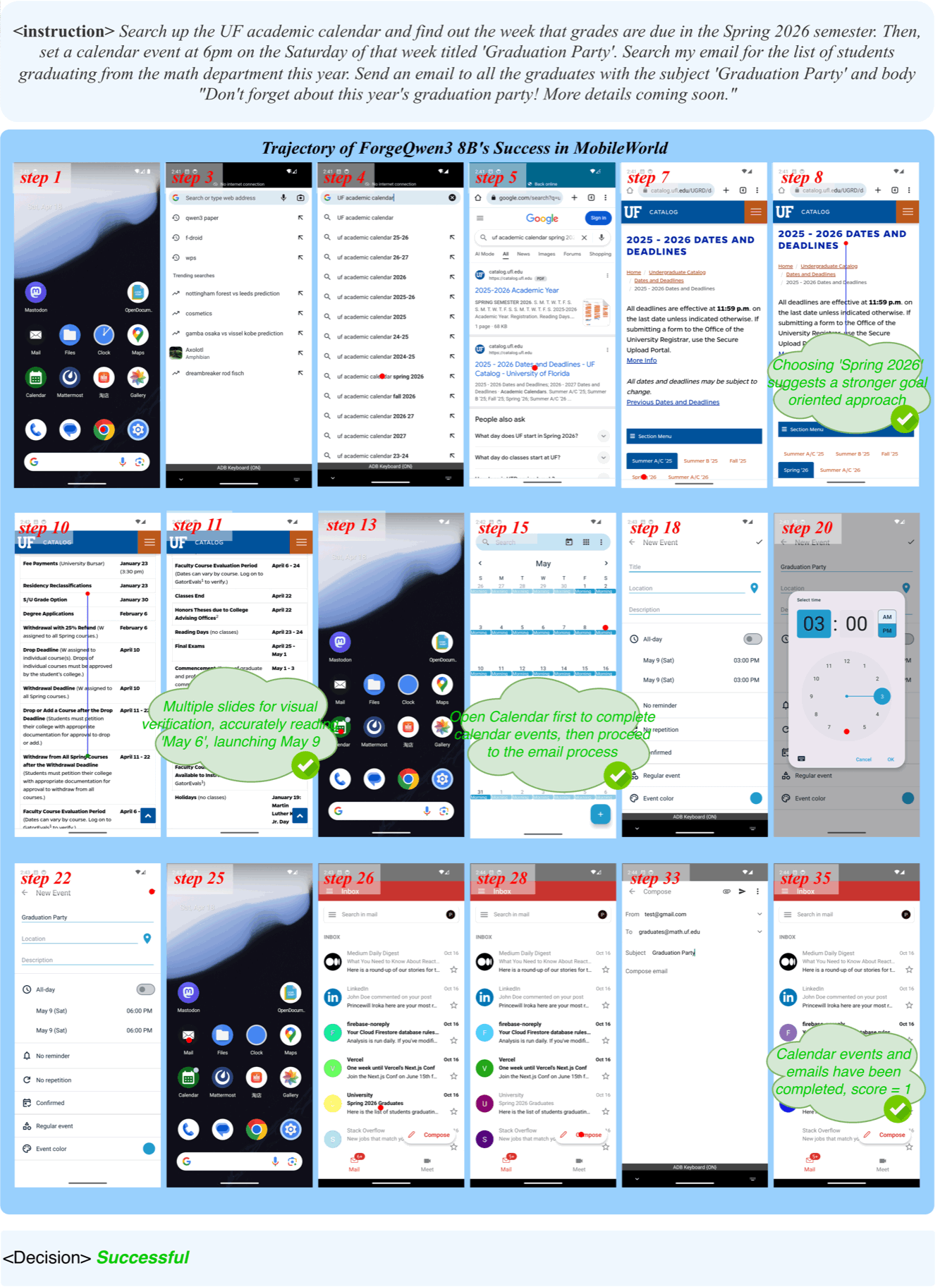}
  \caption{\llmname{ForgeQwen3-8B}: successful}
\end{subfigure}
\caption{\textbf{MobileWorld track-completion comparison for \llmname{Qwen3-VL-8B}.}
On the same multi-app academic-calendar and email task, the base model selects an underspecified semester result and then skips the calendar subtask, while \llmname{ForgeQwen3-8B} verifies the Spring 2026 deadline window, creates the calendar event, and proceeds to the email workflow.}
\label{fig:track_completion_mobileworld_qwen}
\end{figure*}

\begin{figure*}[!p]
\centering
\begin{subfigure}[t]{0.49\textwidth}
  \centering
  \includegraphics[width=\linewidth,height=0.68\textheight,keepaspectratio]{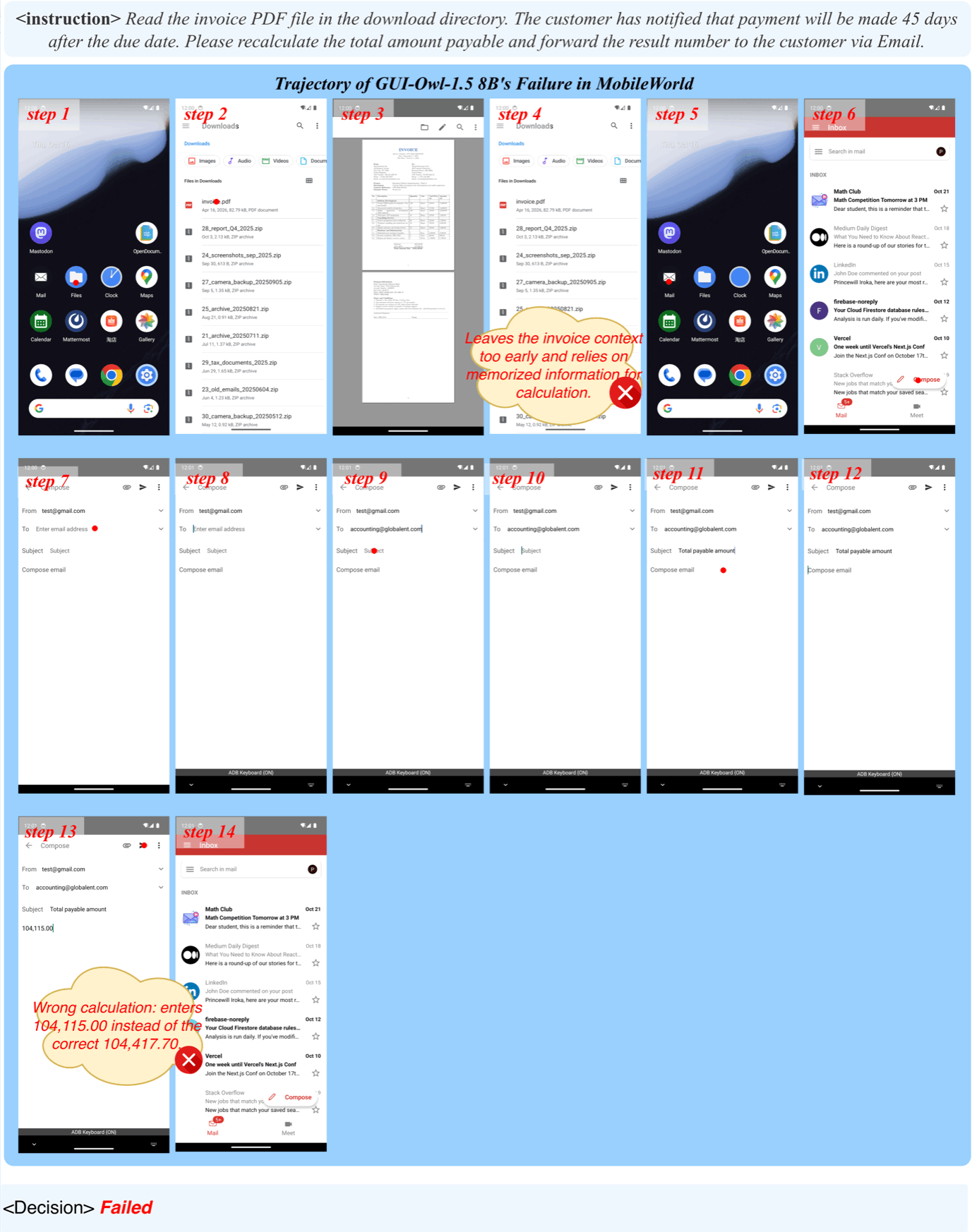}
  \caption{\llmname{GUI-Owl-1.5-8B} base: failed}
\end{subfigure}
\hfill
\begin{subfigure}[t]{0.49\textwidth}
  \centering
  \includegraphics[width=\linewidth,height=0.68\textheight,keepaspectratio]{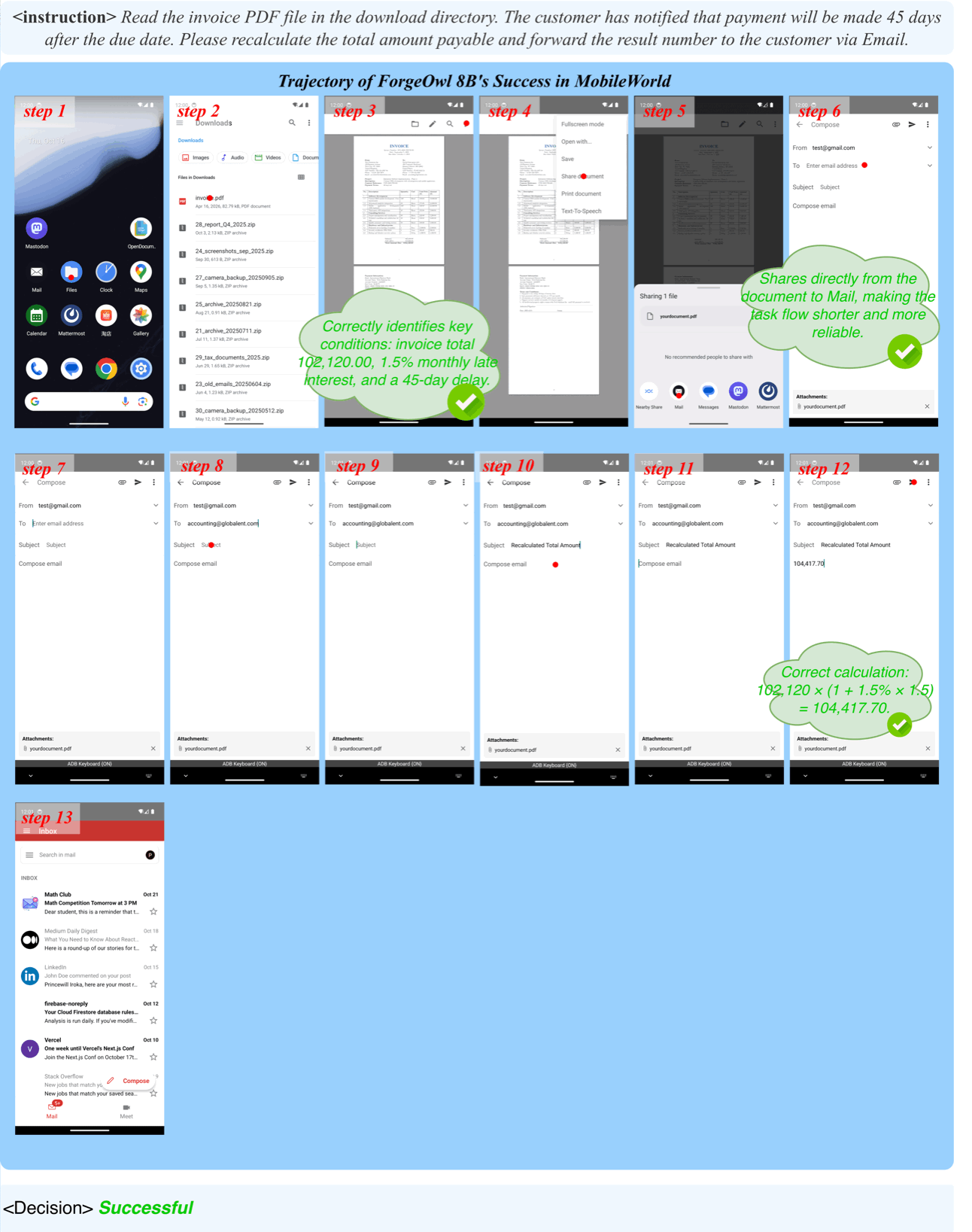}
  \caption{\llmname{ForgeOwl-8B}: successful}
\end{subfigure}
\caption{\textbf{MobileWorld track-completion comparison for \llmname{GUI-Owl-1.5-8B}.}
On the same invoice recalculation and email task, the base model leaves the invoice context too early and sends an incorrect amount, while \llmname{ForgeOwl-8B} preserves the key invoice conditions, shares the document into Mail, and sends the correct recalculated total.}
\label{fig:track_completion_mobileworld_guiowl}
\end{figure*}

\end{document}